\documentclass[fleqn,usenatbib]{mnras}

\usepackage{newtxtext,newtxmath} 
\usepackage[T1]{fontenc} 
\usepackage{ae,aecompl} 
\usepackage{graphicx}	
\usepackage{amsmath}	
\usepackage{amssymb}	
\usepackage{breqn}  
\usepackage{wasysym} 
\usepackage{cleveref}
\crefname{section}{\S}{\S\S}

\usepackage{etoolbox}
\makeatletter
\patchcmd\@combinedblfloats{\box\@outputbox}{\unvbox\@outputbox}{}{\errmessage{\noexpand patch failed}}
\makeatother

\title[L-GALAXIES vs. TNG]{Comparing galaxy formation in the L-GALAXIES semi-analytical model and the IllustrisTNG simulations}

\author[M. Ayromlou et al.]{Mohammadreza Ayromlou,$^{1}$\thanks{E-mail: ayromlou@mpa-garching.mpg.de}
Dylan Nelson,$^{1}$
Robert M. Yates,$^{1}$ 
\newauthor
Guinevere Kauffmann,$^{1}$
Malin Renneby,$^{2,3}$
Simon D. M. White$^{1}$
\\\\
$^{1}$Max Planck Institute for Astrophysics, Karl-Schwarzschild-Str. 1, 85741 Garching bei M{\"u}nchen, Germany\\
$^{2}$HEP/ALCF Divisions, Argonne National Laboratory, 9700 S. Cass Avenue, Lemont, IL 60439, USA\\
$^{3}$Kavli Institute for Cosmological Physics, University of Chicago, Chicago, IL 60637, USA
}

\date{}
\pubyear{2020}

\begin{document}
\label{firstpage}
\pagerange{\pageref{firstpage}--\pageref{lastpage}}
\maketitle

\begin{abstract}
We perform a comparison, object-by-object and statistically, between the Munich semi-analytical model, \textsc{L-Galaxies}, and the IllustrisTNG hydrodynamical simulations. By running \textsc{L-Galaxies} on the IllustrisTNG dark matter-only merger trees, we identify the same galaxies in the two models. This allows us to compare the stellar mass, star formation rate and gas content of galaxies, as well as the baryonic content of subhaloes and haloes in the two models. We find that both the stellar mass functions and the stellar masses of individual galaxies agree to better than $\sim0.2\,$dex. On the other hand, specific star formation rates and gas contents can differ more substantially. At $z=0$ the transition between low-mass star-forming galaxies and high-mass, quenched galaxies occurs at a stellar mass scale $\sim0.5\,$dex lower in IllustrisTNG than in \textsc{L-Galaxies}. IllustrisTNG also produces substantially more quenched galaxies at higher redshifts. Both models predict a halo baryon fraction close to the cosmic value for clusters, but IllustrisTNG predicts lower baryon fractions in group environments. These differences are due primarily to differences in modelling feedback from stars and supermassive black holes. The gas content and star formation rates of galaxies in and around clusters and groups differ substantially, with IllustrisTNG satellites less star-forming and less gas-rich. We show that environmental processes such as ram-pressure stripping are stronger and operate to larger distances and for a broader host mass range in IllustrisTNG. We suggest that the treatment of galaxy evolution in the semi-analytic model needs to be improved by prescriptions which capture local environmental effects more accurately.
\end{abstract}

\begin{keywords}
galaxies: formation -- galaxies: evolution -- large-scale structure of Universe -- methods: analytical -- methods: numerical
\end{keywords}

\section{Introduction}
\label{sec: introduction}

In the framework of $\Lambda$CDM, the Universe is described with a set of only six parameters which can be derived from observations of the cosmic microwave background  \citep[e.g.][]{komatsu2011seven,planck2015_xiii}. The subsequent formation and growth of cosmological structures, i.e. haloes and galaxies, is a rich and evolving topic. Although the evolution of the cold dark matter (CDM) component occurs due only to gravitational interactions, modelling the relevant physics of baryonic matter, i.e. stars, gas, and supermassive black holes, requires a more complex and comprehensive theory. 

The standard picture of galaxy formation postulates that baryonic matter falls into the gravitational potential wells of forming dark matter haloes, then cools and collapses to form stars, ultimately leading to the formation of galaxies \citep{white1978core,white1991galaxy}. Theoretical and observational studies have highlighted the many physical processes affecting the formation and evolution of galaxies \citep[see][for reviews]{mo2010galaxy,benson2010galaxy,somerville15}. The complex interplay of the physical phenomena imply that it is impossible to derive the properties of galaxies with a purely analytical approach. As a result, numerical and semi-numerical approaches such as semi-analytical modelling \citep{kauffmann1993formation,somerville1999semi,cole2000hierarchical,springel2001populating,de2006formation} or hydrodynamical simulations \citep{hernquist89,gottlober07,Schaye2015eagle,dave17} are required.

A semi-analytical model (SAM) is a phenomenological tool that uses a set of simplified equations to account for the key baryonic physical processes involved in formation and evolution of galaxies. Early SAMs that were coupled to merger trees derived using analytical approaches such as Press-Schechter \citep[PS,][]{press_Schechte1974} and extended PS formalism \citep{Bond1991,Sheth_Mo_Tormen2001} were able to produce galaxy populations with properties comparable with observations \citep{kauffmann1993formation,kauffmann1999clustering,somerville1999semi,cole2000hierarchical}. Improved simulation techniques \citep{Springel2001Gadget,Springel2005Gadget} and the completion of larger dark matter only (DMO) simulations such as Millennium \citep{springel2005simulations} enabled new generations of modern SAMs to run on halo merger trees from those DMO simulations \citep{de2006formation,croton2006many}. Today, most SAMs run on top of merger trees generated from simulations and implement a wide variety of physical processes including gas cooling, disc and bulge formation, stellar and black hole feedback and environmental effects \citep{Guo2013Galaxy,lacey2016unified,croton2016semi,lagos2018shark,cora2018semi,henriques2020galaxies}.

Hydrodynamical simulations offer a sophisticated methodology to model galaxy formation and evolution. They solve the equations of gravity, (magneto)hydrodynamics, and thermodynamics for dark matter, gas, and stars \citep{teyssier02,springel2010pur,bryan14,hopkins15,wadsley17}. The treatment of gas is one of the fundamental differences between hydrodynamical simulations and SAMs. In a hydrodynamical simulation the cosmic baryon fluid is modelled through a numerical, discretised solution to the equations of (magneto)hydrodynamics. Therefore, the distribution of gas mass and its thermodynamical properties is a direct outcome of the simulation.

On the other hand, in a semi-analytical model, the gas is partitioned into several discrete components such as: hot, cold and ejected reservoirs. Gas mass is exchanged between these components, but its spatial distribution (and other properties, such as temperature or kinematic structure) is unspecified, and must be derived from simplified models if required \citep{Henriques2013Simulations,Yates+17}. Recently, semi-analytical models have begun to spatially resolve galaxies by discretising their gas and stellar discs into radial 
rings \citep{henriques2020galaxies}, probing down to sub-kpc scales \citep[see also][]{fu13,stevens16}.

The drawback of hydrodynamical simulations is their computational expense, and large projects such as IllustrisTNG can take tens of millions of CPU cores hours to complete \citep[see e.g.][]{nelson2019First}. In contrast, SAMs are computationally much less expensive, typically orders of magnitude faster than hydrodynamical simulations. Given their relative strengths and weaknesses, it is useful to compare and contrast their results in order to make simultaneous improvements on both techniques.

These two approaches to modelling galaxies have been compared in the past. \cite{Yoshida2002Gas} contrast the cooling of gas and its condensation into galaxies between a SAM and a hydrodynamic simulation using the smoothed particle hydrodynamics (SPH) technique. \cite{Saro2010Gas} compare galaxies formed within a massive cluster using both the semi-analytic and hydrodynamic approaches, finding statistically similar results, but significant differences on an object-by-object level, particularly for the star formation history of the central galaxy. \cite{Knebe2018Cosmic} and \cite{Asquith2018Cosmic} carry out a statistical comparison of nine galaxy formation models, including eight SAMs and one halo occupation distribution (HOD) model, with different model calibrations, all run on the same dark matter simulation.

Most similar to our study, \cite{Guo2016Galaxies} compare global statistical properties of the \cite{Guo2013Galaxy} version of \textsc{L-Galaxies}, and the \cite{Gonzalez-Perez2014How} version of the \textsc{Galform} SAM, with the \textsc{Eagle} hydrodynamical simulation \citep{Schaye2015eagle}. They find that statistical properties such as stellar mass functions and star formation rates are similar while galaxy sizes are significantly different. \cite{Mitchell2018Comparing} continue this work by comparing \textsc{Galform} with \textsc{Eagle}, focusing on baryon cycling, angular momentum and feedback. They also study the ratio of stellar masses of the galaxies between the two models, as a function of stellar mass and star formation history. They find that overall, \textsc{Eagle} produces more realistic results when compared to observations, motivating key improvements to \textsc{Galform}. Similarly, \cite{Wang2018TheDearth} compare the star formation quenching from observations with the \cite{henriques2015galaxy} version of the \textsc{L-Galaxies} SAM and the \textsc{Eagle} hydrodynamical simulation. Finally, \cite{Wang2019Comparing} compare galaxy morphologies between the Illustris simulation \citep{vogelsberger2014a,genel2014introducing} and the \cite{guo2011dwarf} version of \textsc{L-Galaxies}, finding that late type galaxies are broadly similar, whereas early types exhibit larger differences.

In this work, we perform a comparison between the \textsc{L-Galaxies} semi-analytical model and IllustrisTNG magnetohydrodynamic simulation. We run the \textsc{L-Galaxies} model on merger trees taken from the IllustrisTNG simulation and our analysis is based on the comparison of physical quantities of the same, matched set of haloes and subhaloes. This enables us to understand their similarities and differences not only statistically, but also on an object-by-object basis. This paper focuses on implications for the physical processes and methods employed in the \textsc{L-Galaxies} and IllustrisTNG models, rather than on a detailed comparison to observational data.

This paper is structured as follows: In \S \ref{sec: Methodology} we describe the IllustrisTNG simulations, the Munich semi-analytical model \textsc{L-Galaxies}, our method to match galaxies and haloes, and model details for the physical processes most relevant to our study. In \S \ref{sec: Galaxies_and_Subhalos} we compare, object-by-object and statistically, several properties of galaxies and haloes produced by \textsc{L-Galaxies} and IllustrisTNG. Section \S \ref{sec: gal_vs_dis} is dedicated to a comparison of the role of environment in galaxy evolution in the two models. Finally, we summarise and discuss our results in \S \ref{sec: summary}.


\section{Methodology} \label{sec: Methodology}

\subsection{IllustrisTNG simulations}
\label{subsec: TNG_sim}

The next generation of the Illustris simulation is the IllustrisTNG project \citep[TNG;][]{nelson18a,springel2018first,pillepich18b,marinacci2018first,naiman2018first}\footnote{\href{https://www.tng-project.org/}{www.tng-project.org}}, a set of graveto-magnetohydrodynamical simulations that model the physical processes most relevant to the formation and evolution of galaxies in cosmological volumes. The TNG model is described in \cite{weinberger17} and \cite{pillepich18a} and is based on the original Illustris model \citep{vogelsberger13,torrey14}, which uses the \textsc{arepo} code \citep{springel2010pur}\footnote{\href{https://arepo-code.org/}{www.arepo-code.org}} to solve the coupled equations of self-gravity and magnetohydrodynamics  \citep{pakmor2011magnetohydrodynamics,pakmor2013simulations}. The TNG model implements key physical processes for galaxy formation, including gas cooling, star formation, stellar evolution, and stellar feedback \citep{pillepich18a}. Processes pertinent to supermassive black holes (SMBH) include seeding/formation, accretion, mergers and thermal and kinetic feedback \citep{weinberger17}.

The TNG model has been used to run three different simulation volumes to date. The largest, TNG300, has a volume of $\sim\rm(300\, Mpc)^3$ with a dark matter resolution of $m_{\rm DM} = 5.9\times 10^{7}\rm M_{\odot}$ and average gas cell mass of $m_{\rm gas} \simeq 1.1\times 10^{7}\rm M_{\odot}$. The intermediate, TNG100, has a volume of $\sim\rm(100\, Mpc)^3$ and dark matter and average gas resolutions of $m_{\rm DM} = 7.5\times 10^{6}\rm M_{\odot}$ and $m_{\rm gas} \simeq 1.4\times 10^{6}\rm M_{\odot}$. We use the publicly available data from both TNG100 and TNG300 \citep{Nelson2019public}. In this work we do not employ the TNG50 simulation, with the highest resolution but smallest volume \citep{pillepich19,nelson2019First}.

For every TNG hydrodynamic simulation, there is a companion dark matter only (gravity only) simulation which has the same initial conditions, box size and resolution. Comparisons between these dark matter only simulations (hereafter TNG-DMO) and the full hydrodynamical runs can shed light on baryonic effects on the underlying dark matter structure \citep{springel2018first}. The DMO simulations can also be used as input to semi-analytic models, as we do here. All the TNG and TNG-DMO simulations start at $z=127$ with initial conditions consistent with a Planck $\Lambda \rm CDM$ cosmology \citep[][$h = 0.673$]{planck2015_xiii}. Haloes are identified using the friends-of-friends (FOF) algorithm \citep{Davis1985TheEvolution}. Bound objects and substructures, i.e. subhaloes and galaxies, are identified and characterised using the \textsc{Subfind} algorithm \citep{springel2001populating}, with a minimum of 20 total particles/cells per object.

The TNG model is calibrated using a number of observations of the galaxy population, including the observed star formation rate density, stellar mass function (SMF), and stellar to halo mass ratio at $z=0$; in addition, the black hole-stellar mass relation, halo gas fraction, and the stellar sizes of galaxies are considered. This calibration is carried out at the fiducial TNG100-1 resolution \citep{pillepich18a}. Model parameters are kept fixed between different boxes (resolutions), the only exception being gravitational softening lengths. As a result, the TNG model has non-trivial numerical convergence behaviours, which must be assessed between different resolution levels as a function of the galaxy property of interest. This leads to important differences with respect to \textsc{L-Galaxies}, as we will discuss in \ref{subsec: combine_100_300_boxes}.

\subsection{\textsc{L-Galaxies} semi-analytical model} \label{subsec: LGal_SAM}

\textsc{L-Galaxies} is a semi-analytical model of galaxy formation typically run on the subhalo merger trees from the Millennium \citep{springel2005simulations} and Millennium-II \citep{boylan2009resolving} N-body simulations, both of which assume a $\rm \Lambda CDM$ cosmology. A number of model branches and updates have been developed over time \citep{springel2005simulations,croton2006many,de2006formation,bertone2007recycling,guo2011dwarf,yates2013modelling,henriques2015galaxy,henriques2020galaxies}. In this paper, we take the publicly released \citealt{henriques2015galaxy} (hereafter "H15") version of the \textsc{L-Galaxies} model for our study\footnote{\href{https://lgalaxiespublicrelease.github.io}{lgalaxiespublicrelease.github.io}}, unless stated otherwise.

\textsc{L-Galaxies} is designed to run on a dark matter only simulation. It contains a set of simplified equations and recipes which model various physical processes involved in the baryon physics of galaxy formation and evolution. These include gas infall and cooling, star formation, disc and bulge formation, stellar and black hole feedback, and the environmental effects of tidal and ram-pressure stripping (RPS). The H15 version of \textsc{L-Galaxies} models the baryon content of each subhalo with seven baryonic components: hot gas, cold (star-forming) gas, bulge stars, disc stars, halo stars, supermassive black holes and ejected material. We note that the halo stars are those which belong to the galaxy's subhalo but do not reside within its disc or bulge, e.g. the stars in the intracluster medium (ICM). In the rest of this paper, we define the galaxy stellar mass as the sum of disc and bulge stellar masses. In addition, the total subhalo stellar mass is the sum of the galaxy stellar mass plus halo stars.

Each subhalo, at its initial formation, begins with a near cosmic mean fraction of diffuse gas. This hot gas cools into the cold gas component from which stars are born. With the death of stars, energy release heats the cold gas, moving it into the hot component. The remaining energy, if there is any, affects the hot gas and moves it into the ejected reservoir. The ejected gas may return to the subhalo at a later time. Black hole feedback is the major mechanism for quenching star formation in massive galaxies. The energy released due to the AGN feedback prevents hot gas from cooling and therefore suppresses star formation. For satellites, the model continues to trace galaxies after their dark matter subhaloes are no longer present, e.g. due to tidal disruption. These are called orphan galaxies (or `type 2' satellites), and are mostly found near the centre of haloes.

\textsc{L-Galaxies} model parameters are calibrated against the observed stellar mass function and fraction of red galaxies at four different redshifts from $z=3$ to $z=0$. These eight observational constraints are equally weighted in a Monte Carlo Markov Chain (MCMC) optimisation \citep{henriques2009monte}. In this work, we employ the fiducial, unchanged model parameters of \cite{henriques2015galaxy}. A full description of the model and its physical processes is given in the supplementary materials of \cite{henriques2015galaxy}.

\subsection{Physical processes most relevant to this study}
\label{subsec: important_processes}

In this section we describe the physical model differences between \textsc{L-Galaxies} and TNG most relevant to our work.

\begin{table*}
    \centering
    \caption{The numbers of subhaloes, galaxies and the fraction of systems matched between the IllustrisTNG simulations and the L-Galaxies runs on the corresponding DMO volumes. Subhaloes in our mass range of interest are almost always matched. On the other hand, orphan galaxies in the L-Galaxies do not reside in subhaloes, and we do not attempt to match these to existent TNG subhaloes, bringing down the matched percentages for less luminous galaxies.}
	\label{tab: matching_table}
    \begin{tabular}{|*{5}{c|}}
    \hline \hline
    Model -> & \textsc{L-Galaxies} (TNG300-DMO) & TNG300 hydrodynamic & \textsc{L-Galaxies} (TNG100-DMO) & TNG100 hydrodynamic \\ \hline \hline
    
    $N_{\rm sub}$ (\textsc{Subfind}) & 154469 ($M_{\rm sub}/\rm M_{\odot}>10^{11.5}$) & 145451 ($M_{\rm sub}/\rm M_{\odot}>10^{11.5}$) & 61263 ($M_{\rm sub}/\rm M_{\odot}>10^{10.5}$) & 52335 ($M_{\rm sub}/\rm M_{\odot}>10^{10.5}$) \\ \hline
    Matched fraction & $\sim 98.4\%$ & $\sim 98.1\%$ & $\sim 97.3\%$ & $\sim 97.3\%$
    
    \\ \hline \hline
    $N_{\rm gal}$ & 211511 ($M_{\star}/\rm M_{\odot}>10^{9.5}$) & 158112 ($M_{\star}/\rm M_{\odot}>10^{9.5}$) & 28196 ($M_{\star}/\rm M_{\odot}>10^{8.5}$) & 33677 ($M_{\star}/\rm M_{\odot}>10^{8.5}$) \\ \hline
    Matched fraction & $\sim 81.3\%$ & $\sim 83.7\%$ & $\sim 77.5\%$ & $\sim 76.4\%$
    
    \end{tabular}
\end{table*}

\subsubsection{Supermassive black hole feedback}
\label{subsec: important_processes_AGN}

Supermassive black hole (SMBH or AGN) feedback is invoked as the physical mechanism for quenching massive galaxies in both models, although their implementations are significantly different. 

TNG's black hole feedback scheme has two modes, thermal and kinetic, depending on the accretion rate of the SMBH; these do not operate at the same time. At high accretion rates, black holes heat the surrounding gas thermally. This thermal energy is injected in a small local environment around the black hole which initially increases the temperature of neighbouring gas cells. This will affect the future evolution and thermodynamical properties of gas. At low accretion rates, on the other hand, black holes inject kinetic energy. These randomly oriented, high-velocity outflows displace interstellar star-forming gas \citep{nelson2019First} as well as circumgalactic medium gas \citep{truong20,davies20}. They are also preventative \citep{terrazas20}, modulating the cooling properties of the hot halo \citep{Zinger2020Ejective} and effectively suppressing star formation \citep{weinberger18}. SMBH driven outflows can also push gas entirely beyond the halo $R_{200}$, which both prevents re-accretion and suppresses the halo baryon fraction (see \S \ref{subsec: halo_baryonFrac}). Black hole feedback switches from thermal to kinetic mode as a function of the BH accretion rate, which is proportional to the black hole mass \citep[see \S 2.1 of][]{weinberger17}.

In \textsc{L-Galaxies}, there are two modes of black hole accretion; quasar mode and radio mode. In the quasar mode, black holes grow by cold gas accretion, and during mergers, where the rate depends on the total cold gas mass of both merging galaxies and the ratio of their total baryon masses (see \S S.10.1 of \citealt{henriques2015galaxy}). This accretion mode is therefore more efficient for major mergers and less efficient for minor mergers. No explicit feedback mechanism is attributed to the quasar mode in \textsc{L-Galaxies}. In the radio mode, galaxies can accrete gas from the hot gas of their host subhaloes, producing radio-mode AGN feedback. In this state black holes inject energy into the hot gas, which suppresses cooling and as a result, star formation. The amount of energy injected into the hot gas in the radio mode is proportional to the black hole mass and the host subhalo hot gas mass. Therefore, the radio mode is most effective for massive galaxies, whose black holes reside in high-mass haloes where hot gas dominates the baryon budget. We note that the two accretion modes in \textsc{L-Galaxies} can coexist.

\subsubsection{Supernova feedback}
\label{subsec: important_processes_supernova}

The other key process which affects star formation, particularly for lower mass star-forming galaxies, is supernova feedback. In the TNG model, supernovae launch galactic winds from the dense interstellar medium (ISM), and these winds are modelled as collisionless particles ejected from star-forming gas cells, where the rate of total energy released from each cell is proportional to its instantaneous star formation rate \citep{pillepich18a}. Wind particles are ejected in random directions and take their properties from the gas cell from which they are ejected. These outflows remove mass from the ISM \citep{nelson2019First}, and as we show below, they can also push gas out of the dark matter halo itself, suppressing the halo baryon fraction.

In \textsc{L-Galaxies}, supernova feedback injects energy into the cold star-forming gas, heating some fraction through transfer to the hot gas. The remaining energy, if any, pushes hot gas into the ejected reservoir, from whence it can return to the hot gas on a reincorporation timescale, thus becoming available for cooling again (see \S S1.7 of \citealt{henriques2015galaxy}). Energy is injected by supernovae in proportion to the mass of stars formed. The reincorporation timescale is inversely proportional to the total halo $M_{200}$ so that ejected gas returns to massive galaxies more quickly.

\subsubsection{Environmental effects}
\label{subsec: important_processes_Env}

Hydrodynamical simulations such as IllustrisTNG self-consistently capture gas-dynamical processes including tidal and ram-pressure stripping, to the degree allowed by the numerical resolution. The strength of ram-pressure stripping, for instance, depends on the background gas density and so the efficiency of the process is modulated by the halo gas fraction. In contrast, these processes need to be explicitly added to semi-analytic models.

In the \cite{henriques2015galaxy} version of \textsc{L-Galaxies}, which has been used in this paper, tidal and ram-pressure stripping affect the hot gas of satellite galaxies within the halo boundary, $R_{200}$. While tidal stripping acts for all host halo masses, RPS only occurs within clusters of $M_{200} > 1.2\times 10^{14}\,\rm M_{\odot}$. Therefore, there is no stripping for galaxies beyond the halo boundary and no RPS for satellites in groups or low-mass haloes. We note that this threshold value of $1.2\times 10^{14}\,\rm M_{\odot}$ is a numerical fix and we do not consider it a physical threshold. Even more importantly, no cold gas stripping of any kind is implemented \cite[in contrast see][]{Luo2016Resolution}. 

These limitations are addressed and improved in our recent study \cite{ayromlou2019new}, but we do not incorporate these model improvements here, focusing on the publicly available H15 model.

\subsection{Matching galaxies between TNG simulation and \textsc{L-Galaxies} run on TNG-DMO}
\label{subsec: Matching}

\subsubsection{Object-by-object matching}
\label{subsec: obj-by-obj_matching}

We aim to make an object-by-object comparison between \textsc{L-Galaxies} and TNG by running \textsc{L-Galaxies} on the TNG-DMO simulations. These have the same initial conditions (i.e. phases) as the full hydrodynamical simulations, producing very nearly the same dark matter halo populations.

We then match subhaloes between the baryonic and DMO simulations following \cite{nelson15a}, using the \textsc{LHaloTree} algorithm. By comparing unique dark matter particle IDs, matched subhaloes are defined as those with the highest fractions of common particles. The match must be bidirectional, i.e. the same starting from either of the two runs. When necessary, FOF haloes are matched based on their central subhalo\footnote{The catalogue of L-Galaxies run on TNG-DMO is publicly available at \href{https://www.tng-project.org/ayromlou20}{www.tng-project.org/ayromlou20} for all the TNG snapshots.}.

The subhalo matching catalogue also provides, by definition, a matching of the galaxies which reside in those subhaloes. The notable exception is for orphan galaxies, which are not hosted by identifiable dark matter subhaloes (see \S \ref{subsec: LGal_SAM}). However, unmatched galaxies/subhaloes are usually low mass and form a small fraction of the objects in our stellar mass range of interest. The number of subhaloes, galaxies and the corresponding matched fractions are given in Table \ref{tab: matching_table}; the great majority of subhaloes are successfully matched. We perform two kinds of analysis in this work: i) statistical and ii) object-by-object. For the statistical analyses (e.g. stellar mass functions) we use all subhaloes/galaxies, while for the object-by-object analyses (e.g. the ratio of the stellar masses) we consider only successfully matched objects.

\subsubsection{Combining the 100 Mpc and 300 Mpc boxes}
\label{subsec: combine_100_300_boxes}

We have analysed both the $\rm (300\,Mpc)^{3}$ and $\rm (100\,Mpc)^{3}$ simulation boxes, although we often focus on just one simulation that is more appropriate for the aspect of the galaxy population being studied. In practice, we compare TNG300 with \textsc{L-Galaxies} run on TNG300-DMO (hereafter LGal300) and we compare TNG100 with \textsc{L-Galaxies} run on TNG100-DMO (hereafter LGal100). The $\rm 300\,Mpc$ box provides better statistics for massive dark matter haloes, while the $\rm 100\,Mpc$ box has a higher resolution for lower mass galaxies. When we combine the results of these two boxes, we have verified that the resolution difference does not introduce any biases in our interpretation.

The quantitative results of the TNG model depend on resolution. In particular, the stellar masses of galaxies at $z=0$ are roughly 40\% higher, at fixed total halo mass, in the higher resolution TNG100 as compared to TNG300 \citep{pillepich18b}. It can be useful to re-scale stellar masses to compensate for the lower resolution of the TNG300 simulation (labelled `rTNG300'). We do so only in Fig. \ref{Fig: SMF}, for stellar mass functions and stellar mass to halo mass ratios. We follow the method of \cite{pillepich18b} and derive a correction factor equal to the ratio of TNG100 to TNG100-2\footnote{The TNG100-2 is a TNG simulation with resolution equal to TNG300 but performed in a 100\,Mpc box.} stellar mass in bins of fixed halo mass. On average this multiplicative correction factor is $\sim 1.4$ at $z=0$, and lower at higher redshifts. For simplicity, we avoid re-scaling stellar masses in any other case, and likewise do not re-scale any other properties of galaxies. A similar step is not required for \textsc{L-Galaxies} since stellar masses are converged over the mass range of interest \citep[see][]{guo2011dwarf}. 

\subsubsection{Deriving galaxy properties}
\label{subsec: gal_prop_definitions}

Because of finite numerical resolution, care is needed to compare some galaxy properties. We enforce a minimum for the star formation rates (SFR) of galaxies, choosing $\rm SFR_{\rm min} = 10^{-3}M_{\odot} yr^{-1}$. This allows us to analyse galaxies with zero star formation rate, which occurs in hydrodynamical simulations below some resolution-dependent threshold \citep{donnari19}. The value of $\rm SFR_{\rm min}$ ensures that for galaxies with $\log_{10}(M_{\star}/\rm M_{\odot}) > 8$ the quenched fraction is unaffected, as galaxies with $\rm log_{10}(sSFR/yr^{-1}) < -11$ are considered as quenched at $z=0$. A limit is also imposed for gas masses at $\log_{10}(M_{\rm gas,min}/\rm M_{\odot}) = 5$ which similarly allows us to include very gas-poor objects without biasing our results.

In the \textsc{L-Galaxies} model calibration, a random Gaussian centred at zero with $\sigma = 0.08(1+z)$ is added to all the logarithmic stellar masses when comparing to data \citep[see Fig. 2 of][]{henriques2015galaxy}. This accounts for the uncertainties in observational stellar mass determinations \citep[also see][]{d2015massive,ilbert2013mass}. In this work, we only apply this modification when plotting the stellar mass functions for \textsc{L-Galaxies} (top panel of Fig. \ref{Fig: SMF}), leaving all other stellar masses in our analysis unchanged.

Throughout, unless stated otherwise, we take the stellar mass definition to be the subhalo stellar mass (in the case of \textsc{L-Galaxies}) or the stellar mass within twice the stellar half-mass radius (for TNG). The $M_{200}$ ($M_{500}$) of a FOF halo is the total mass within the $R_{200}$ ($R_{500})$ of the halo, the radius within which the matter density equals 200 (500) times the critical density of the Universe.


\section{General properties of galaxies and haloes}
\label{sec: Galaxies_and_Subhalos}

\begin{figure*}
\centering
    \includegraphics[width=0.49\textwidth]{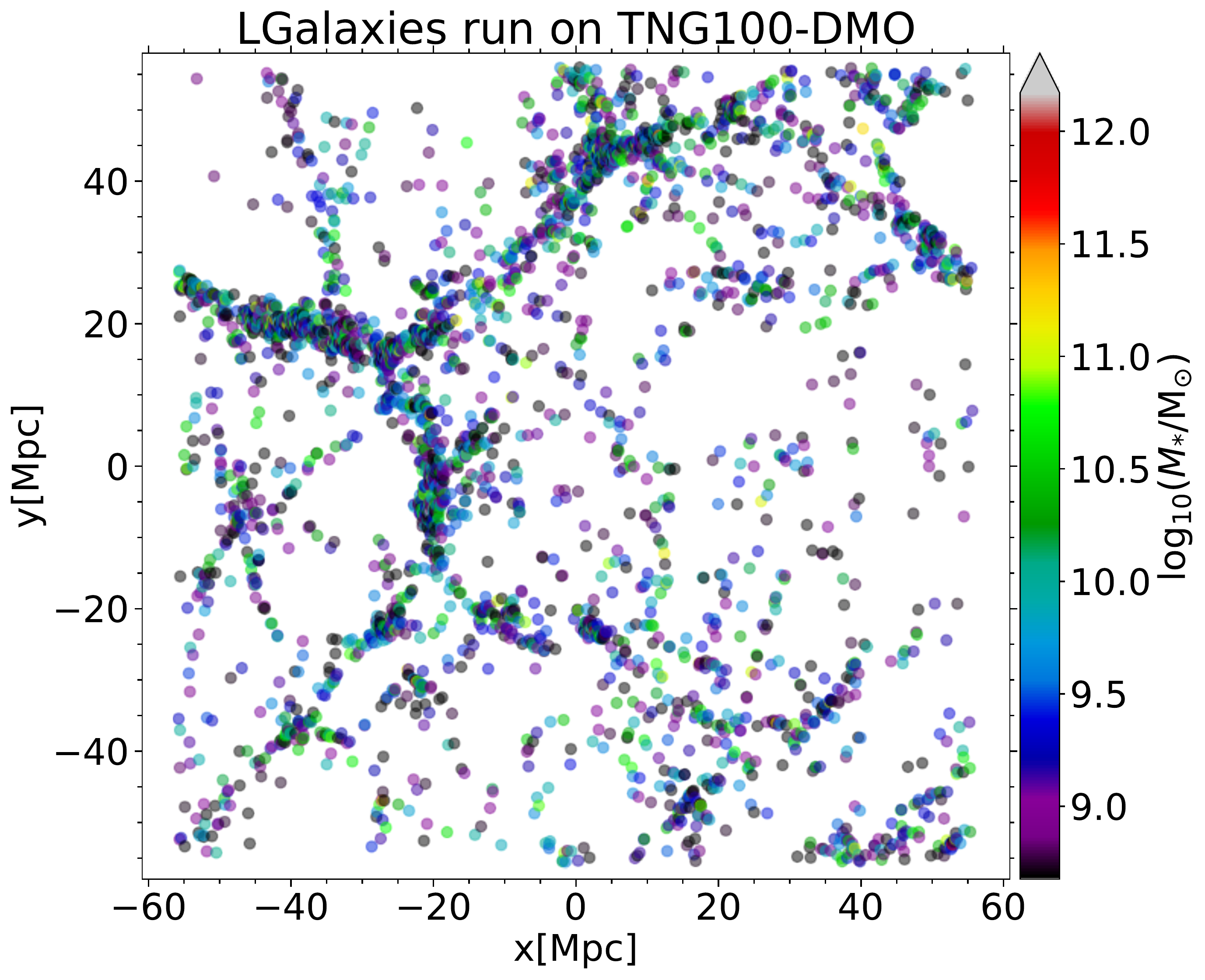}
    \includegraphics[width=0.49\textwidth]{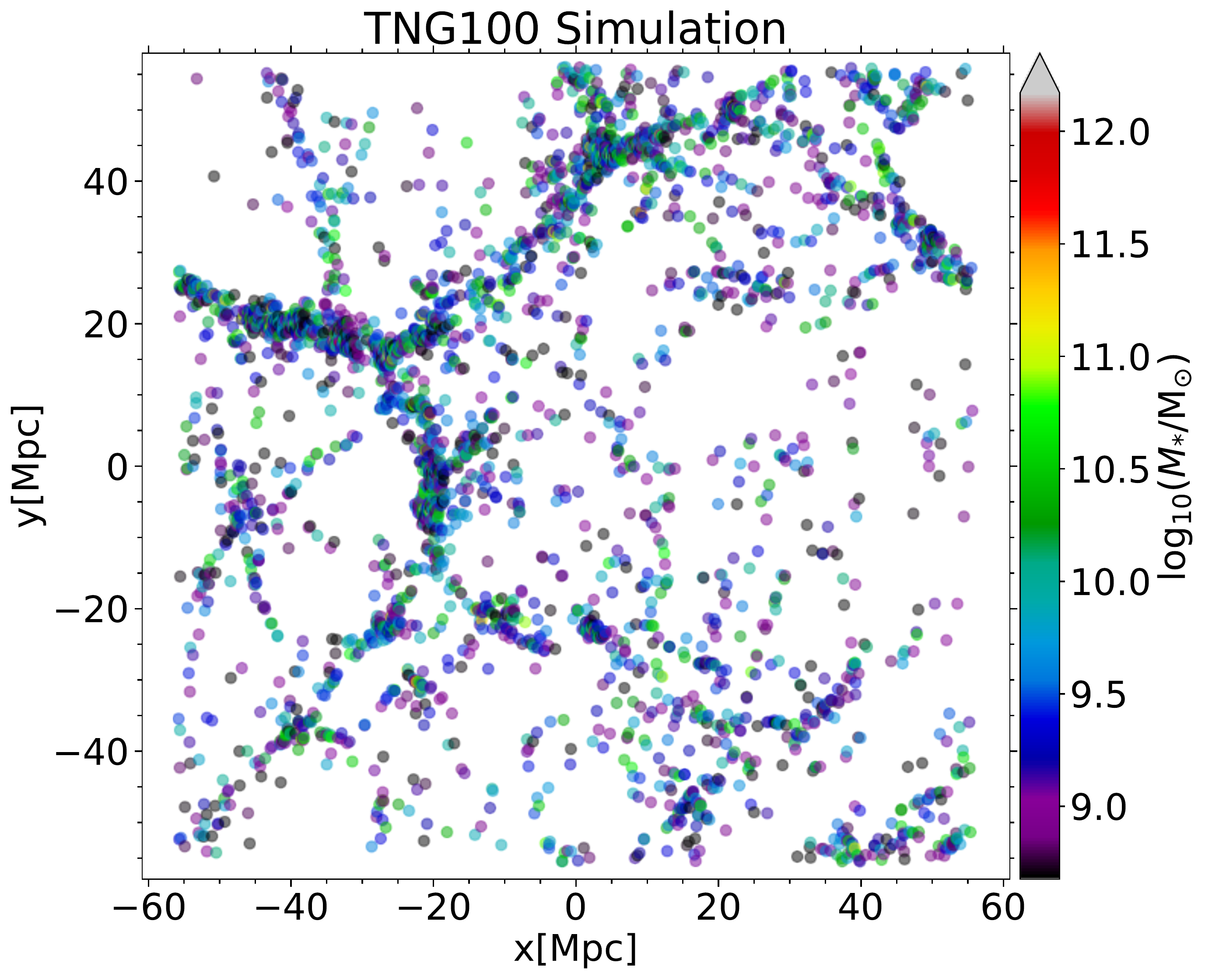}
    \includegraphics[width=0.49\textwidth]{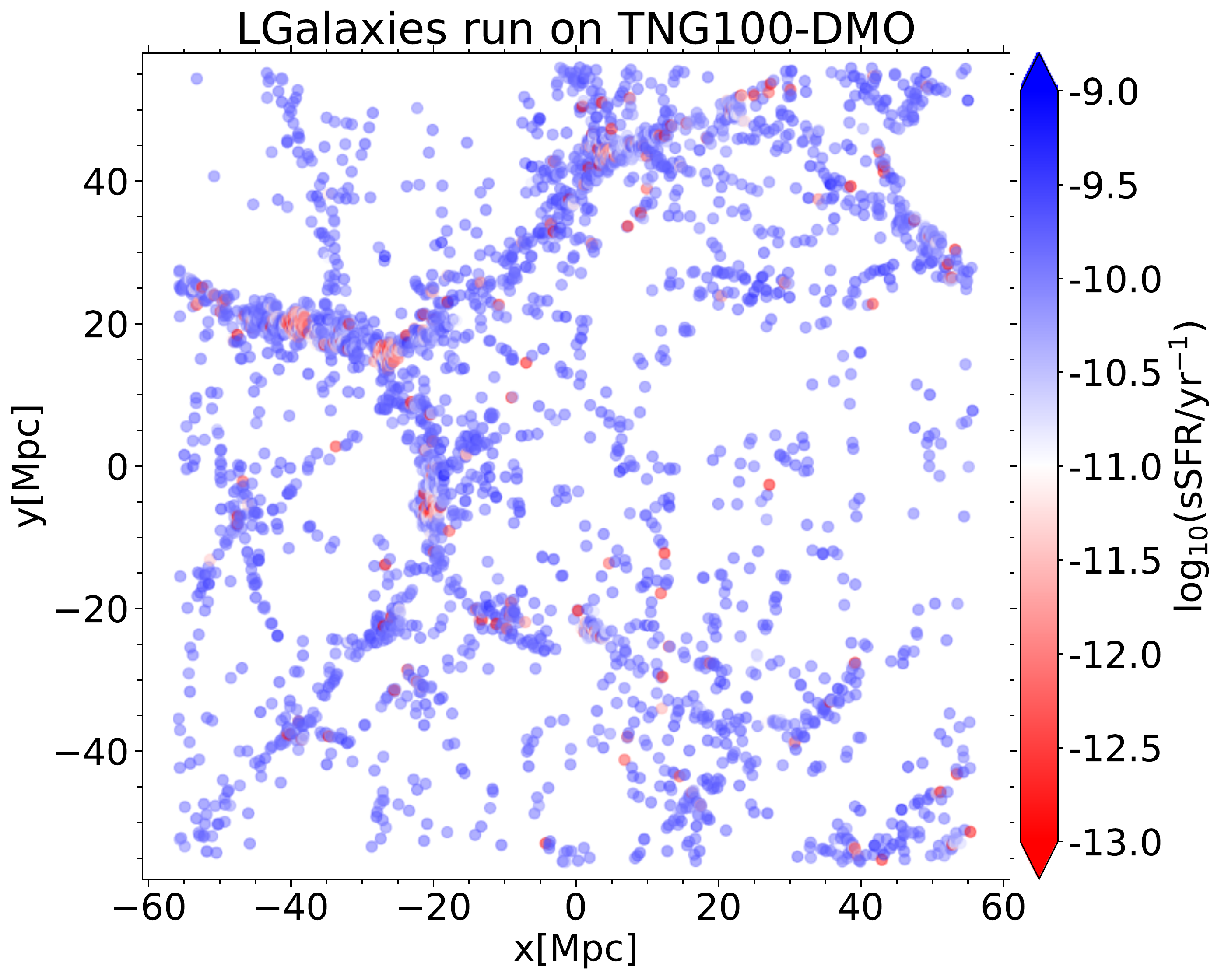}
    \includegraphics[width=0.49\textwidth]{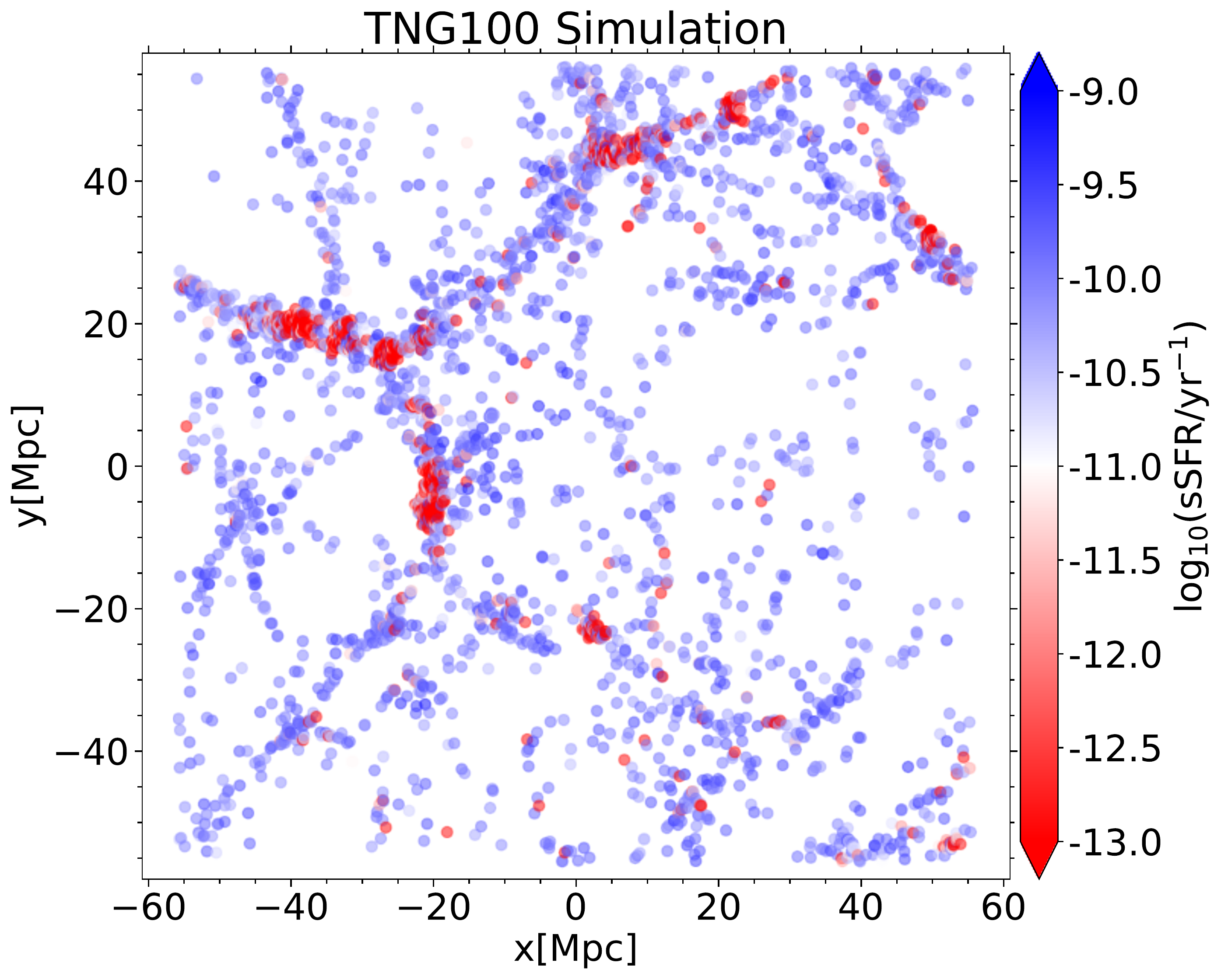}
    \caption{Visual overview of galaxy properties, comparing the results of \textsc{L-Galaxies} (left columns) versus TNG100 (right columns) at $z=0$. We project the $\rm 100\,Mpc$ simulation volume through a depth of $\rm 10\,Mpc$, and each circle corresponds to a galaxy. The colour shows galaxy stellar mass (top panels) or specific star formation rate (bottom panels). In the bottom panel, systems are separated based on sSFR, where galaxies with $\rm log_{10}(sSFR/yr^{-1})>-11$ are considered as star-forming and are shown in blue, while galaxies with $\rm log_{10}(sSFR/yr^{-1})<-11$ are considered to be quenched and are shown in red.}
\label{Fig: Box_Schematic}
\end{figure*}

We begin by comparing the properties of galaxies and haloes\footnote{We note that baryonic physics leads to small changes in the total mass of subhaloes, but this has a negligible impact on our analyses. For instance, \cite{springel2018first} shows that halo masses ($M_{200}$) are up to 20 per cent lower in baryonic versus DMO runs. We similarly find that subhalo masses are usually smaller in the hydrodynamical run, by twenty percent or less, depending on the subhalo mass, with a typical scatter of $0.1-0.2$ dex.} between TNG and \textsc{L-Galaxies}. To give a visual overview, Fig. \ref{Fig: Box_Schematic} illustrates the distribution of galaxies in a $\rm 10\,Mpc$ thick slice of the $\rm 100 \,Mpc$ box, as simulated by the \textsc{L-Galaxies} (left column) and TNG (right column) models. Qualitatively, the overall cosmic web structure is nearly identical in the two plots and the predicted stellar masses are very similar (the top panels). On the other hand, the specific star formation rates (sSFR = SFR/$M_{\star}$, bottom panel) start to exhibit differences. Particularly in nodes where filaments come together, the sites of massive groups and clusters, galaxies are overall less star-forming in TNG.

\subsection{Comparing stellar masses}
\label{subsec: stellar_masses}

\begin{figure*}
    \includegraphics[width=1\columnwidth]{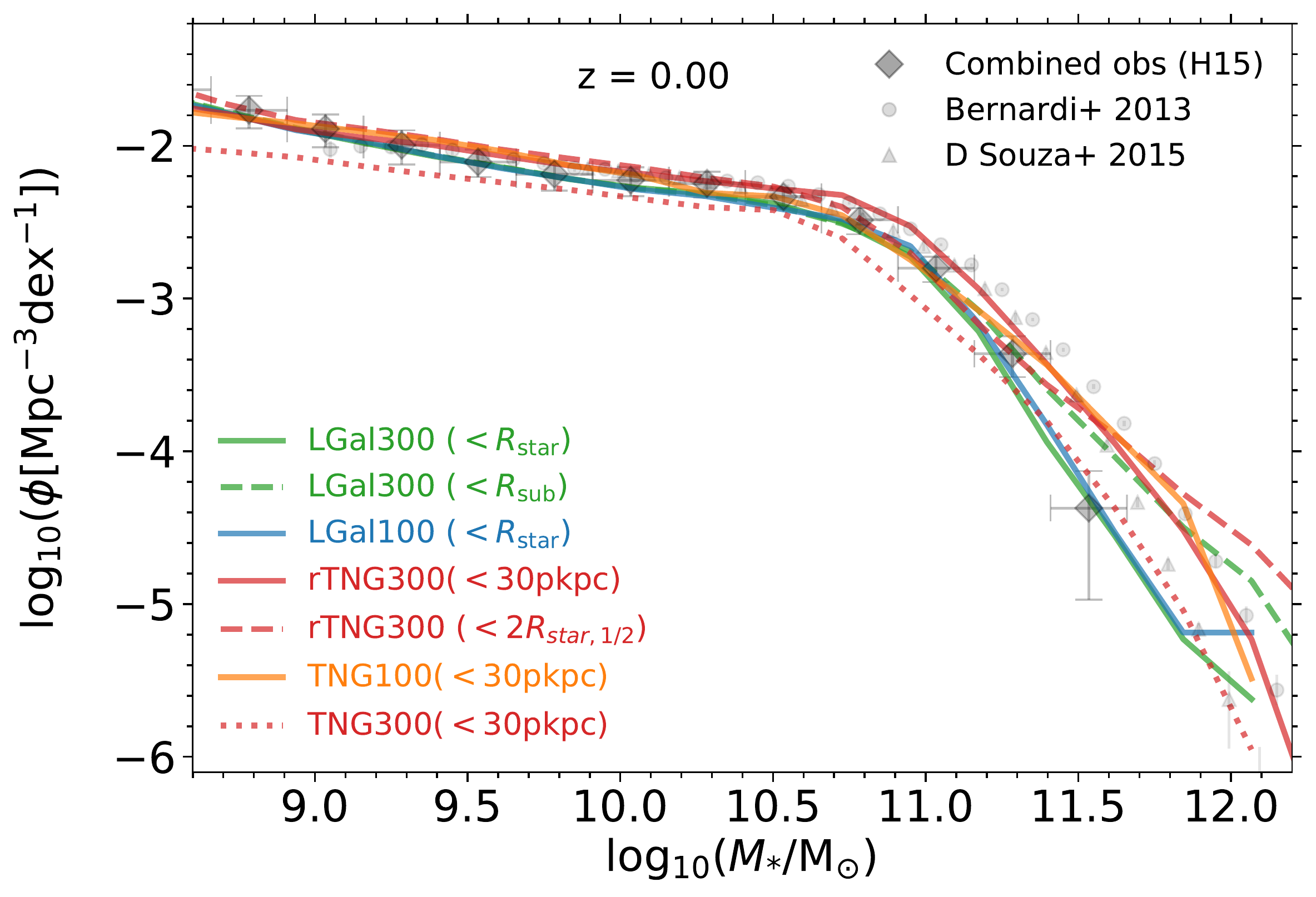}
    \includegraphics[width=1\columnwidth]{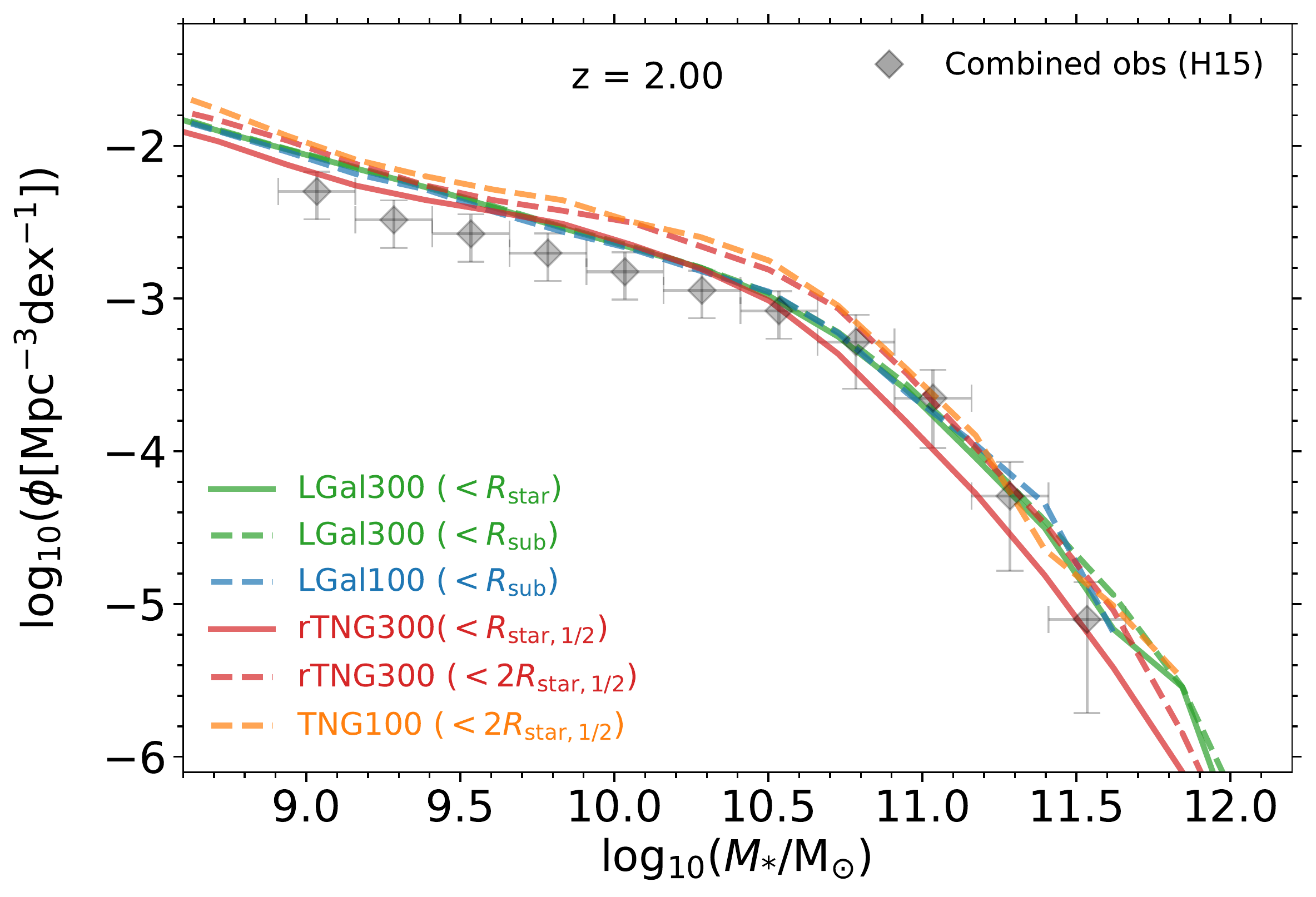}
    \includegraphics[width=1\columnwidth]{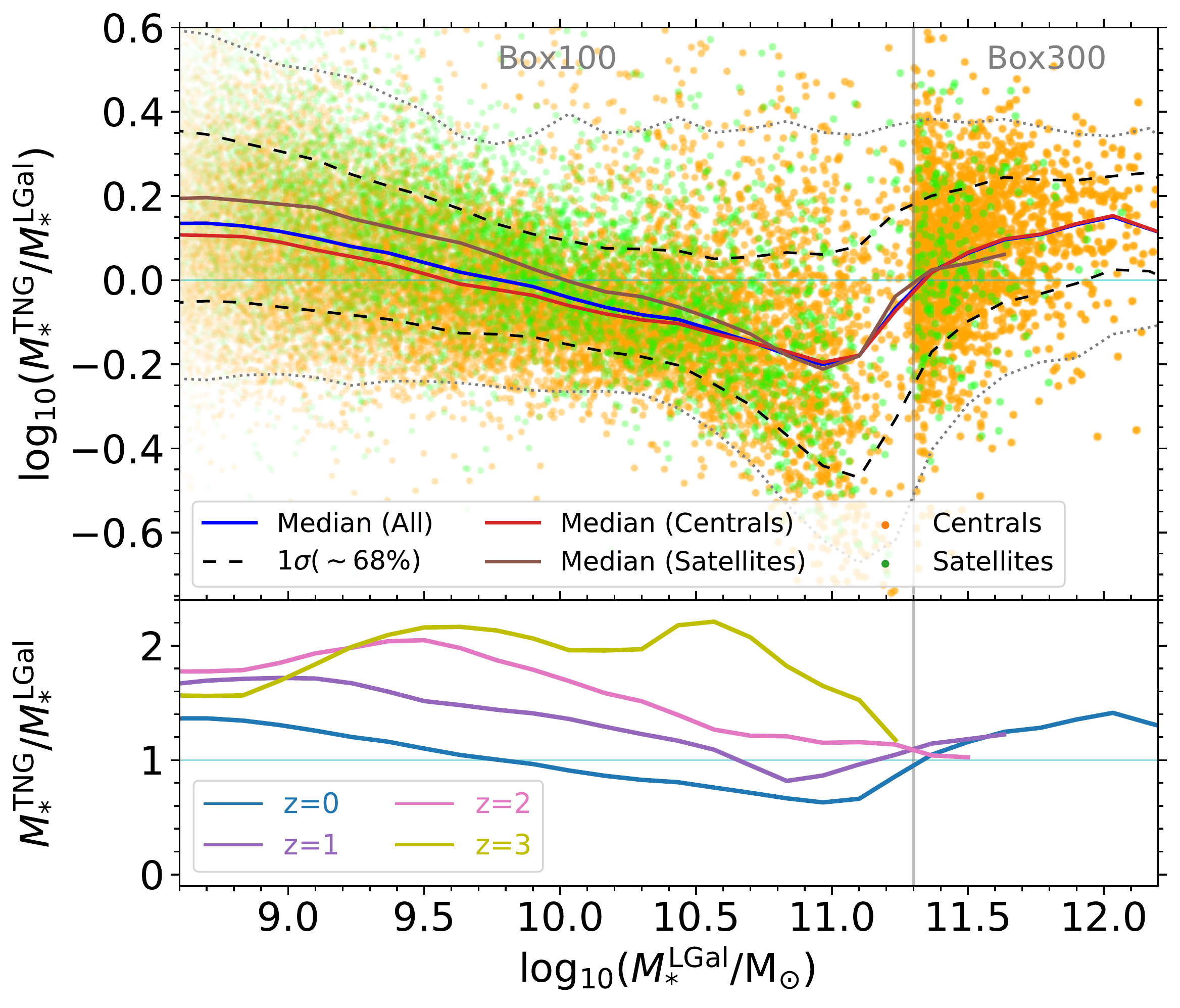}
    \includegraphics[width=1\columnwidth]{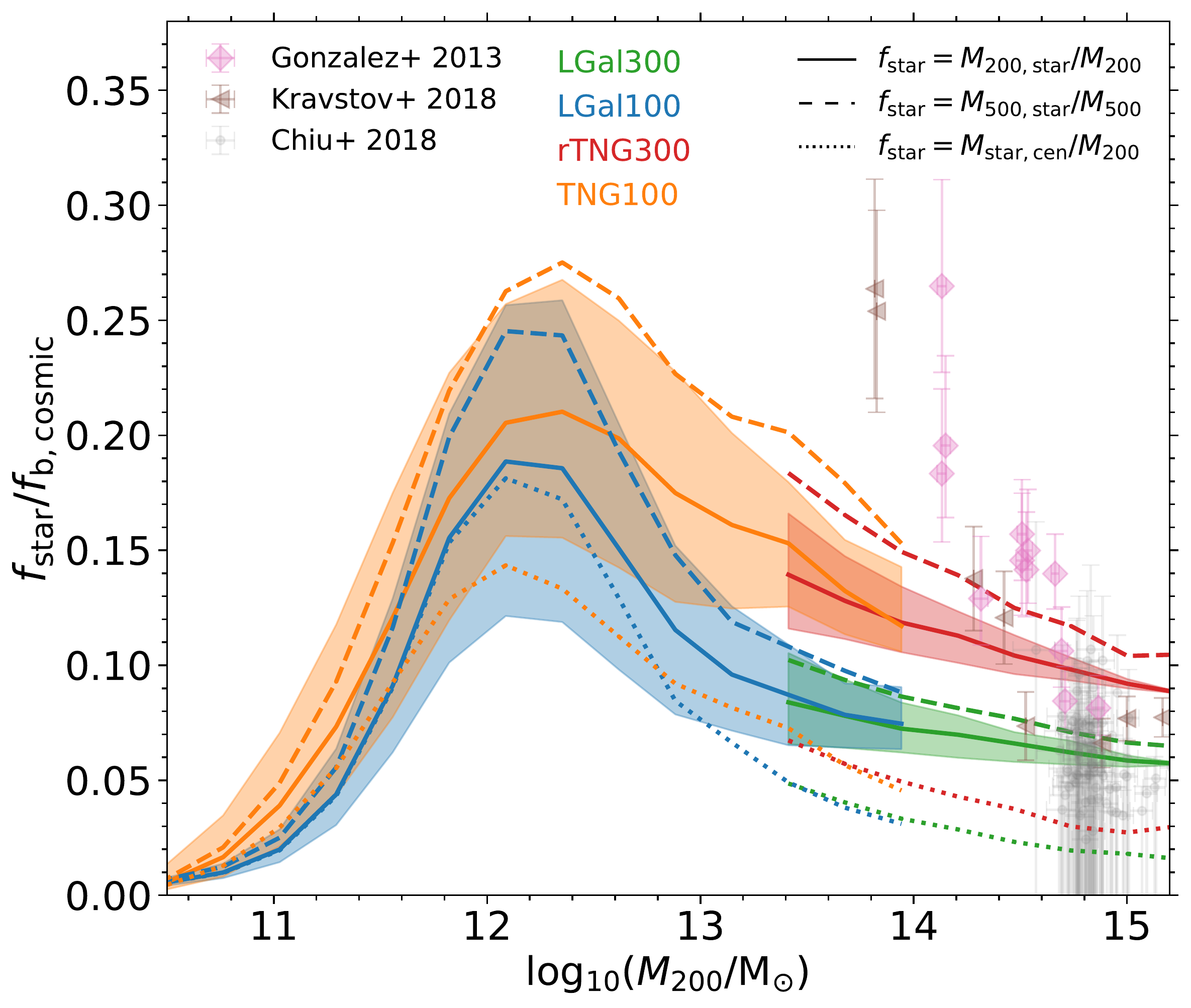}
    \caption{The top panels show the stellar mass functions from \textsc{L-Galaxies} and TNG at $z=0$ and $z=2$. Observational data are taken from \protect\cite{baldry2008galaxy,baldry2012galaxy,li2009distribution} (together labelled as "Combined obs", used in H15), and \protect\cite{bernardi2013massive,d2015massive} for $z=0$ and \protect\cite{marchesini2009evolution,marchesini2010most,ilbert2010galaxy,ilbert2013mass,dominguez2011evolution,muzzin2013evolution,tomczak2014galaxy} (together labelled as "Combined obs", used in H15) for $z=2$. Solid, dashed and dotted lines correspond to different definitions of stellar mass (see the legends). The bottom left panel shows the object-by-object ratio of TNG stellar masses to \textsc{L-Galaxies} stellar masses. Here we use the subhalo stellar mass for \textsc{L-Galaxies} and the stellar mass within twice the half stellar mass radius in TNG (dashed lines in the top panel). Each dot corresponds to one galaxy, where centrals and satellites are orange and green, respectively. The dashed and dotted lines show the $1\sigma$ and $2\sigma$ scatter of the distribution. The lower sub-panel shows the median relations at different redshifts. The bottom right panel shows the ratio of stellar mass to halo mass (normalised by the cosmic baryon fraction). The solid and dashed lines correspond to SMHM ratio within $R_{200}$ and $R_{500}$, respectively. The dotted line shows the ratio of the stellar mass of the central galaxy to its host halo $M_{200}$. The shaded regions correspond to the $1\sigma$ scatter for the SMHM ratio within $R_{200}$. Observational data are taken from \protect\cite{gonzalez2013galaxy,kravtsov2018stellar,chiu2018baryon} and show $f_{\rm star}$ within $R_{500}$ (to be compared with dashed lines). The `rTNG300' curves are re-scaled TNG values (see \S \ref{subsec: combine_100_300_boxes}).}
\label{Fig: SMF}
\end{figure*}

\begin{figure}
    \includegraphics[width=1\columnwidth]{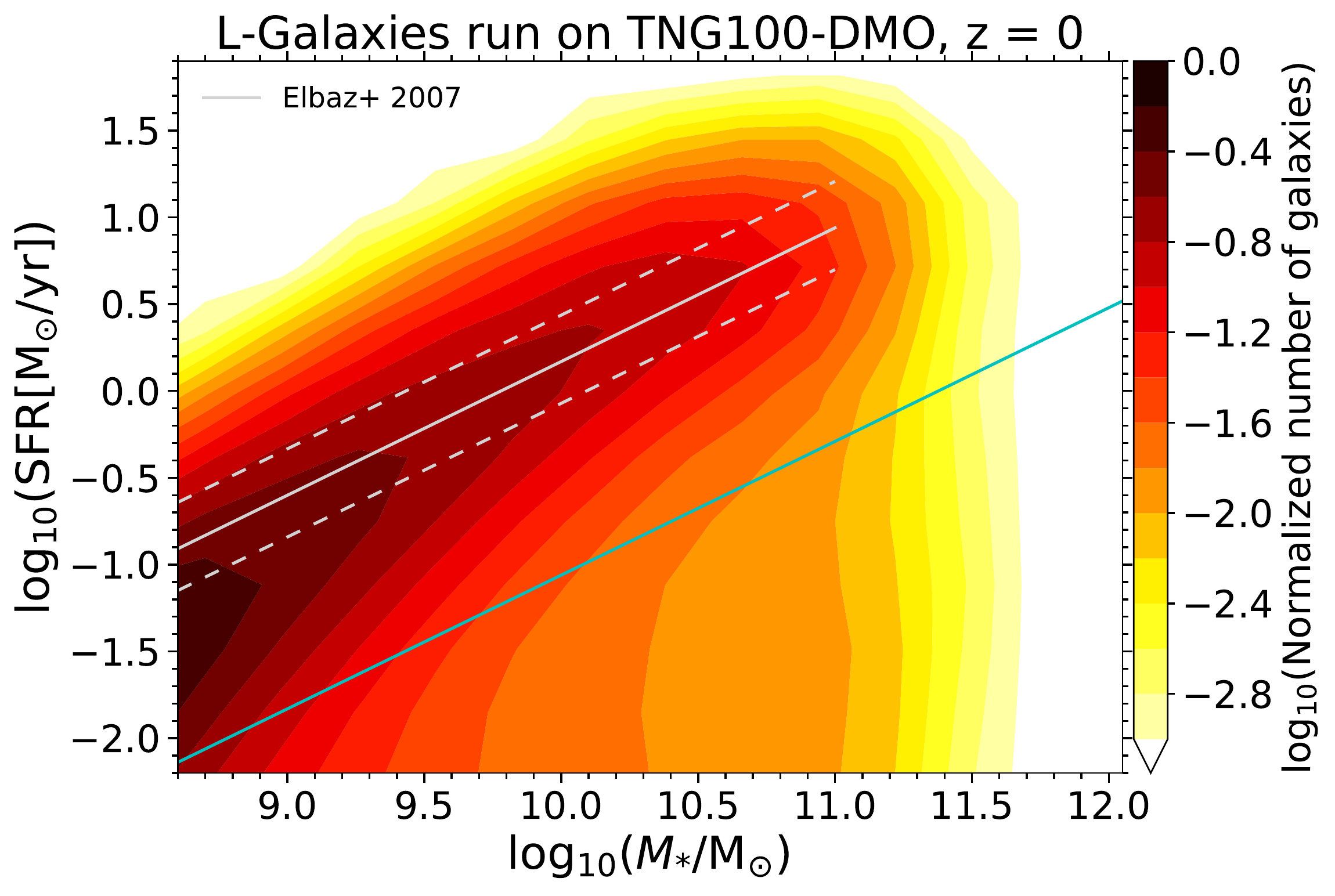}
    \includegraphics[width=1\columnwidth]{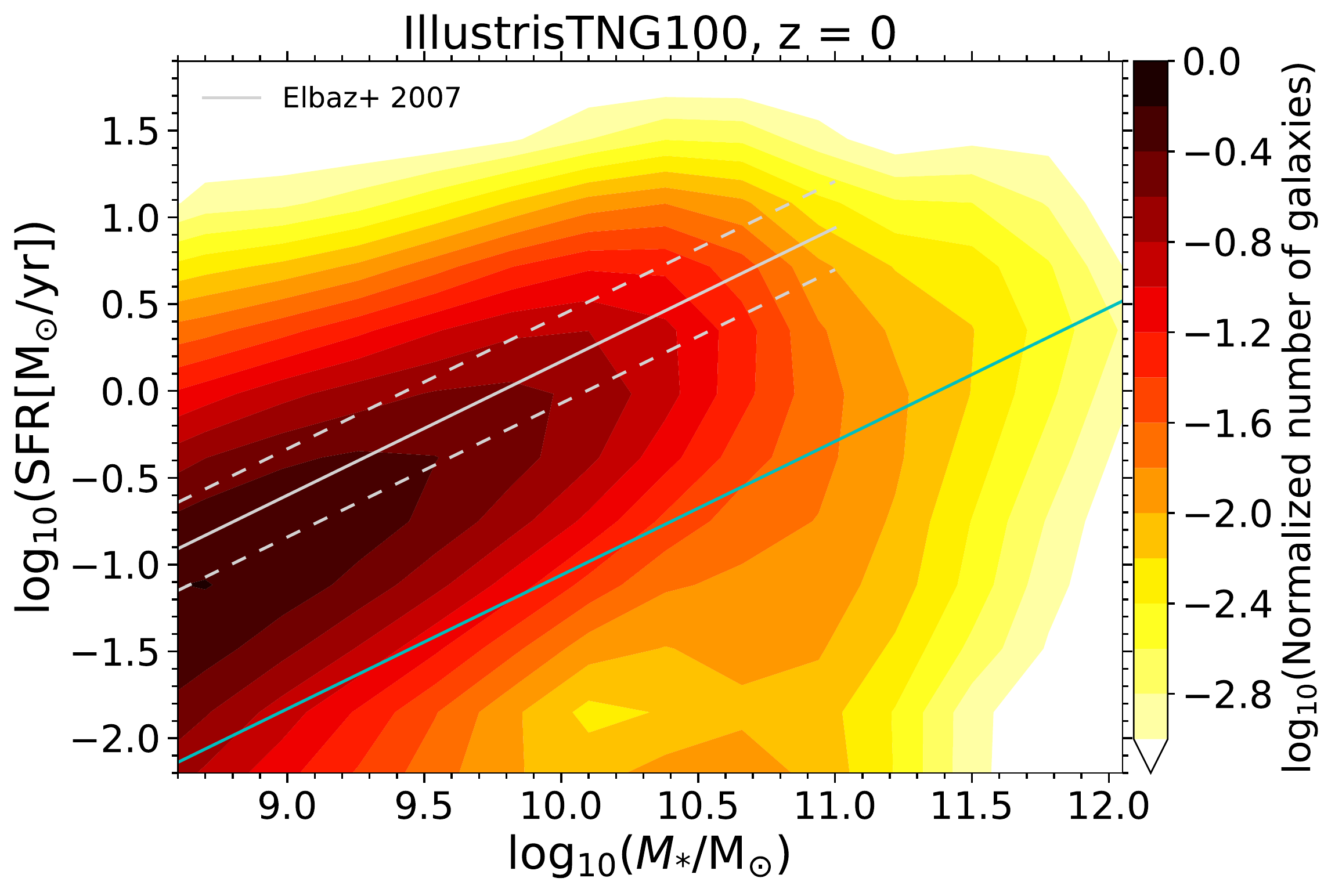}
    \caption{Occupation of the SFR-stellar mass plane by galaxies produced in \textsc{L-Galaxies} (top) and TNG100 (bottom) at $z=0$. Colours illustrate contour levels, normalised to the maximum number of galaxies found in a bin. The solid and dashed gray lines correspond to the median and $1\sigma$ scatter of the main sequence of star-forming galaxies by \protect\cite{elbaz2007reversal} from SDSS data. For reference, the solid green lines are 1 dex below the main sequence, a possible definition of quiescence.}
\label{Fig: SFRs}
\end{figure}

We start with a quantitative comparison of the stellar content of galaxies in \textsc{L-Galaxies} and TNG. The top panels of Fig. \ref{Fig: SMF} show stellar mass functions from the two models at redshifts $z=0$ and $z=2$. We consider two definitions of galaxy stellar mass for the TNG galaxies at $z=0$. The first is the stellar mass within twice the half-stellar-mass-radius, $R<2R_{\rm star,1/2}$, and the second is the stellar mass within 30 physical kiloparsecs, $R<\rm 30 \, pkpc$. For \textsc{L-Galaxies}, we show both galaxy and subhalo stellar masses. 

Regardless of the redshift, the stellar mass functions of \textsc{L-Galaxies} and TNG agree relatively well with each other and with observations. As discussed in \S \ref{sec: Methodology}, both models attempt to calibrate their free parameters to fit the observed stellar mass function at $z=0$ (\textsc{L-Galaxies} also uses data at $z = 0,1,2,3$). As a result, it is not surprising that they are broadly consistent with the data at $z=0$, and the agreement at other redshifts is similarly good. At the high mass end, $\log_{10}(M_{\star}/\rm M_{\odot}) \gtrsim 11$, the definition of stellar mass starts to become critical, as galaxy sizes increase and more stars are found in the extended intracluster component of haloes. In all models, definitions which integrate to larger apertures produce a shallower high-mass end of the SMF (red dashed line above red solid; green dashed above green solid). Whereas \textsc{L-Galaxies} SMFs are consistent between resolutions, modulo box volume effects (green versus blue solid lines), the TNG100 SMF is above the TNG300 due to resolution convergence (orange solid line versus red dotted line), but in agreement with rTNG300 \citep[red dotted line; see][]{pillepich18b}. Overall, the agreement between the two models is statistical, and does not necessarily imply that the stellar masses of individual objects also agree.

We therefore compare at the object-by-object level, contrasting model predictions for the stellar masses of individual galaxies. The results are given in the bottom left panel of Fig. \ref{Fig: SMF}, where we show the ratio of stellar masses between the two models, for matched galaxies, at $z=0$. Each dot corresponds to a galaxy; centrals and satellites are shown with orange and green dots, respectively. The sub-panel shows the redshift evolution of the median relation since $z=3$, in linear units. There is remarkably good agreement between the stellar masses of individual objects since $z=3$. At $z=0$, the median rarely differs from unity by more than 40\%, which is relatively small considering the wide range of stellar masses we consider and the differences in the physical recipes and general approach used in each model. In addition, the $1\sigma$ scatter of the distributions (dashed lines) is typically smaller than $0.2-0.3$ dex. We find that the scatter decreases with redshift, from $\sim 0.1$ dex at $z=3$ to $\sim 0.2$ dex by $z=0$ (not shown). The median line at $z=0$ shows that for both low and high stellar mass ranges, i.e. $\log_{10}(M_{\star}/\rm M_{\odot}) \lesssim 10$ and $\log_{10}(M_{\star}/\rm M_{\odot}) \gtrsim 11.5$ TNG predicts higher stellar masses, while for the intermediate stellar mass range, $10 \lesssim \log_{10}(M_{\star}/\rm M_{\odot}) \lesssim 11.5$, \textsc{L-Galaxies} produces more stars.

\begin{figure}
    \includegraphics[width=1\columnwidth]{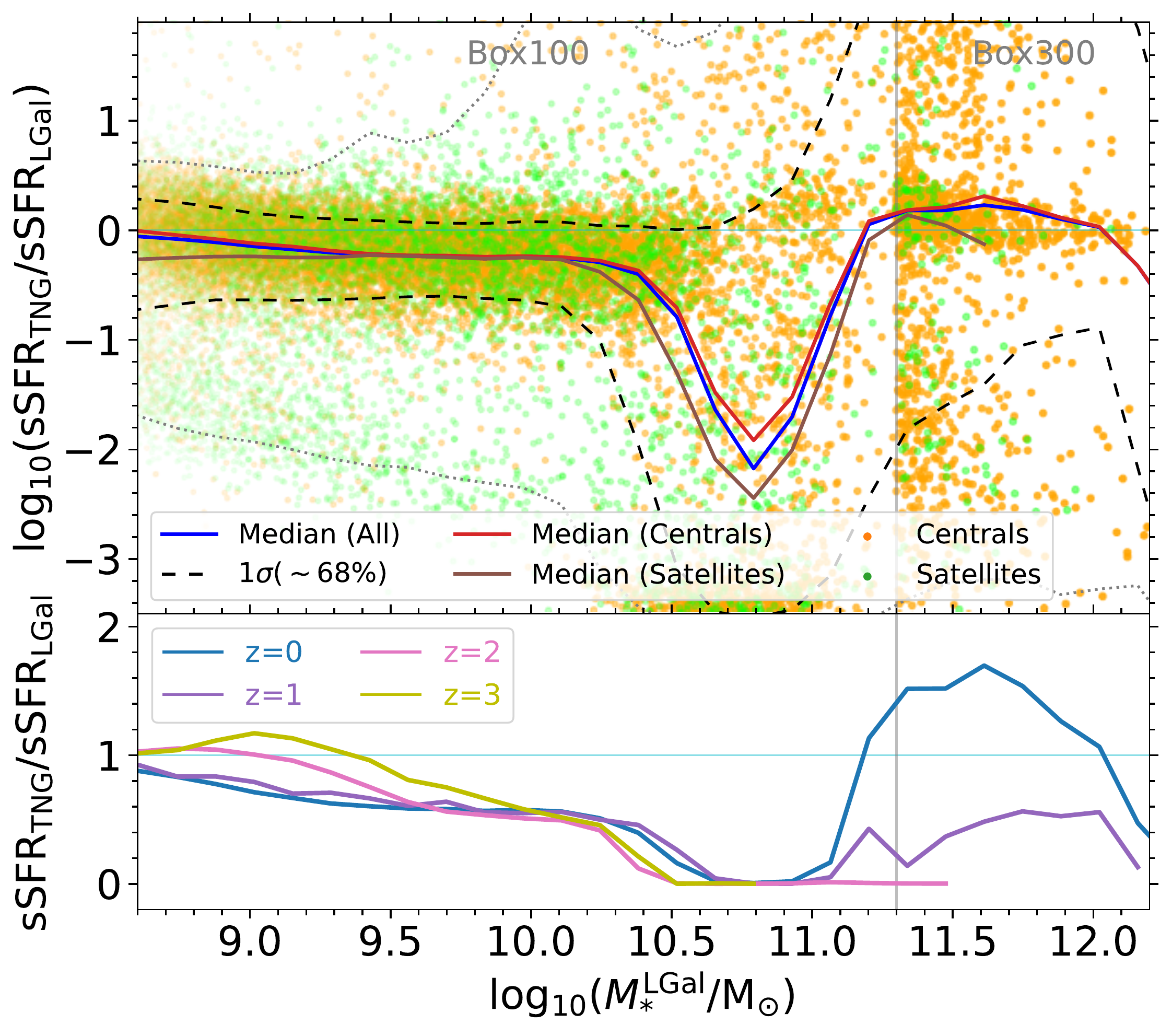}
    \caption{The ratio of specific star formation rates for galaxies in TNG and \textsc{L-Galaxies}. The top panel shows the object-by-object ratio at $z=0$. Each dot corresponds to one galaxy, where centrals and satellites are shown as orange and green dots, respectively. The blue solid line gives the median relation for all galaxies, while red and brown solid lines show medians for central and satellite galaxies alone. In addition, the dashed and dotted lines indicate the $1\sigma$ and $2\sigma$ scatter. The lower sub-panel shows median lines at different redshifts in linear units.}
\label{Fig: ratio_SSFRs}
\end{figure}

We have also plotted medians of central (red solid lines) and satellite galaxies (brown solid lines). Satellite galaxies are more sensitive to the difference in physical prescriptions between models, and in general their ratio $M_{\star}^{\rm TNG} / M_{\star}^{\rm LGal}$ is larger than for centrals.

To compare the star formation efficiency between \textsc{L-Galaxies} and TNG haloes, the bottom right panel of Fig. \ref{Fig: SMF} shows the fraction of stars within haloes -- the stellar mass to halo mass (SMHM) ratio. Solid lines indicate the median stellar-to-total mass ratio within $R_{200}$, normalised by the cosmic baryon fraction, while the $1\sigma$ scatter is shown as shaded regions. The dashed lines correspond to the same quantity, measured with respect to $R_{500}$ and $M_{500}$ instead. Finally, the ratio of the stellar masses of central galaxies (i.e. excluding halo stars) to $M_{200}$ are shown as dotted lines.

\begin{figure*}
    \includegraphics[width=0.30\textwidth]{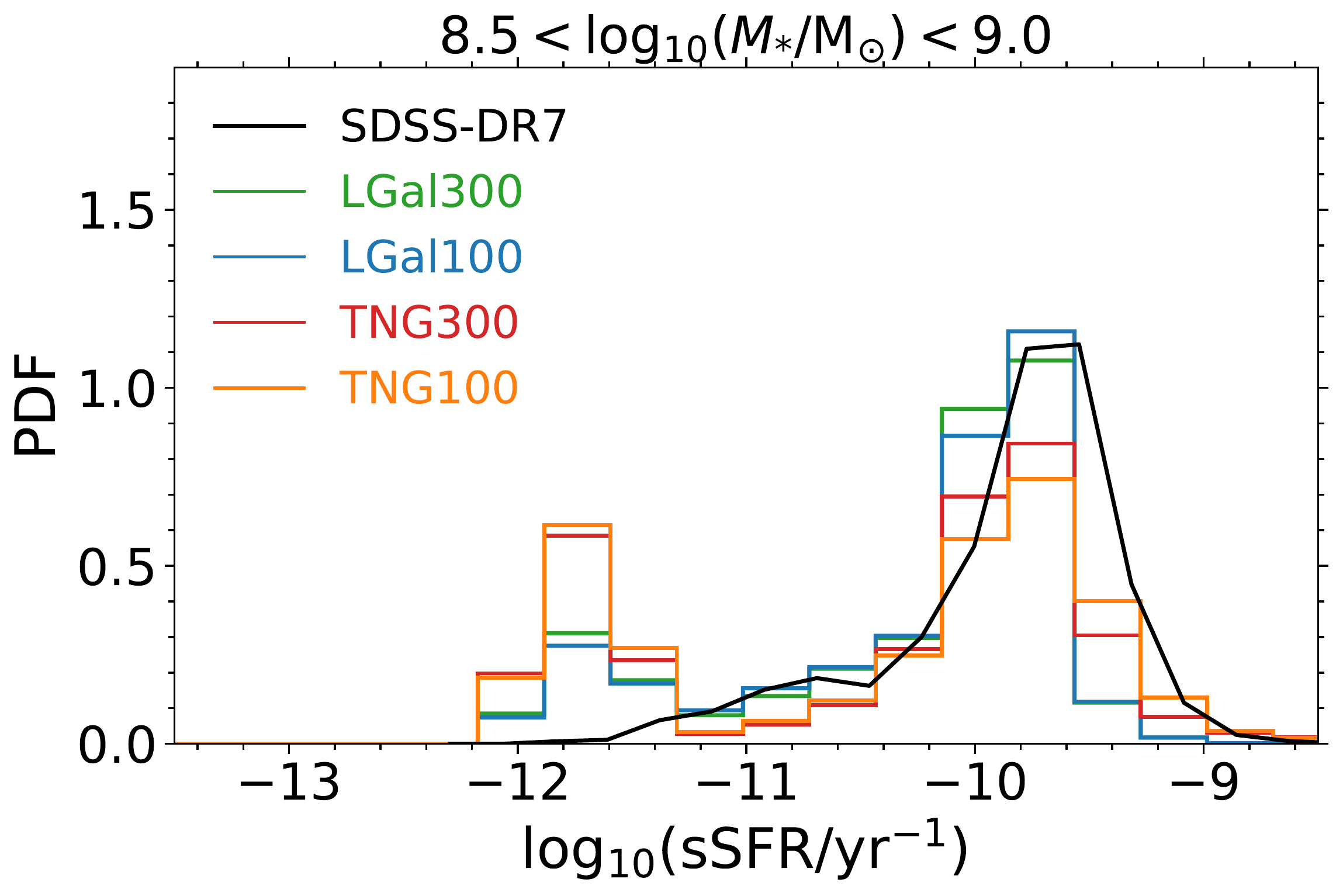}
    \includegraphics[width=0.30\textwidth]{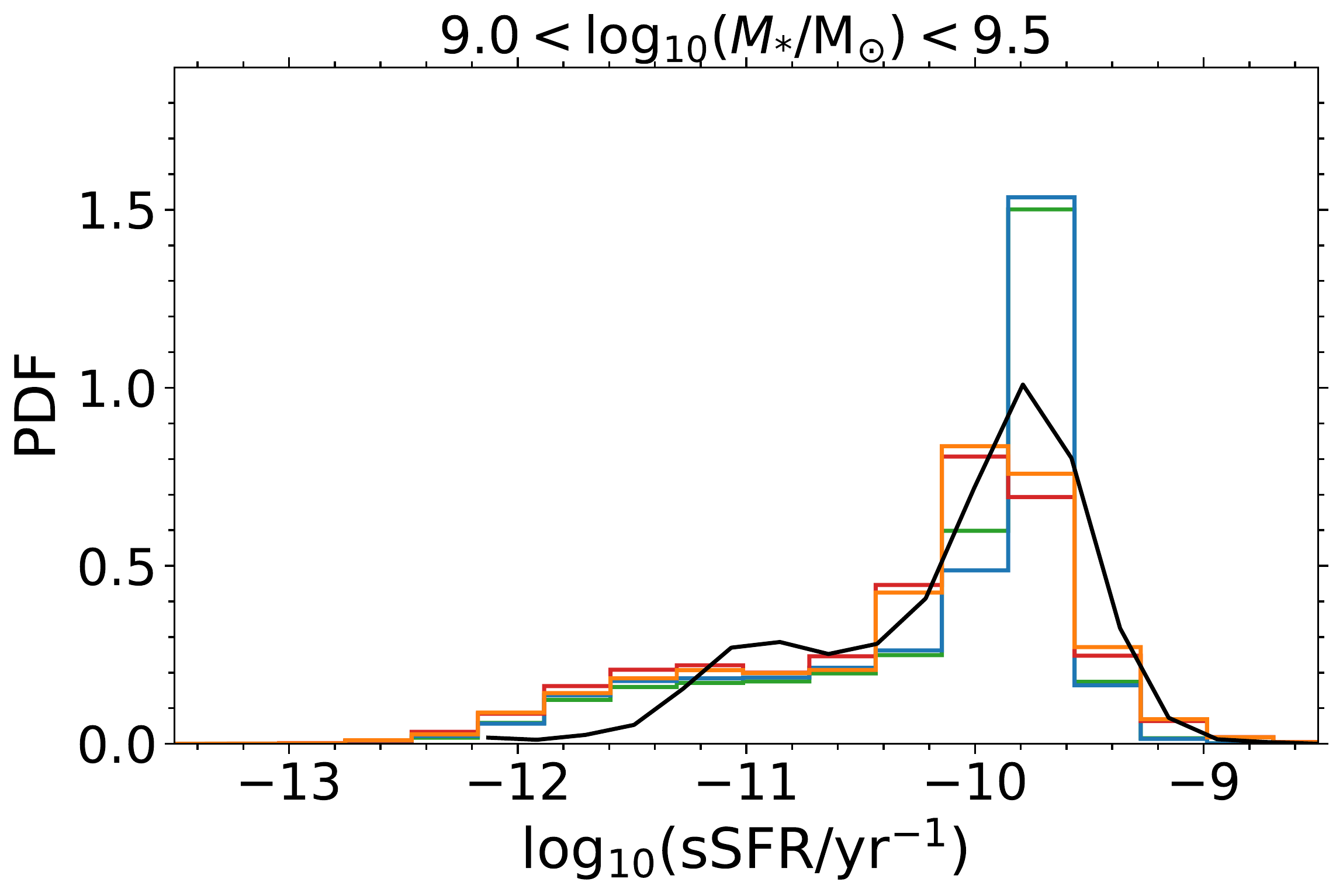}
    \includegraphics[width=0.30\textwidth]{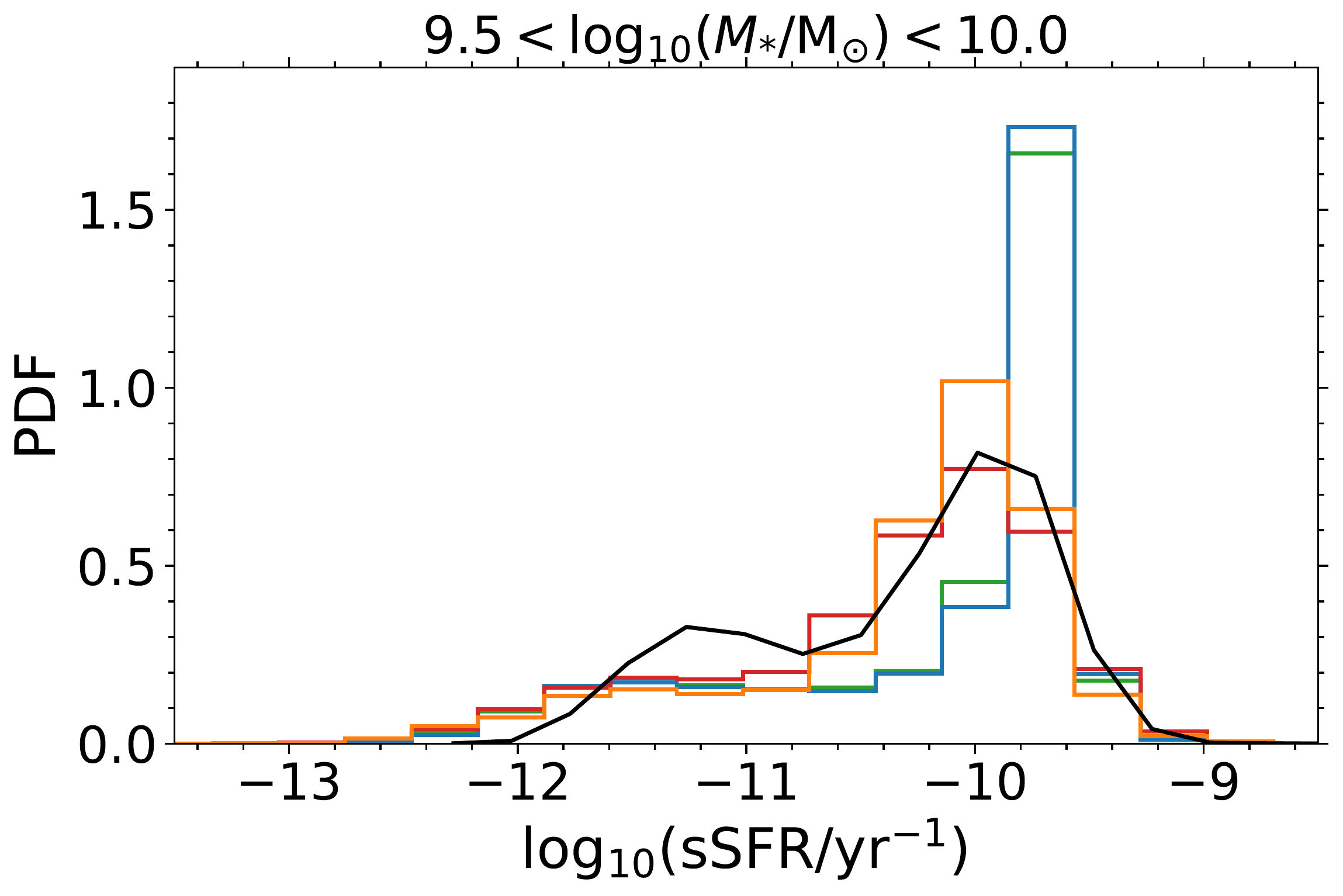}
    \includegraphics[width=0.30\textwidth]{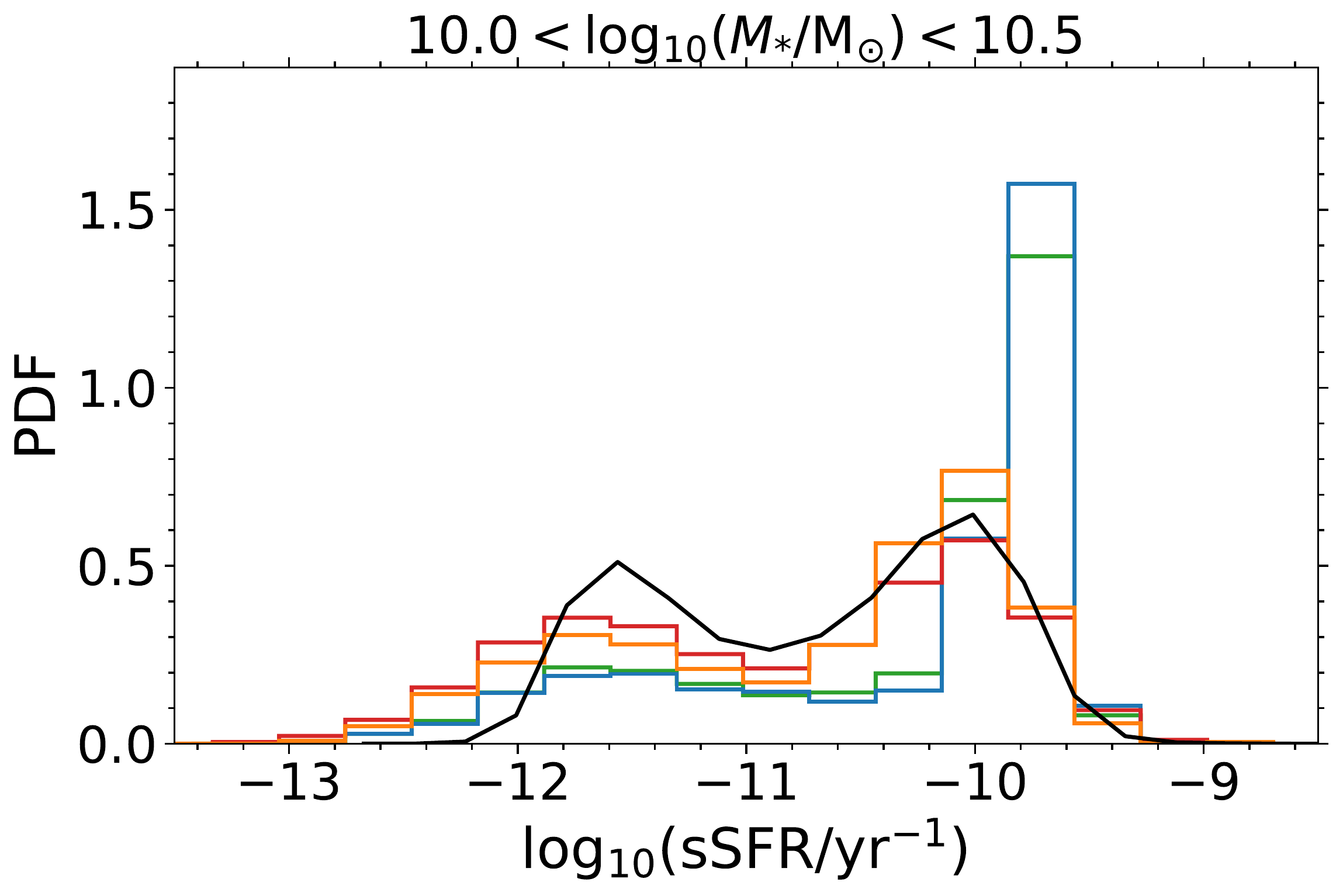}
    \includegraphics[width=0.30\textwidth]{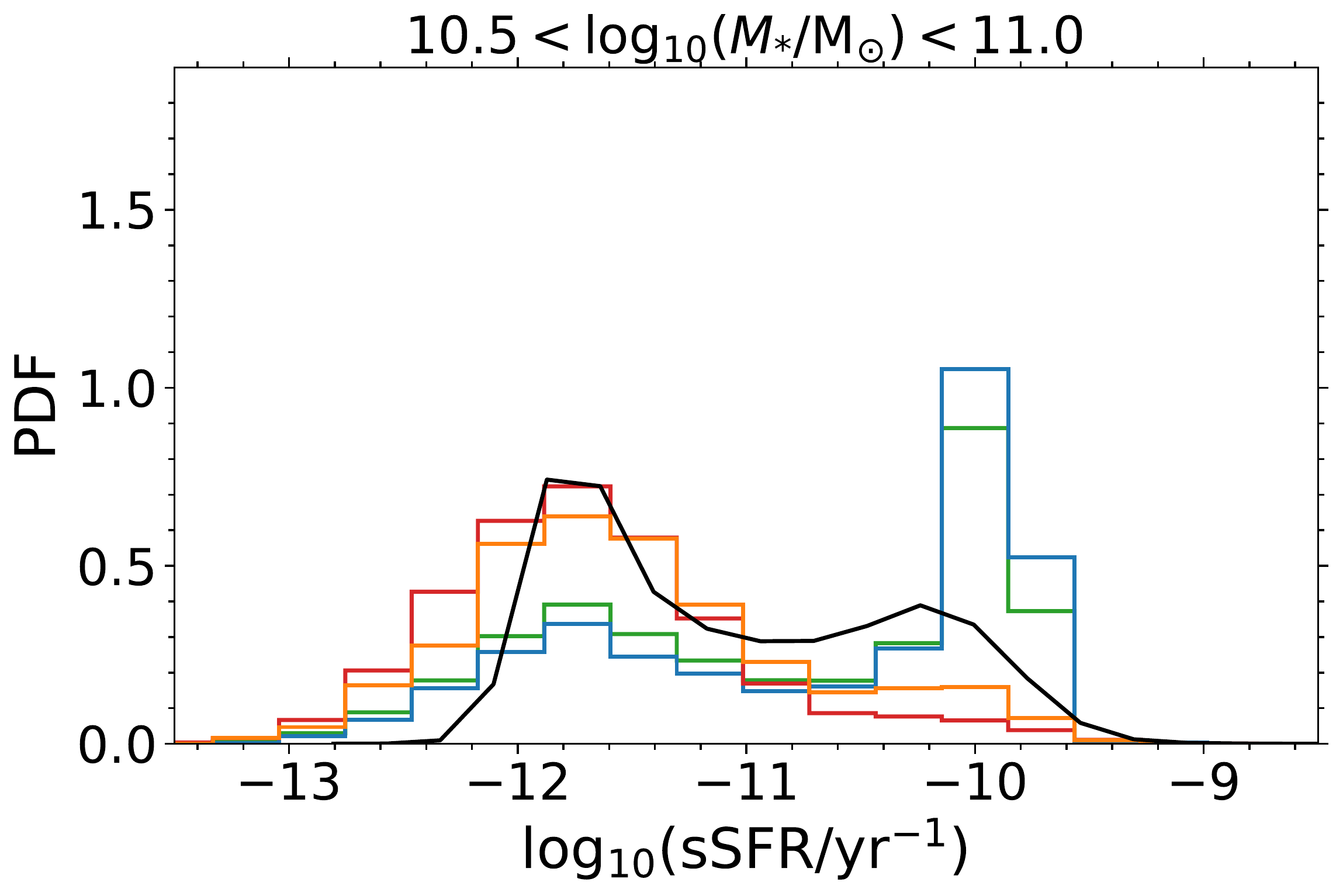}
    \includegraphics[width=0.30\textwidth]{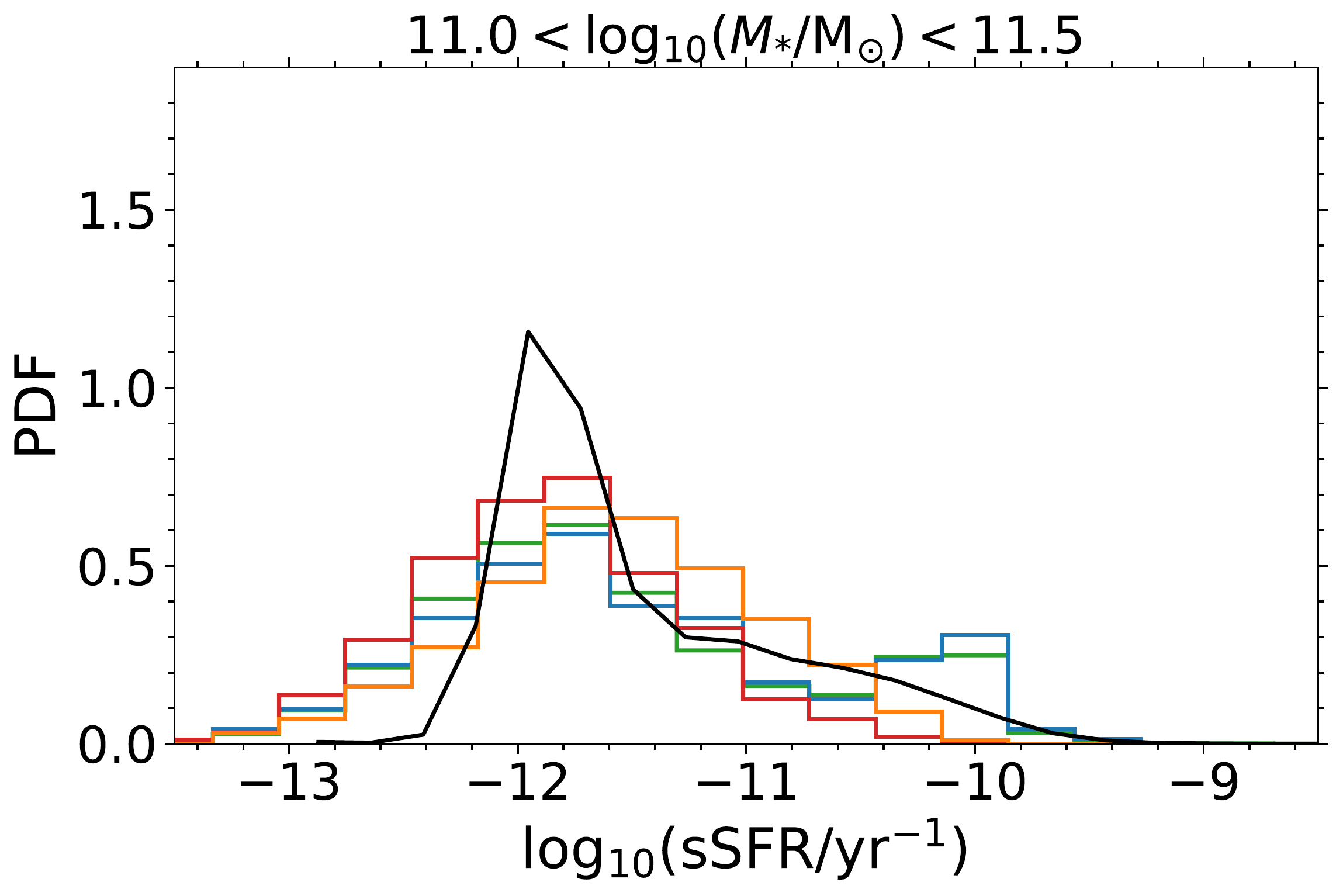}
    \caption{Histograms of sSFR for galaxies in \textsc{L-Galaxies}, TNG, and observations. Each panel corresponds to a particular stellar mass interval. Note that a minimum value of $\rm SFR=10^{-3}M_{\odot}/yr$ is enforced for the star formation rates of simulated galaxies (see \S \ref{subsec: gal_prop_definitions} for more detail). The simulated galaxies with $\rm \log_{10}(sSFR/yr^{-1}) < -12$ are assigned a value from a random Gaussian with mean and dispersion  $\mu = \log_{10}
    \langle{\rm sSFR}\rangle = -0.3\log_{10}(M_\star) - 8.6$ and $\sigma = 0.5$ in order to account for measurements of observed galaxies with no emission lines assigned a value for SFR based on SED fitting (see text). SDSS data are taken from \protect\cite{henriques2015galaxy,henriques2020galaxies} based on \protect\cite{brinchmann2004physical} with the corrections of \protect\cite{salim2007uv}.}
\label{Fig: ssfr_hist_Obs}
\end{figure*}

In general, for haloes with $\log_{10}(M_{200} / \rm M_{\odot}) \gtrsim 10.5$, TNG has a higher $f_{\rm star}/f_{\rm b,cosmic}$ than \textsc{L-Galaxies}, and this is true both within $R_{200}$ and $R_{500}$. Nevertheless, the difference between TNG and \textsc{L-Galaxies} is always less than a factor of two. The model predictions fall on top of each other at low-mass, but are much more discrepant at high mass. They overlap at $\log_{10}(M_{200} / \rm M_{\odot}) \lesssim 13$, but even the $2\sigma$ scatter is disjoint for clusters with $\log_{10}(M_{200}/\rm M_{\odot})\gtrsim 14$.

In this panel we also include several observational datasets for comparison. \cite{gonzalez2013galaxy} and \cite{kravtsov2018stellar} provide the total stellar mass within $R_{500}$, and so are comparable with the dashed lines from the simulations. In addition, \cite{chiu2018baryon} measures the ratio of stellar mass of the central galaxy versus host halo $M_{500}$, and should be compared with the dotted lines. As we report SMHM ratio as a function of $M_{200}$, any $M_{500}$ values are converted to $M_{200}$ using the ratio of these two quantities from TNG. This only shifts the x-axis values, leaving the SMHM ratios on the y-axis unchanged. The results from \textsc{L-Galaxies} are in better agreement with \cite{kravtsov2018stellar}, while the results from TNG are in a better agreement with \cite{gonzalez2013galaxy}. Both models agree reasonably well with \cite{chiu2018baryon}. We note that the scatter between different observations is much larger than the intrinsic scatter in either model.

\subsection{Galaxy star formation activity}
\label{subsec: ssfr}

Although \textsc{L-Galaxies} and TNG galaxies have similar stellar masses, these models have different star formation rates at $z=0$. Fig. \ref{Fig: SFRs} shows the density of galaxies in terms of star formation rate as a function of stellar mass, for \textsc{L-Galaxies} (top panel) and TNG (bottom panel). The star formation main sequence can be seen in both models, in qualitative agreement with observations. Overall, TNG galaxies are somewhat less star-forming than in \textsc{L-Galaxies}.

\subsubsection{Specific star formation rates}
\label{subsec: ssfr2}

The upper panel of Fig. \ref{Fig: ratio_SSFRs} illustrates the ratio of sSFR between TNG and \textsc{L-Galaxies} for individual galaxies, as a function of stellar mass at $z=0$. The lower sub-panel again shows the redshift evolution of the median relations, for $0 < z < 3$. At $z=0$, there are three distinct regimes: low-mass galaxies with $\log_{10}(M_{\star}/\rm M_{\odot}) \lesssim 10.5$, intermediate-mass galaxies with $10.5\lesssim \log_{10}(M_{\star}/\rm M_{\odot}) \lesssim 11.2$, and massive galaxies with $\log_{10}(M_{\star}/\rm M_{\odot}) \gtrsim 11.2$. The sSFRs of the first and the third sets of galaxies, i.e. low-mass and high-mass, are in reasonable agreement at a level of better than $0.3$ dex, depending on the stellar mass. The $1\sigma$ of scatter of the distribution (dashes lines) is $0.2-0.3$ dex for the low-mass galaxies but larger for the massive systems. 

On the other hand, the sSFRs of intermediate-mass galaxies differ substantially between \textsc{L-Galaxies} and TNG. Intermediate-mass galaxies are mostly quenched in TNG while a large fraction of them are still star-forming in \textsc{L-Galaxies}. This is caused primarily by differences in the implementation of SMBH feedback in the two models, namely the mass scale where strong feedback begins to quench galaxies. In TNG this occurs at $\log_{10}(M_{\star} / \rm M_{\odot}) \sim 10.5$, while in \textsc{L-Galaxies} quenching via AGN happens at characteristically higher stellar masses, $\log_{10}(M_{\star} / \rm M_{\odot}) \sim 11$.

The sSFR ratio between the models has a weak trend with redshift, as seen in the lower part of Fig. \ref{Fig: ratio_SSFRs}. The ratio $\rm sSFR^{TNG}/sSFR^{LGal}$ decreases with cosmic time for galaxies with $\log_{10}(M_{\star} / \rm M_{\odot}) \leq 10$. This implies that low-mass TNG galaxies are initially slightly more star-forming and their star formation rates decrease with time more rapidly than in \textsc{L-Galaxies}. The same is not true for massive galaxies, which have lower SFRs in TNG at $z=1$ but higher SFRs by $z=0$.

To understand the star formation activity of galaxies in more detail, Fig. \ref{Fig: ssfr_hist_Obs} shows sSFR histograms in six different stellar mass ranges, compared with observations from SDSS DR7 \citep{brinchmann2004physical}. Here we modify the intrinsic SFRs of simulated galaxies from both models to account for observed galaxies with no emission lines that were assigned a value of SFR based on SED fitting. This is done following the prescription of \cite{henriques2015galaxy}, where the sSFRs of simulated galaxies with $\rm \log_{10}(sSFR/yr^{-1})<-12$ are modelled by a random Gaussian with mean and dispersion $\mu = \log_{10} \langle{\rm sSFR}\rangle = - 0.3 \log_{10}(M_\star / \rm M_{\odot}) - 8.6$ and $\sigma = 0.5$.

\begin{figure*}
    \includegraphics[width=0.4\textwidth]{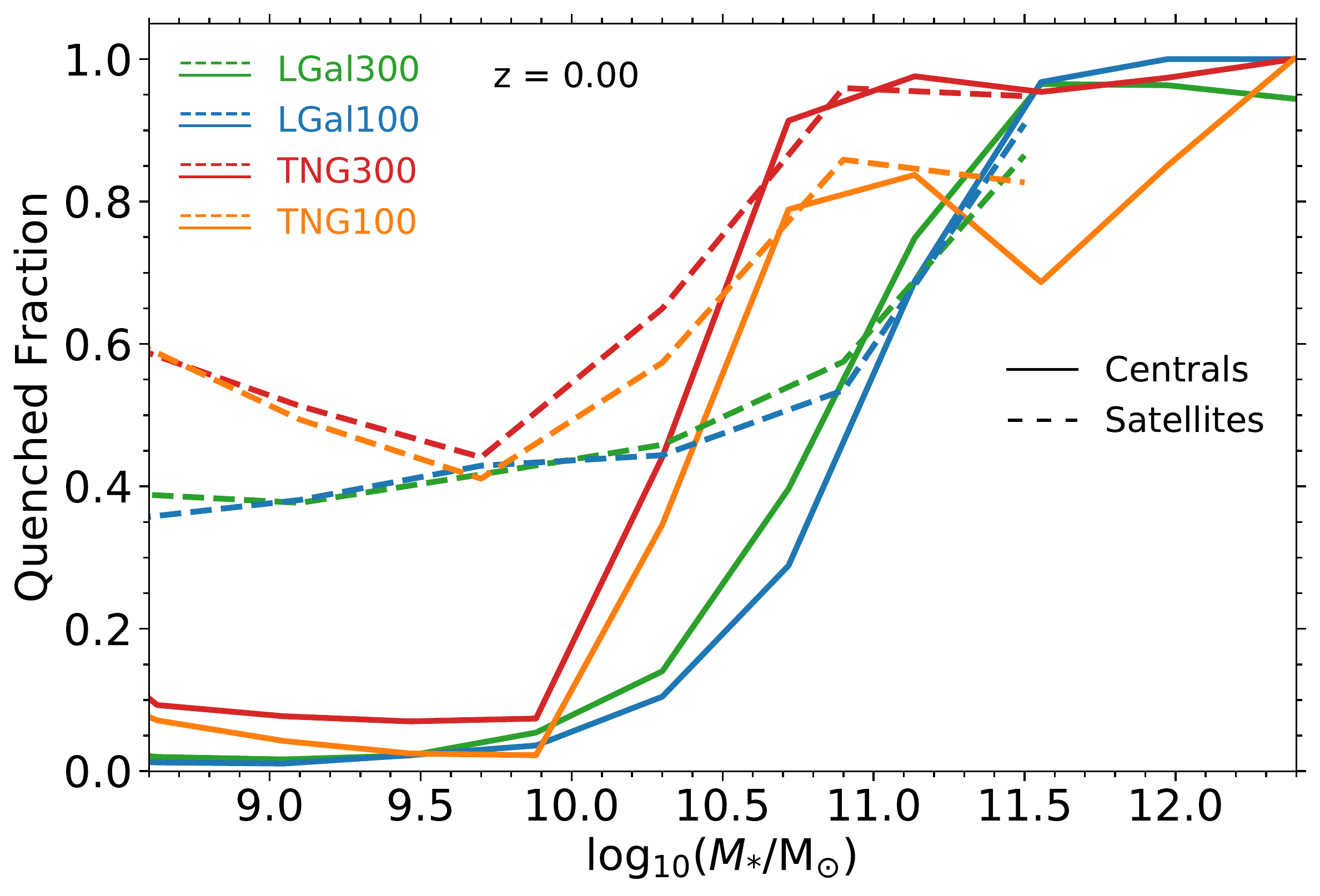}
    \includegraphics[width=0.4\textwidth]{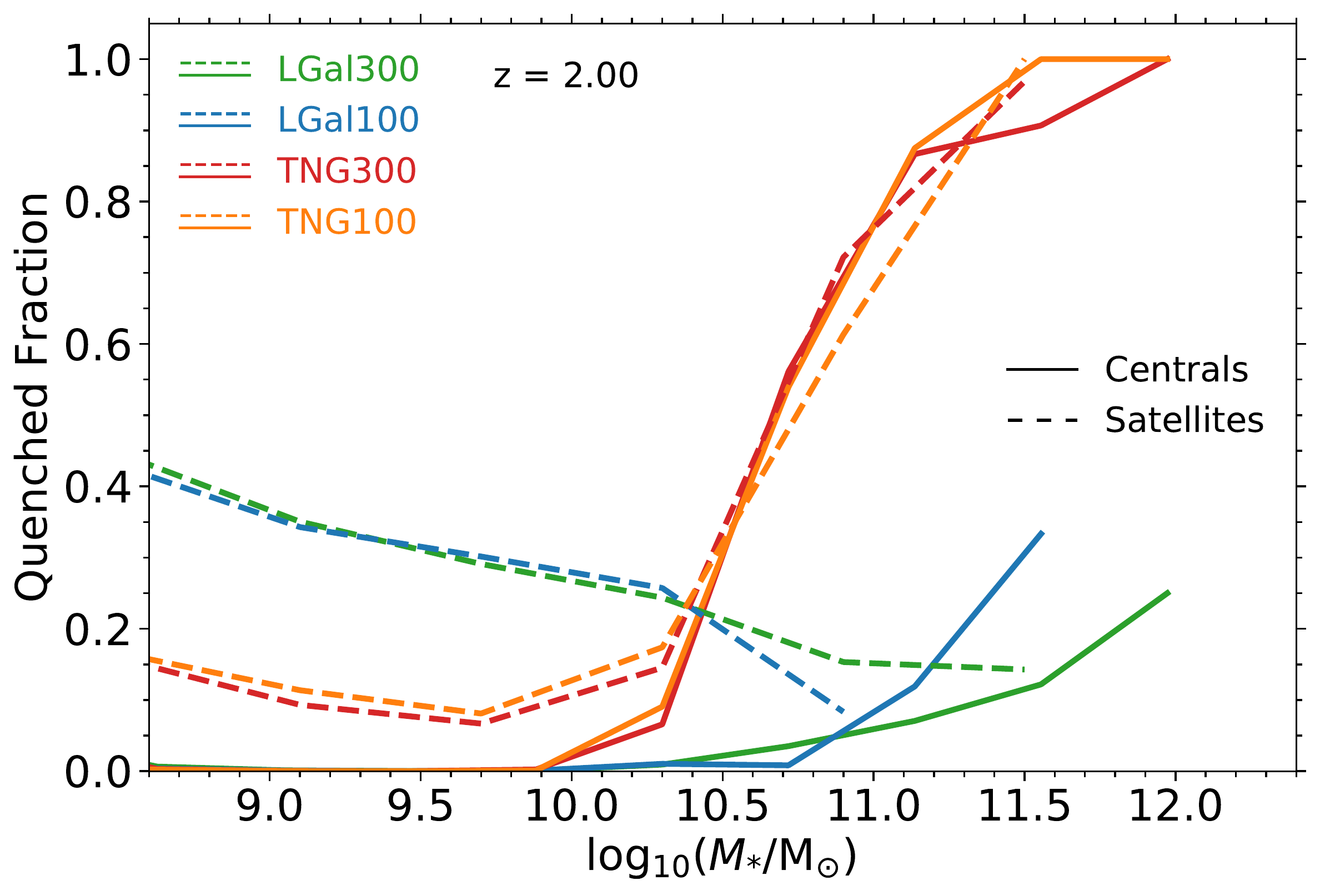}
    \caption{Fraction of quenched galaxies from \textsc{L-Galaxies} and TNG simulations as a function of stellar mass, at $z=0$ (left) and $z=2$ (right). We adopt the definition that galaxies with $\rm log_{10}(sSFR/yr^{-1})<-11$ and $\rm log_{10}(sSFR/yr^{-1})<-10$ are considered as quenched at $z=0$ and $z=2$, respectively.}
\label{Fig: quenchedfrac}
\end{figure*}

The joint distribution of (SFR,\,$M_\star$) is a stringent constraint on theory. Both models qualitatively reproduce trends seen in the SDSS data, where the main difference is in the transition from mostly star-forming to mostly quenched galaxies, which happens at lower mass in TNG than in \textsc{L-Galaxies}. More than half the galaxies in \textsc{L-Galaxies} are still actively star-forming at $10.5 < \log_{10}(M_{\star} / \rm M_{\odot}) < 11$ and the population only becomes mostly quenched at $11 < \log_{10}(M_{\star} / \rm M_{\odot}) < 11.5$. In contrast, there is a transition from star-forming to quenched in TNG over the range $10.0 < \log_{10}(M_{\star} / \rm M_{\odot}) < 11$. In the middle of this regime the distribution of sSFRs becomes strongly bi-modal, as also seen in SDSS. The star-forming main sequence is narrower, spanning a smaller range of SFRs, in \textsc{L-Galaxies} than in TNG or in the observations, both of which favour a broader main sequence shifted to somewhat lower values of sSFR \footnote{We note that recent developments in \cite{henriques2020galaxies} have lead to improvements to the sSFRs and HI-to-stellar mass ratios.}.

\subsubsection{Fraction of quenched galaxies}
\label{subsec: quenchedFrac}

Adopting definitions for quiescence based on galaxy sSFR, Fig. \ref{Fig: quenchedfrac} shows the quenched fraction of galaxies as a function of stellar mass, at $z=0$ (left) and $z=2$. At $z=0$ we consider galaxies quenched if $\rm log_{10}(sSFR/yr^{-1}) < -11$, while for $z=2$ the threshold moves to $\rm log_{10}(sSFR/yr^{-1}) < -10$ to account for the redshift evolution of the star-forming main sequence. We show quenched fractions for central galaxies (solid lines) and satellites (dashed lines) separately. At $z=0$, low mass central galaxies are mostly star-forming, whereas massive galaxies are predominantly quenched. At the same time, a large fraction of low mass satellite galaxies (about 40-50\%) are quenched at all masses. At all redshifts, satellite galaxies are more quenched than centrals of the same stellar mass, reflecting the strong impact of environmental effects.

In TNG, regardless of redshift, the transition from star-forming to quenched happens at $\log_{10}(M_{\star} / \rm M_{\odot}) \sim 10-11$ for both central and satellite galaxies. Indeed, more than 90\% of the TNG massive galaxies with $\log_{10}(M_{\star} / \rm M_{\odot}) \gtrsim 11$ are quenched. TNG has almost no quenched centrals with $\log_{10}(M_{\star} / \rm M_{\odot}) \lesssim 9.5$, while about 40-50\% (10-15\%) of its satellites are quenched in this mass range at $z=0$ ($z=2$). In \textsc{L-Galaxies}, on the other hand, central quiescence is a strong function of redshift. At $z=2$, roughly 20-30\% of massive centrals, $\log_{10}(M_{\star} / \rm M_{\odot}) \gtrsim 11$, are quenched, dropping to less than 5\% for lower mass galaxies. More than 20\% of satellites are quenched at all masses. At $z=0$, quenched fractions for both centrals and satellites in \textsc{L-Galaxies} are much more similar to TNG, except shifted to higher stellar masses.

The transition from predominantly star-forming to predominantly quenched galaxies for both TNG and \textsc{L-Galaxies} is caused by supermassive black hole feedback, producing similar trends of quenched fraction versus stellar mass. In TNG, the transition point where the quenched fraction equals 50\% occurs at a stellar mass roughly 0.5 dex lower than in \textsc{L-Galaxies}. The most significant difference is at $z=2$, where effective black hole feedback in TNG produces a population of quenched galaxies even at early times \citep[see][]{donnari19}. In contrast, the quenching mechanism in \textsc{L-Galaxies} does not operate  as strongly at high redshifts, which results in a much lower abundance of massive quenched galaxies \citep[see][for an observational perspective]{martis16}.

\begin{figure*}
    \includegraphics[width=1\columnwidth]{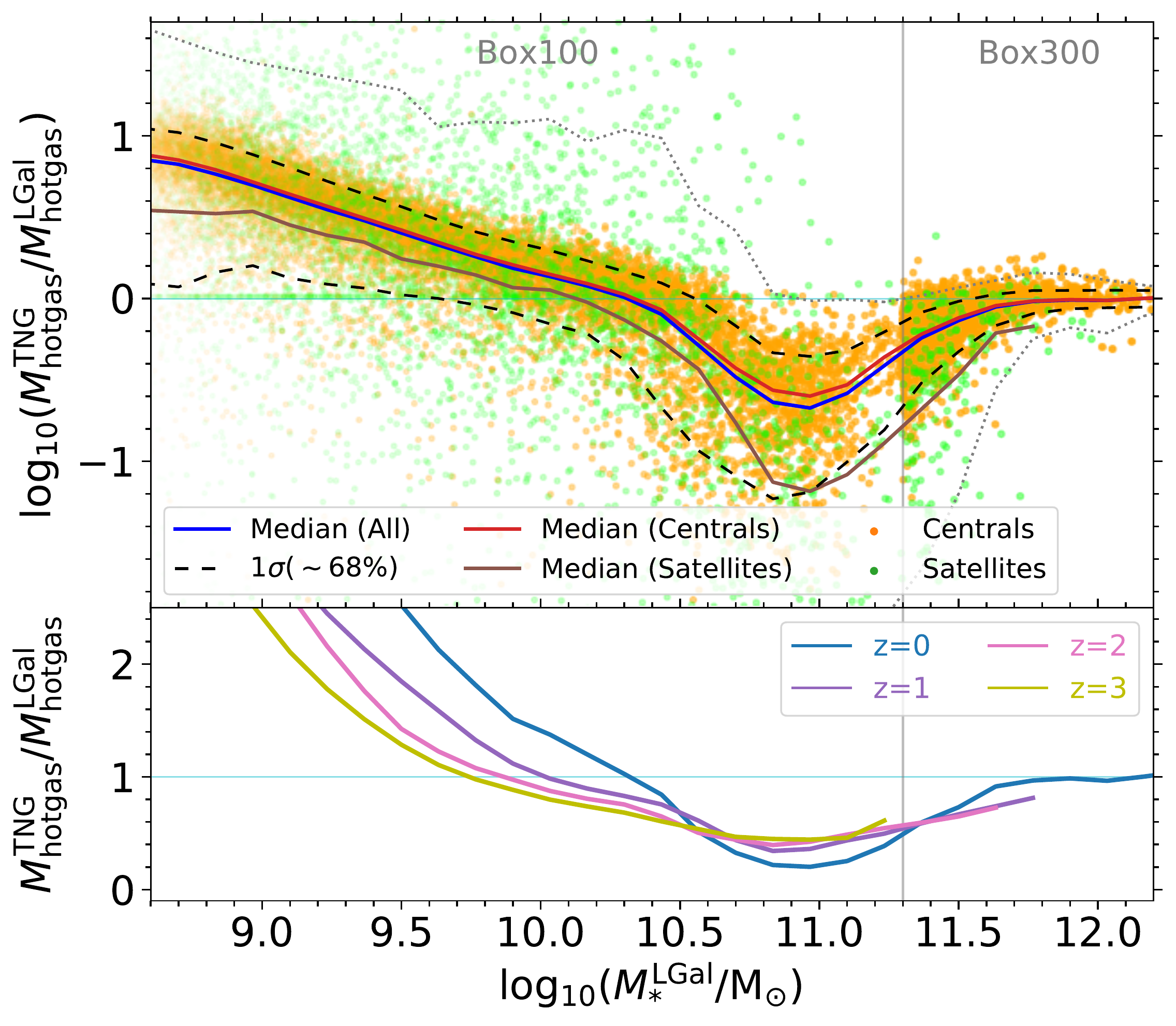}
    \includegraphics[width=1\columnwidth]{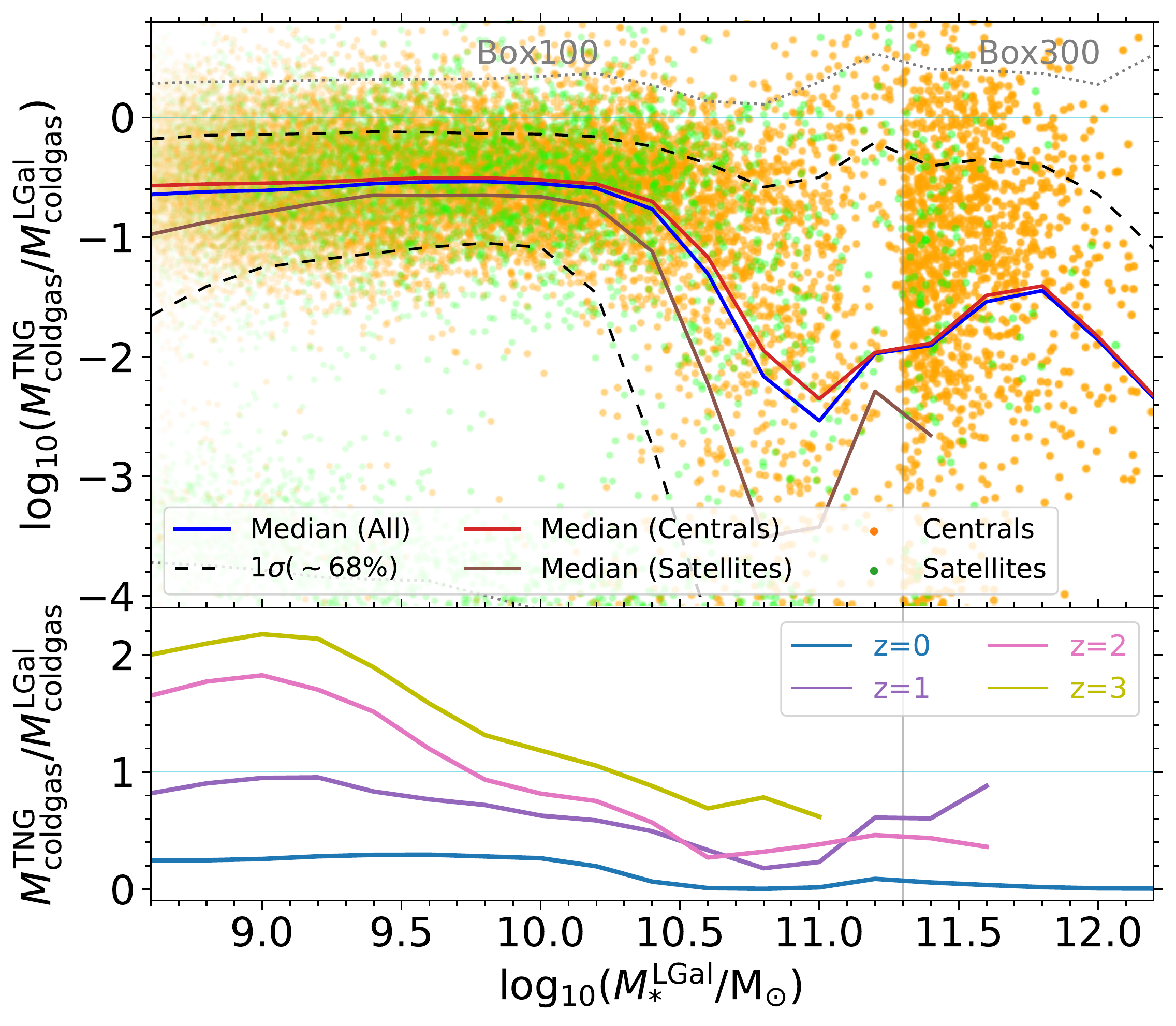}
    \caption{The ratios of hot gas mass (left panel) and cold gas mass (right panel) for individual objects between the two models. Each dot corresponds to a single galaxy; centrals and satellites are shown in orange and green, respectively. Solid lines show medians, while dotted and dashed lines show $1\sigma$ and 2$\sigma$ scatter. The bottom sub-panels emphasise the redshift evolution of the median relations, from $0 < z < 3$. While low-mass TNG galaxies have overall more hot gas, the trend is reversed at high masses. On the other hand, TNG galaxies of all masses have less cold (star-forming) gas than in \textsc{L-Galaxies}, by a factor of a few at low masses, with a maximal difference at $\log_{10}(M_{\star}/\rm M_{\odot}) \sim 11$ of $> 1$ dex, due to a combination of feedback and environmental effects.}
\label{Fig: Hot_Cold_Gas_ratio}
\end{figure*}

\subsection{Gas content of galaxies and subhaloes}
\label{subsec: gas_content}

The way in which cosmic gas is treated is one of the principal differences between hydrodynamical simulations such as TNG and semi-analytical models including \textsc{L-Galaxies}. In order to compare the gas contents of galaxies between these two models, we divide the gas cells in TNG subhaloes into two categories, to mimic the hot and cold gas components modelled in \textsc{L-Galaxies}. We consider, for each subhalo, the hot gas to be the sum of all the bound gas cells with $\rm SFR=0$ and the cold gas to be the sum of the ones with $\rm SFR>0$. By definition, these two subsets are disjoint, and sum to the total mass of gas. We note that in \textsc{L-Galaxies} the third gas reservoir of ejected material is not spatially specified, but for our purposes here we consider it to be entirely outside the (sub)halo.

The ratio of hot gas mass between matched galaxies is given in the left panel of Fig. \ref{Fig: Hot_Cold_Gas_ratio}. The median (blue solid line) relation shows that galaxies with $8.5 \lesssim \log_{10}(M_{\star} / \rm M_{\odot}) \lesssim 10.5$ have more hot gas in TNG than in \textsc{L-Galaxies}. This difference increases towards lower stellar masses, reaching nearly 1 dex by $\log_{10}(M_{\star}/\rm M_{\odot})\sim 8.5$. Across this stellar mass range a large fraction of the hot gas of \textsc{L-Galaxies} subhaloes has been pushed into the ejected reservoir by supernova feedback. On the other hand, galaxies with $10.5 \lesssim \log_{10}(M_{\star} / \rm M_{\odot}) \lesssim 11.5$ are more gas-rich in \textsc{L-Galaxies} than in TNG, because much of the hot gas in TNG galaxies has been pushed out of the halo by AGN feedback. Finally for galaxies with $\log_{10}(M_{\star}/\rm M_{\odot})\gtrsim11.5$ the median ratio approaches unity. These trends are similar for centrals and satellites, although the satellite ratios are always lower than the central ratios by $\gtrsim 0.1$ dex. This shows that, in addition to differences in feedback/internal processes, there are also considerable differences in environmental effects which we will analyse in more detail in \S \ref{sec: gal_vs_dis}.

The left sub-panel of Fig. \ref{Fig: Hot_Cold_Gas_ratio} depicts median individual hot gas ratios as a function of stellar mass at four different redshifts. A clear trend can be seen with time. The ratios are closer to unity at earlier cosmic times, particularly for low-mass galaxies.

Moving to the cold (star-forming) gas component, the ratios of cold gas mass between TNG and \textsc{L-Galaxies} are shown in the right panel of Fig. \ref{Fig: Hot_Cold_Gas_ratio}. The median ratio of all galaxies (blue solid line) shows that galaxies have lower cold gas masses in TNG than in \textsc{L-Galaxies} at $z=0$, and this is true for all stellar masses. A rapid decrease in the median ratio occurs at $\log_{10}(M_{\star} / \rm M_{\odot}) \sim 10.5-11$ as many TNG galaxies become depleted or even devoid of cold star-forming gas. Note that in TNG, the smallest amount of cold gas detectable in a galaxy is roughly the mass of a single gas cell, $\log_{10}(m_{\rm cell}/\rm M_{\odot}) \sim 6-7$. In \textsc{L-Galaxies} there is a cold gas mass threshold, $M_{\rm crit}$, below which no star formation is allowed to occur \citep[see][Eq. S15]{henriques2015galaxy}. As a side effect, quenched galaxies with no ongoing star formation can typically maintain a non-negligible amount of cold gas.

On average, the ratio for satellites (brown solid line) is below that for centrals (red solid line) by at least 0.1 dex. The right sub-panel of Fig. \ref{Fig: Hot_Cold_Gas_ratio} shows median ratios for cold gas as a function of stellar mass at different redshifts. There is a strong trend with redshift: at higher redshifts ($z \sim 3$), low-mass TNG galaxies have more cold gas and $M_{\rm coldgas}^{\rm TNG}/M_{\rm coldgas}^{\rm LGal} > 1$, but this ratio decreases with time and drops below unity for $z < 1$. On the other hand, at the high mass end and $\log_{10}(M_{\star}/\rm M_{\odot}) \gtrsim 10.5$, cold gas masses are always smaller in TNG galaxies for $z < 3$, and this difference becomes slightly more pronounced towards redshift zero.

\subsection{Halo baryon and gas fractions}
\label{subsec: halo_baryonFrac}

\begin{figure*}
    \includegraphics[width=1\columnwidth]{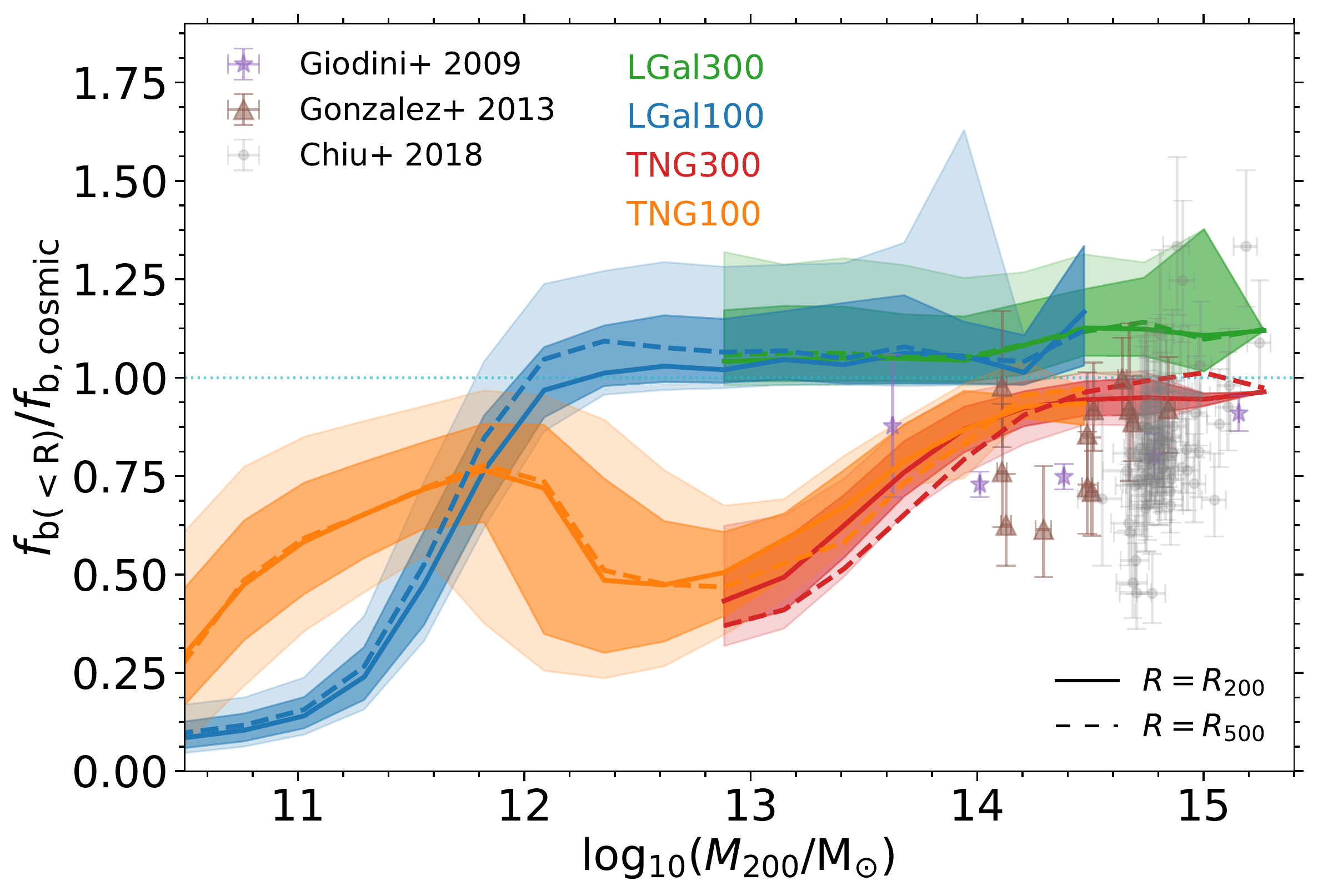}
    \includegraphics[width=1\columnwidth]{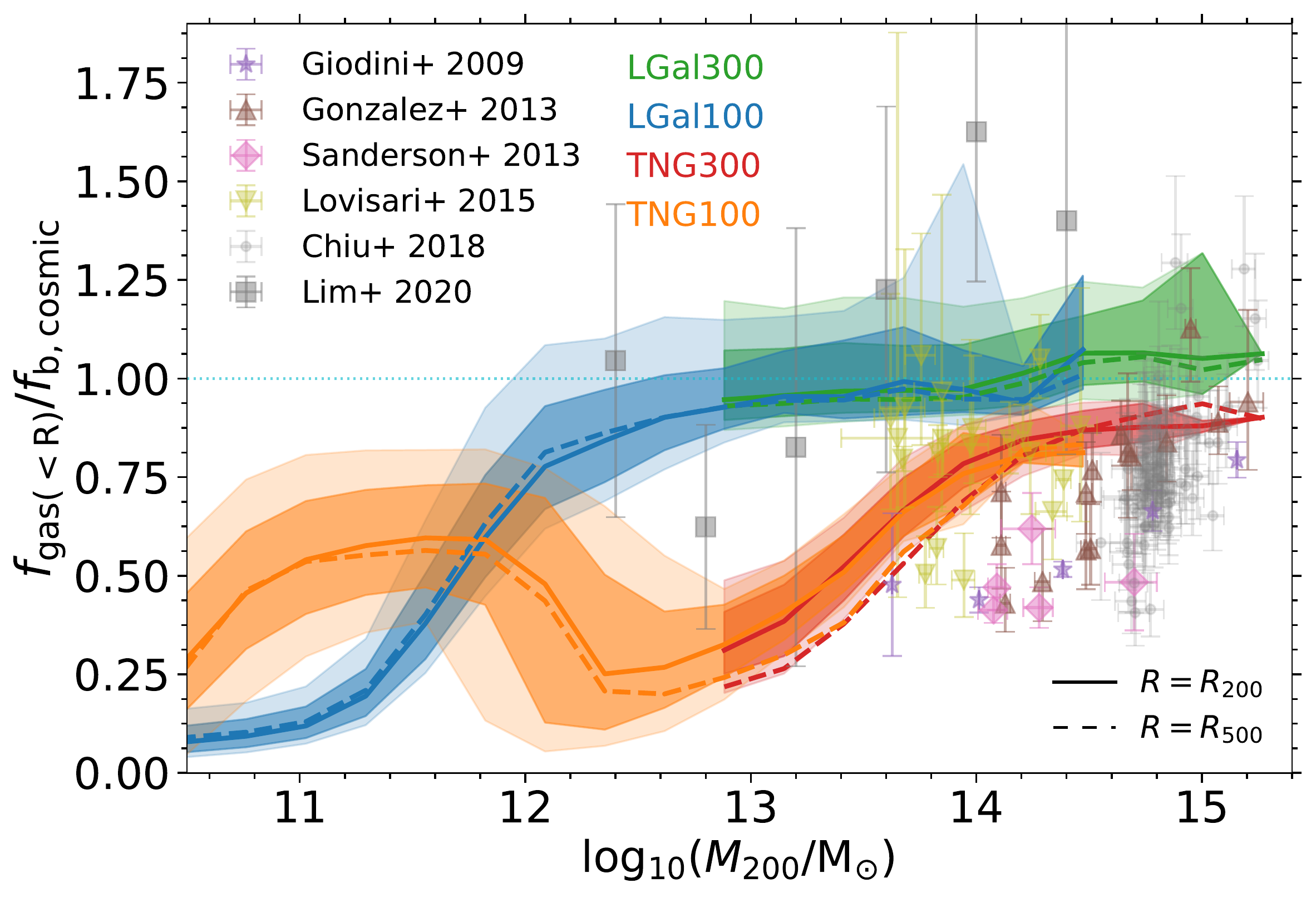}
    \caption{The baryon (left panel) and gas (right panel) fractions against halo $M_{200}$. The solid and dashed lines correspond to the median of the total baryon/gas to halo mass ratio within $R_{200}$ and $R_{500}$ respectively. The dark and light shaded regions illustrate $1\sigma$ and $2\sigma$ of the distribution for within $R_{200}$ case. All the observations illustrate ratios within $R_{500}$ except for \protect\cite{lim2020detection} which corresponds to the gas ratio within $R_{200}$.}
\label{Fig: baryonFrac_LGal_TNG}
\end{figure*}

Recent measurements of the cosmic microwave background show that the cosmic fraction of baryons is $f_{\rm b,cosmic} \sim 0.16$ \citep{2018arXiv180706209P}, and the baryon fraction in haloes is observed to be close to the cosmic value in massive galaxy clusters with $\log_{10}(M_{200} / \rm M_{\odot}) > 14$, \citep[e.g. see][]{sanderson2013baryon,chiu2018baryon}. However, many studies have shown that the baryon fraction in lower mass haloes, $13 < \log_{10}(M_{200}  /\rm M_{\odot}) < 14$, can be much lower than this value \citep[e.g. see][]{david2006hot,lovisari2015scaling}, a manifestation of the `missing baryons' problem. 

Fig. \ref{Fig: baryonFrac_LGal_TNG} compares theoretical predictions for halo baryon fractions in  \textsc{L-Galaxies} versus TNG, and shows how these results stack up against available observations. The observational data are based on a number of different sample selection and gas detection techniques. \cite{lim2020detection} measures gas content within $R_{200}$ using the kinematic Sunyaev–Zel'dovich effect (\citealt{sunyaev1972observations}, hereafter "SZE"). All other datasets measure gas content within $R_{500}$  based on X-ray emission: \cite{giodini2009stellar,gonzalez2013galaxy,lovisari2015scaling} have X-ray selected samples, \cite{sanderson2013baryon} use an optically selected sample and \cite{chiu2018baryon} employ a sample of galaxy clusters selected based on SZE. In Fig. \ref{Fig: baryonFrac_LGal_TNG} we report baryon and gas fractions as a function of $M_{200}$, so we have converted any $M_{500}$ values to $M_{200}$ using the ratio of these two quantities from TNG. We note that this shifts the x-axis values, while baryon and gas fractions on the y-axis are unchanged.

For the models, we show baryon and gas fractions both within $R_{200}$ (solid lines) and within $R_{500}$ (dashed lines). The dark and light shaded regions correspond to the $1\sigma$ and $2\sigma$ scatter of the $R_{200}$ measurements. We note that \textsc{L-Galaxies} only considers total gas within $R_{200}$, and we assume an isothermal $\rho \propto r^{-2}$ profile for hot gas to derive the mass contained within $R_{500}$ for consistency with the choice of profile assumed in the modelling of processes such as cooling and
ram-pressure stripping. Other choices, such as a $\beta$ profile, would change the gas and baryon fractions within $R_{500}$ \citep[e.g. see Fig. 6 of][for groups and clusters]{Yates+17}.

For some haloes, \textsc{L-Galaxies} baryon fractions are somewhat greater than $f_{\rm b,cosmic}$. The reason is due to its implementation of environmental effects such as tidal and ram-pressure stripping, as well as its gas-infall recipe, as discussed in \cite{Yates+17}. Due to tidal effects, galaxies in the infall regions lose dark matter in the DMO simulation on which \textsc{L-Galaxies} run. However, as there is no gas stripping for those galaxies (see \S \ref{subsec: important_processes_Env}), they do not lose any gas, artificially increasing the baryon fraction of their host halo. This is resolved in \cite{ayromlou2019new} with our new gas stripping implementation, which treats all galaxies based on measurements of their local environment. In addition, if the halo $M_{200}$ decreases because of morphology or concentration changes, the gas mass remains unchanged, and this causes a spurious increase in baryon fraction. \cite{Yates+17} resolve this issue by correcting input halo merger trees to prevent $M_{200}$ from decreasing with cosmic time. We will implement similar corrections in future work.

For clusters with $\log_{10}(M_{200} / \rm M_{\odot}) \gtrsim 14$, both \textsc{L-Galaxies} and TNG predict halo baryon fractions close to the cosmic value, in agreement with most observations. Both models have a relatively small scatter. Clusters are the only case where \textsc{L-Galaxies} and TNG produce similar baryon fractions, while otherwise the outcomes are notably different. In lower mass haloes with $10 \lesssim \log_{10}(M_{200} / \rm M_{\odot}) \lesssim 12$, \textsc{L-Galaxies} baryon and gas fractions are lower than in TNG. This is mainly due to strong supernova feedback in \textsc{L-Galaxies}, which pushes material, in particular hot gas, into an ejected reservoir for some time. 

For haloes with $12 \lesssim \log_{10}(M_{200} / \rm M_{\odot}) \lesssim 14$, the TNG baryon fraction is significantly lower than the cosmic value in contrast with \textsc{L-Galaxies}, where these fractions are almost at the cosmic value for most haloes with $\log_{10}(M_{200}/\rm M_{\odot}) \gtrsim 12$. This is primarily caused by different implementations of AGN feedback in the two models, as described in \S \ref{subsec: important_processes_AGN}. Briefly, AGN feedback in TNG is able to push gas out of the halo, while in \textsc{L-Galaxies} it only prevents cooling onto the central galaxy, without changing the total baryonic mass.

Comparing the group mass scale $13 \lesssim \log_{10}(M_{200} / \rm M_{\odot}) \lesssim 14$ against observations, TNG baryon fractions agree better with \cite{lovisari2015scaling}, while \textsc{L-Galaxies} haloes are in a better agreement with \cite{lim2020detection}. For lower mass haloes with $12 \lesssim \log_{10}(M_{200} / \rm M_{\odot}) \lesssim 13$, \textsc{L-Galaxies} predictions are in better agreement with \cite{lim2020detection} who find values close to the cosmic baryon fraction. We note that many subtleties exist in a proper comparison of these quantities. For example, the halo mass itself is computed in simulations using the three-dimensional spherical overdensity value given the total matter distribution, whereas in observations this quantity must be indirectly derived from an observable such as galaxy luminosity or stellar mass \citep[e.g. see][]{yang2007galaxy}. Detailed comparison of halo $f_{\rm gas}(M_{200})$ trends would benefit from mock X-ray and/or SZE measurements from the simulations.


\section{Environmental dependency of galaxy evolution}
\label{sec: gal_vs_dis}

\begin{figure*}
    \includegraphics[width=1\columnwidth]{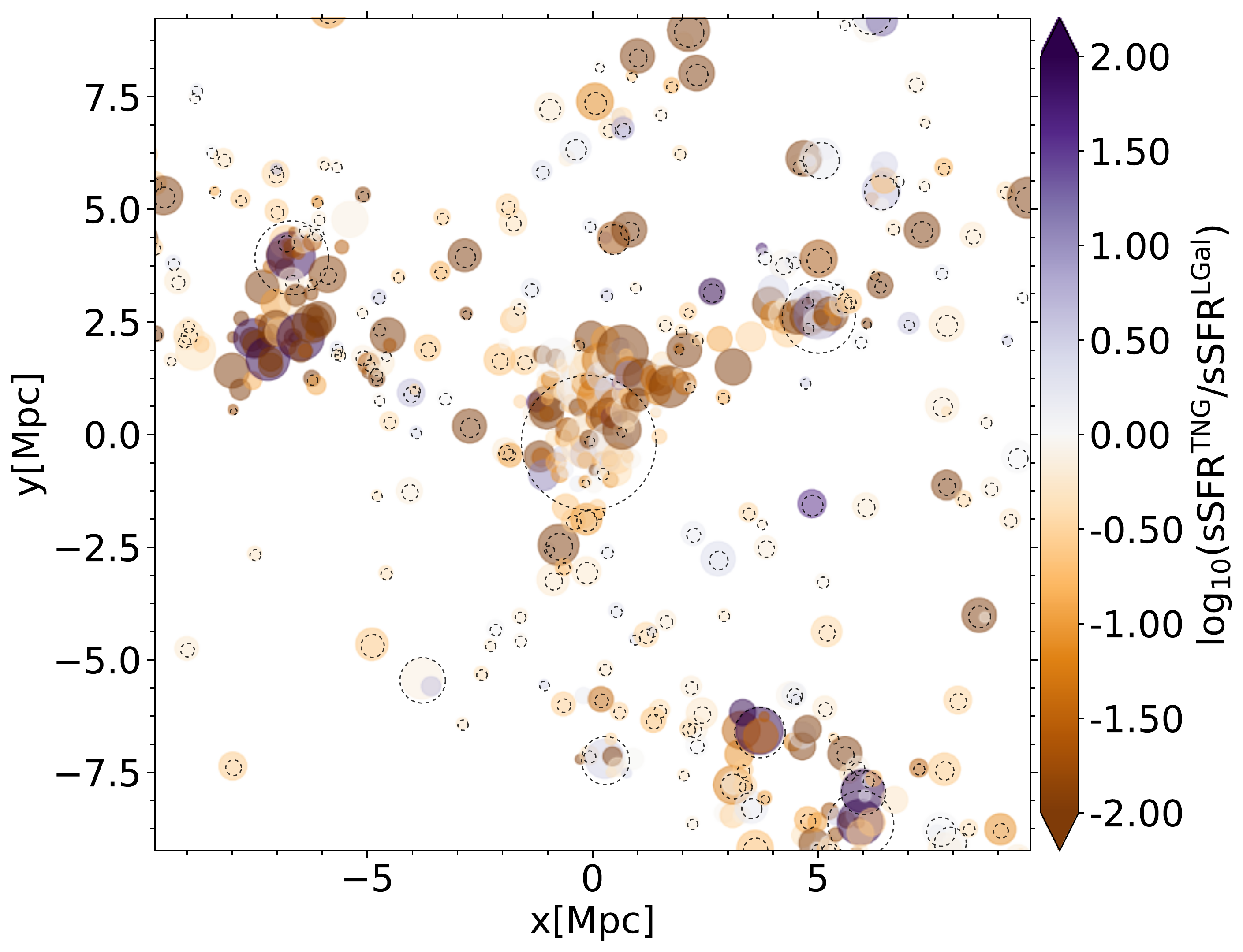}
    \includegraphics[width=1\columnwidth]{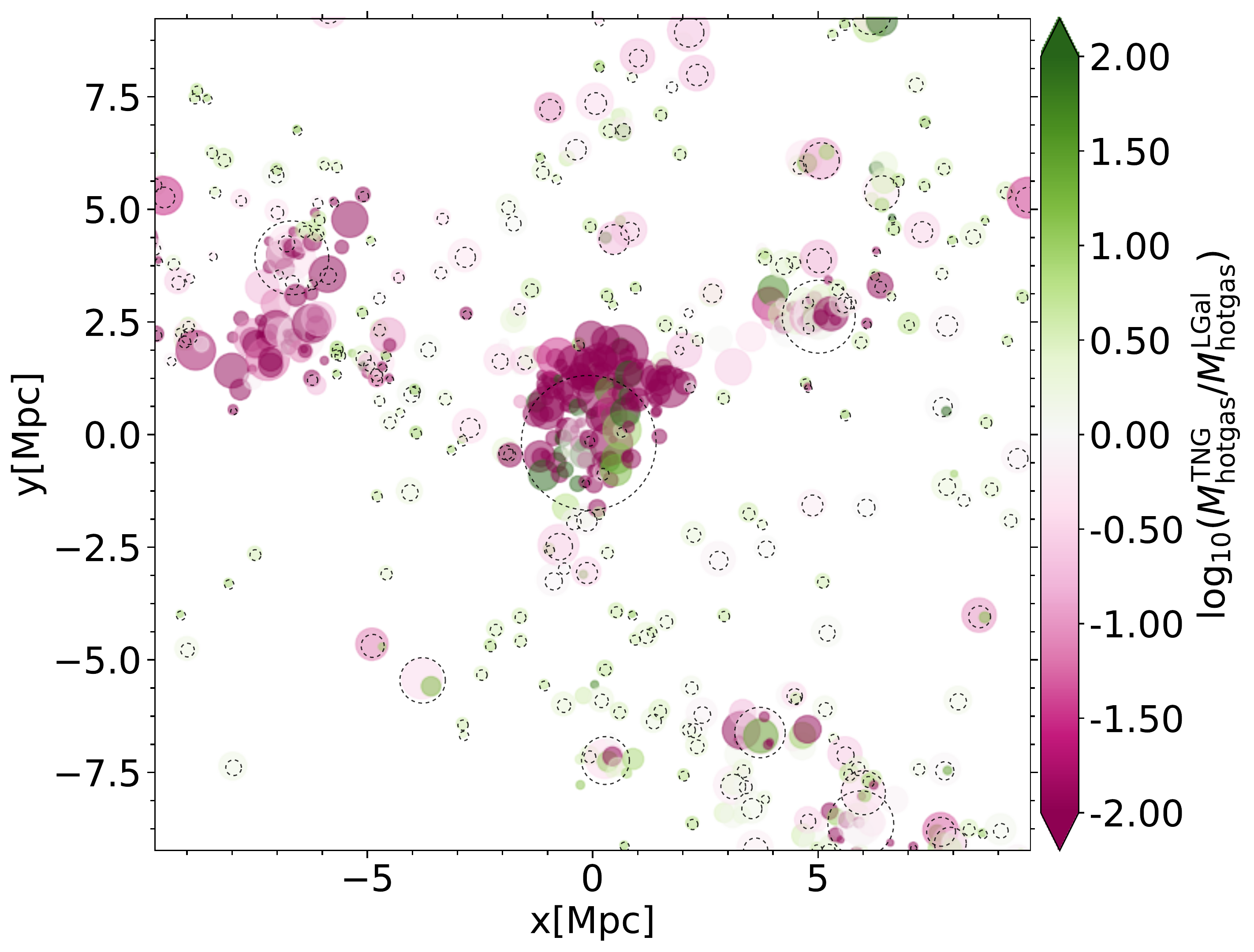}
    \caption{Galaxies within a region of size $\rm 8\, Mpc$ centred on a massive cluster. Each circle shows a galaxy with its size reflecting the galaxy's stellar mass. Colours correspond to the ratio between predictions of TNG and \textsc{L-Galaxies} for sSFR (left panel) and hot gas mass (right panel). The dashed grey circles depict the boundaries, $R_{200}$, for various haloes. The galaxies shown are those matched in the two simulations and all have $\log_{10}(M_{\star} / \rm M_{\odot}) > 8.5$ in \textsc{L-Galaxies}.}
\label{Fig: box_schematic2}
\end{figure*}

As shown in \S \ref{sec: Galaxies_and_Subhalos}, the differences between \textsc{L-Galaxies} and TNG are generally stronger for satellite galaxies than for centrals, due to the different treatment of key environmental processes, which we now study in more detail. We analyse properties of galaxies in and around haloes as a function of host halo mass, from low-mass groups to clusters. As our analysis is focused on more massive haloes, in this section we use the larger volume TNG300 simulation and the corresponding LGal300, unless stated otherwise.

Fig. \ref{Fig: box_schematic2} depicts galaxies surrounding a single cluster with $M_{200}\sim 10^{14} \rm M_{\odot}$, out to $\sim 6R_{200}$ in the $\rm 100 \, Mpc$ box. Several nearby haloes can also be seen in this region, and we mark their $R_{200}$ with grey circles. Every galaxy is denoted by a circle, whose colour indicates the ratio of sSFR (left) or hot gas mass (right) between the TNG and \textsc{L-Galaxies} models. Qualitatively, we see that both sSFR and hot gas mass can differ considerably, and that the differences are larger near massive groups and clusters. Broadly speaking, TNG galaxies in dense environments have less gas and lower star formation rates, showing that environmental effects play out differently in the two models.

\subsection{Stellar masses}
\label{subsec: stellar_masses_dis}

In \S \ref{subsec: stellar_masses} we demonstrated that the satellites in TNG tend to have higher stellar masses than in \textsc{L-Galaxies}. In the top row of Fig. \ref{Fig: StellarMass_SSFR_ratio_dis} we show the ratio of TNG stellar mass to \textsc{L-Galaxies} stellar mass as a function of distance from halo centre. Solid lines show the median for all galaxies at that distance, regardless of whether they are centrals or satellites. We consider three central halo mass bins, with $\log_{10}(M_{\rm 200,host}/\rm M_{\odot})$ between 12 and 13, between 13 and 14, and above 14, as well as three galaxy stellar mass bins (using \textsc{L-Galaxies} masses), from $\log_{10}(M_{\star}/\rm M_{\odot})=9$ to $\log_{10}(M_{\star}/\rm M_{\odot})=12$. Dashed lines represent median ratios for each stellar mass bin for the simulation as a whole. Deviations from this global value can be interpreted as reflecting a difference in environmental effects between the two models. 

A clear trend is present. Apart from massive galaxies (brown solid lines), the closer a galaxy is to halo centre, the larger its TNG stellar mass is compared to its \textsc{L-Galaxies} stellar mass. The effects are greatest for low-mass galaxies ($9<\log_{10}(M_{\star}/\rm M_{\odot})<9.5$) close to the centre of low-mass groups with $12<\log_{10}(M_{200}/\rm M_{\odot})<13$, where the enhancement of the ratio reaches 60\%. Notice also that the effects extend out to several times the $R_{200}$ of the central haloes.

\subsection{Star formation}
\label{subsec: ssfr_dis}

\begin{figure*}
    \includegraphics[width=0.329\textwidth]{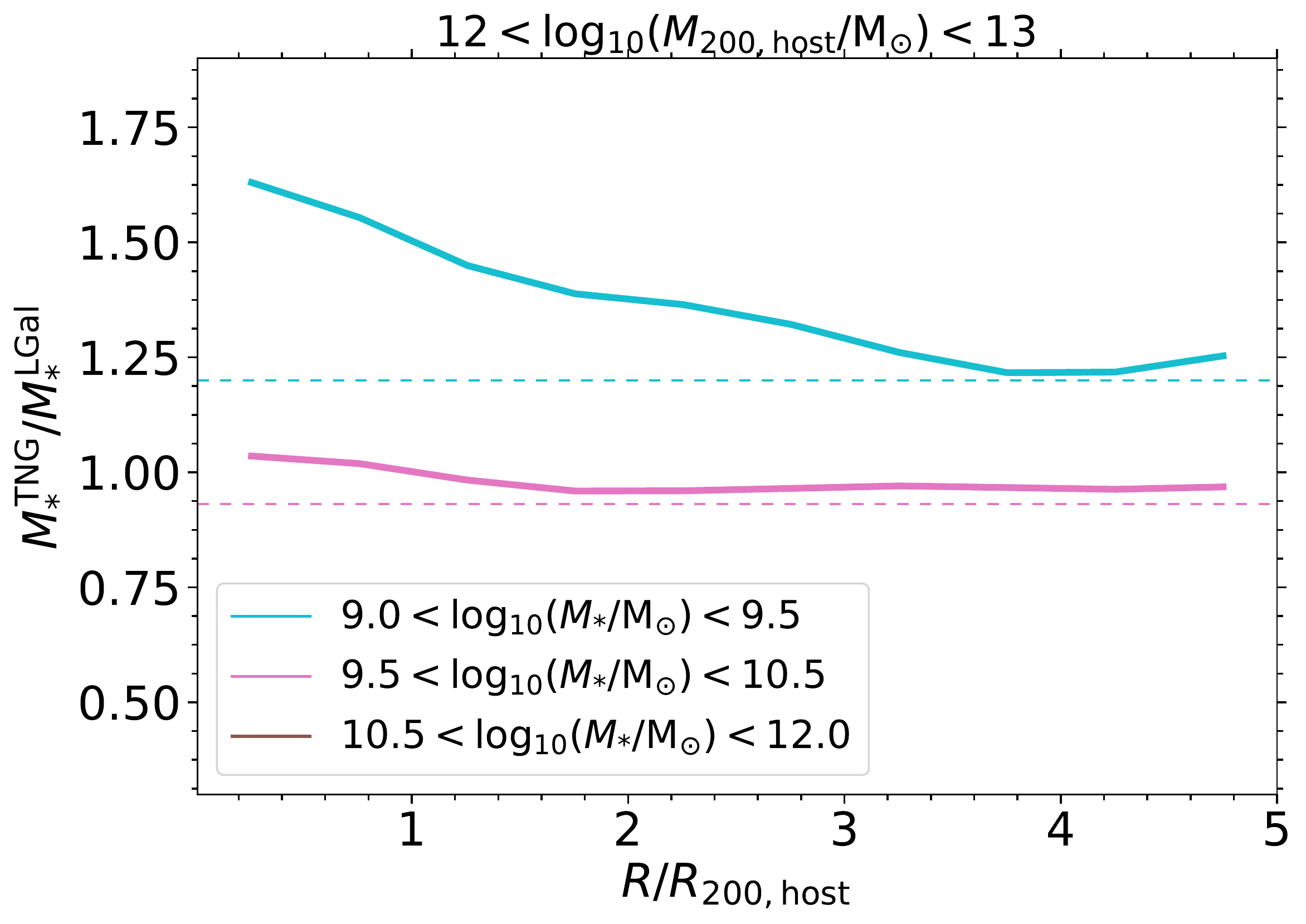}
    \includegraphics[width=0.329\textwidth]{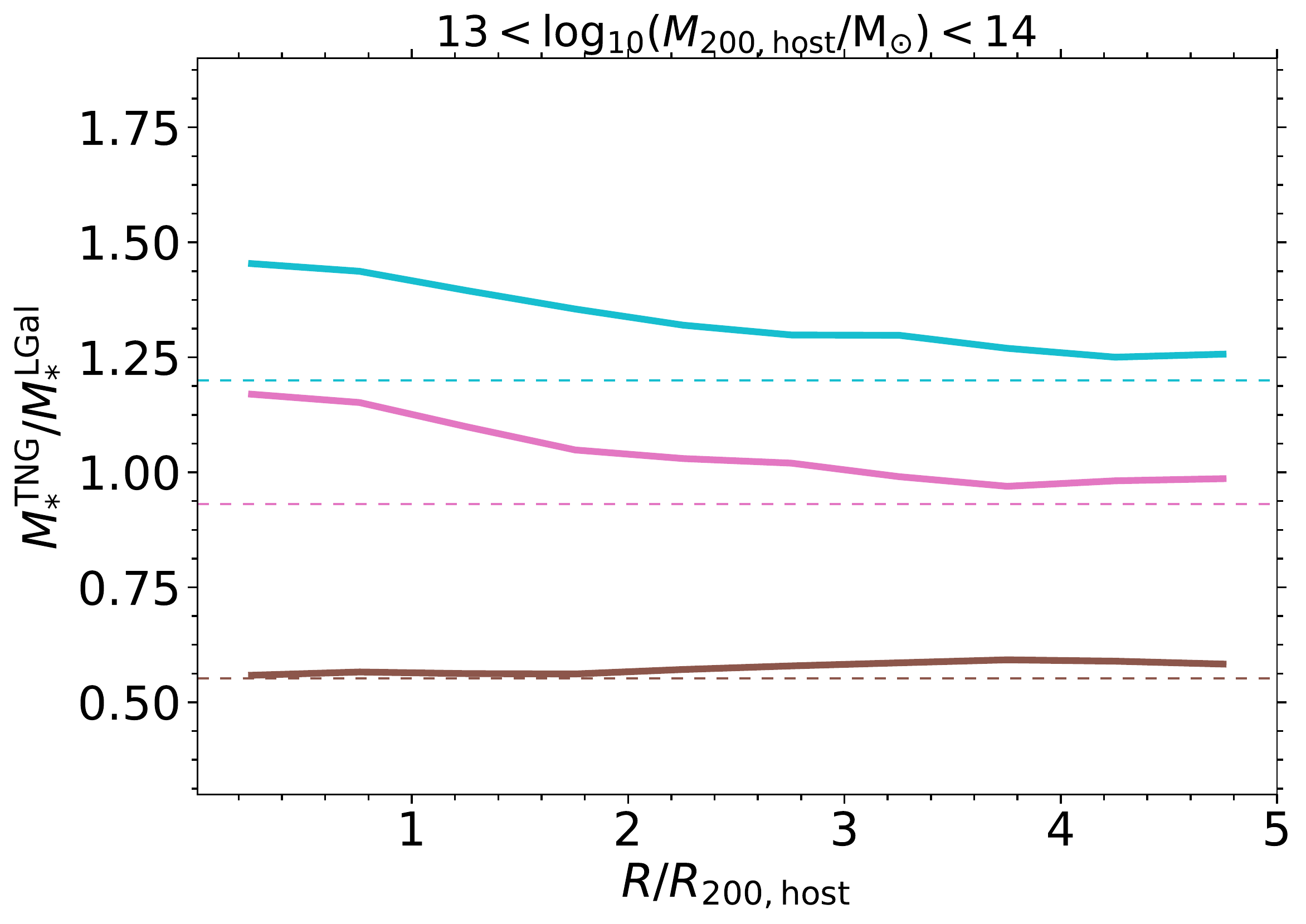}
    \includegraphics[width=0.329\textwidth]{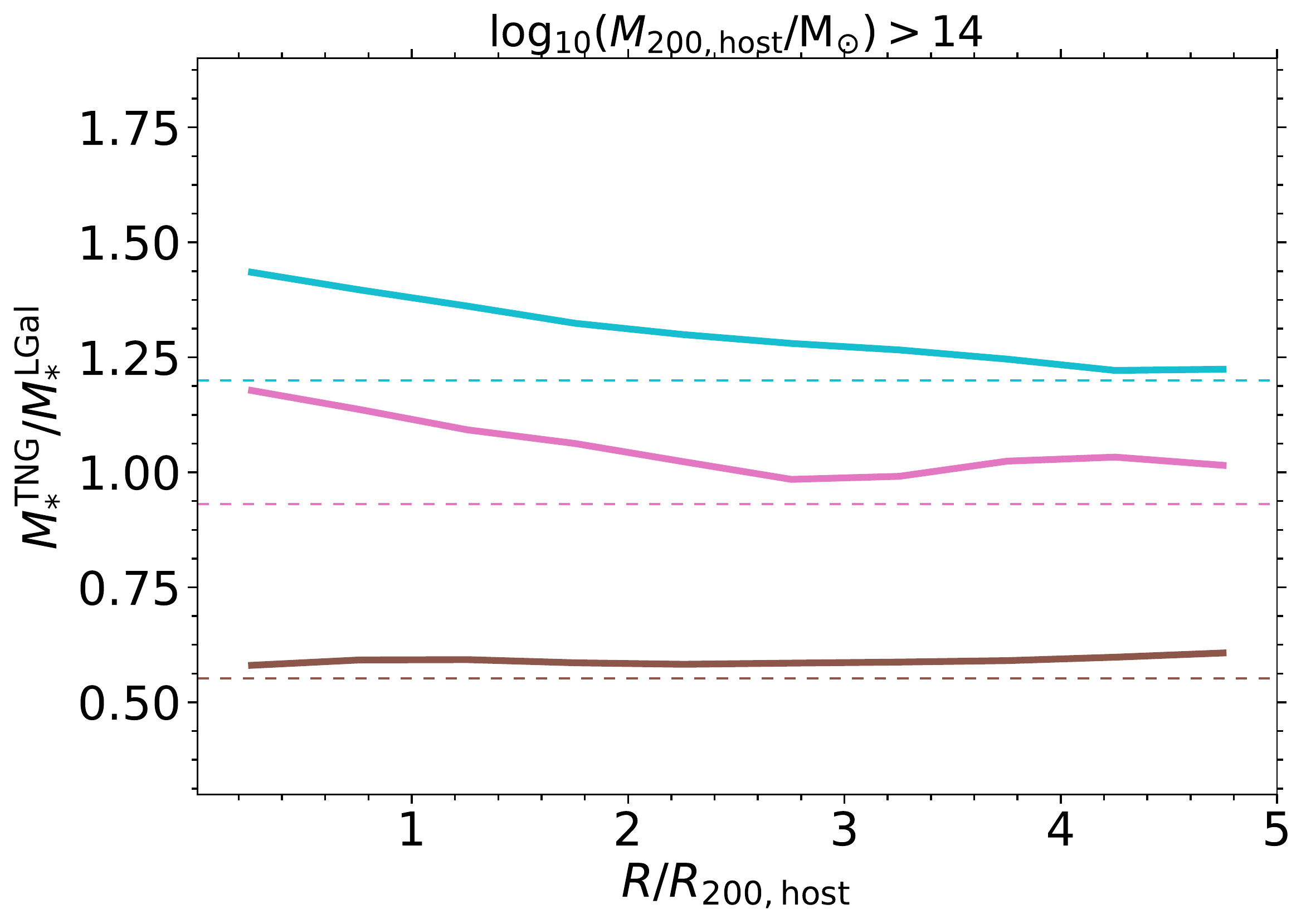}
    \includegraphics[width=0.329\textwidth]{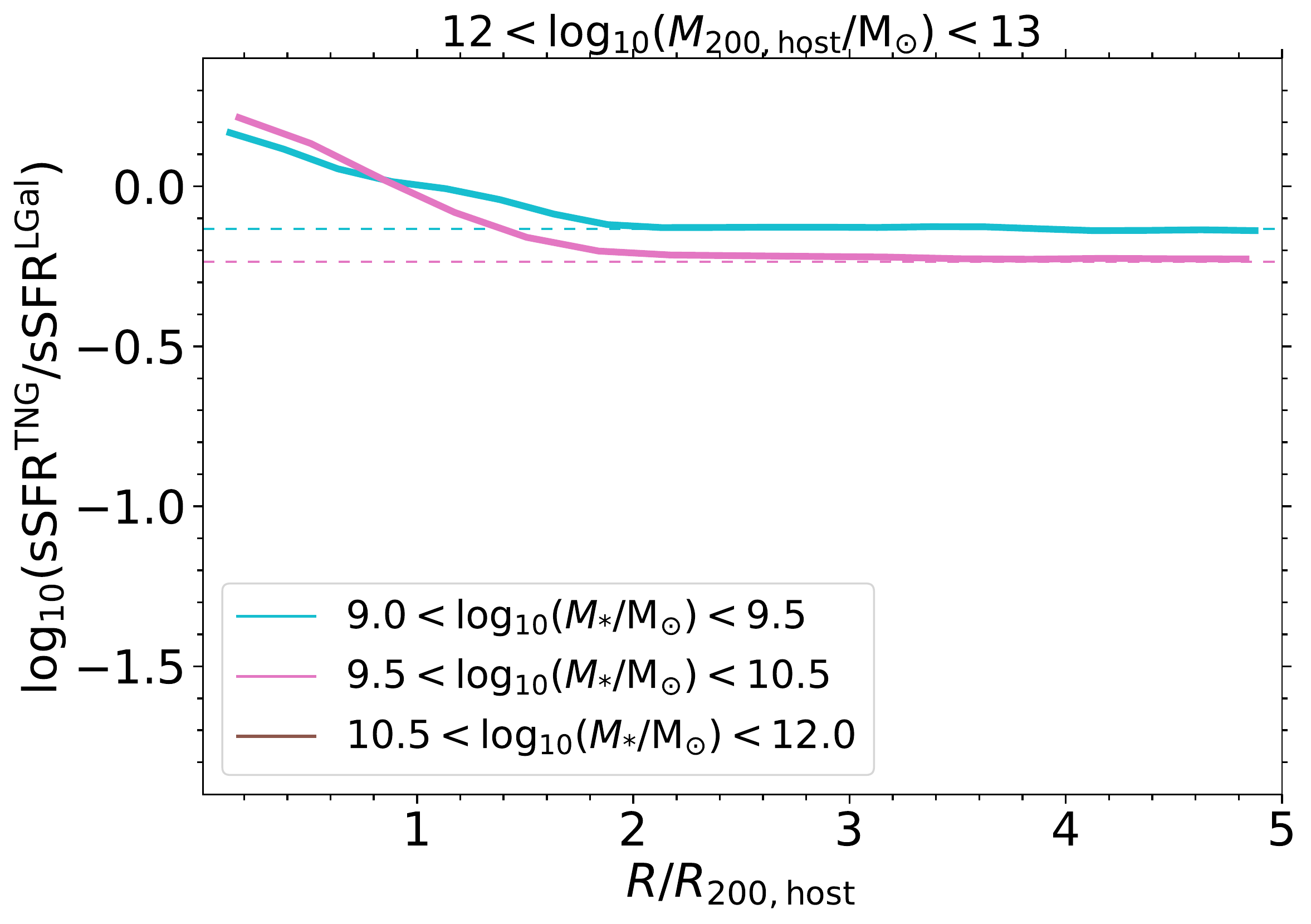}
    \includegraphics[width=0.329\textwidth]{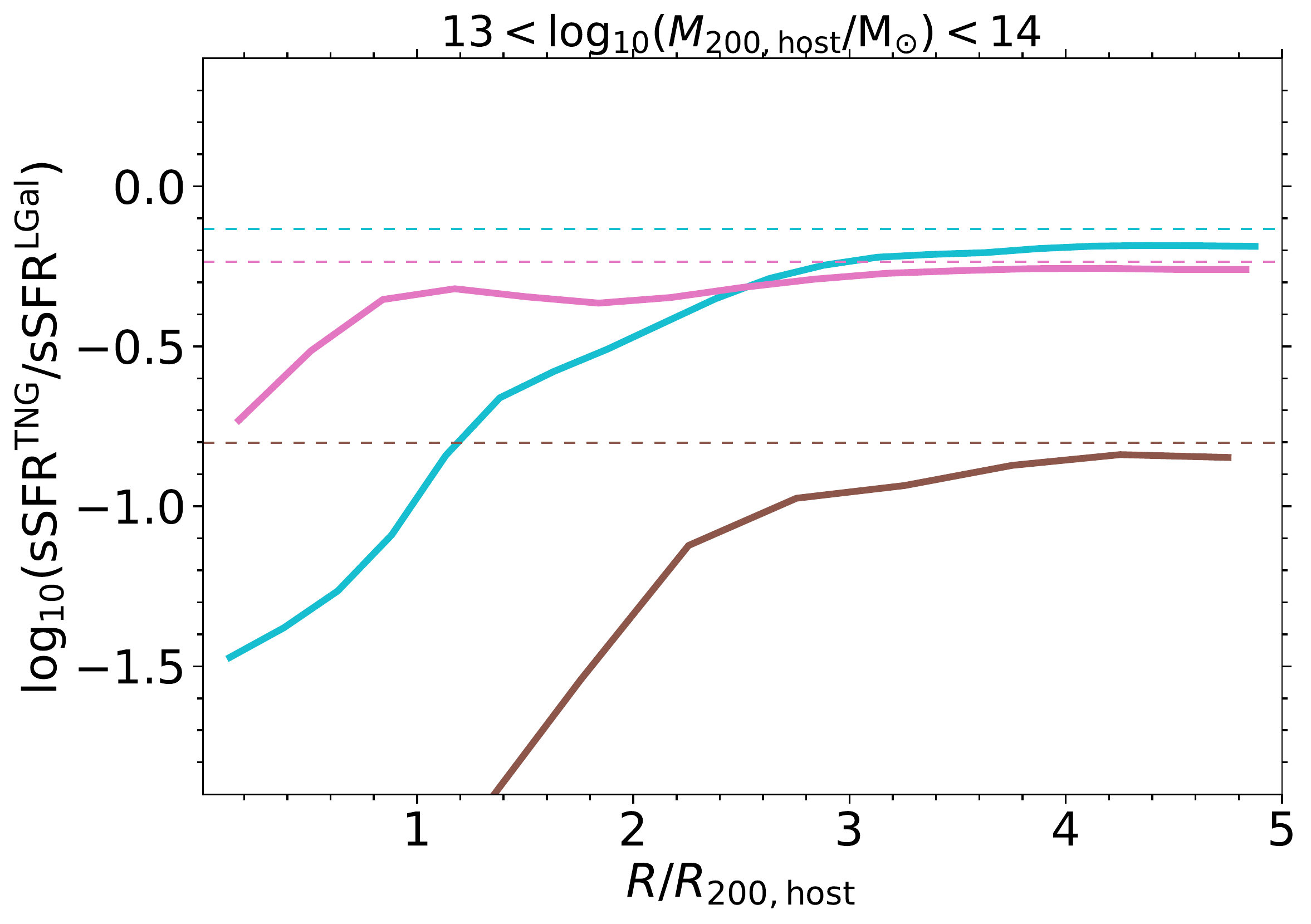}
    \includegraphics[width=0.329\textwidth]{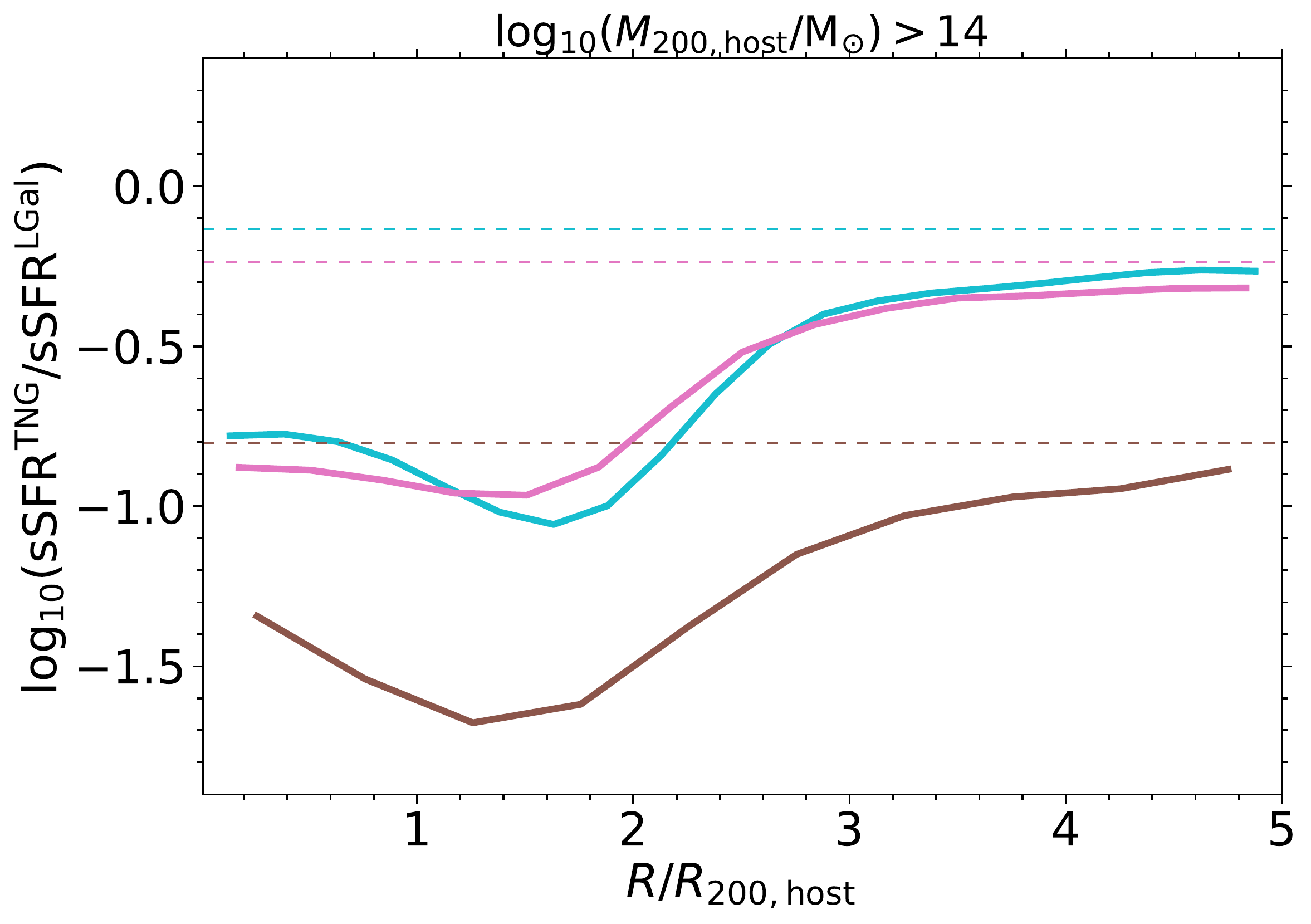}
    \caption{The ratio between the stellar mass (top row) and the sSFR (bottom row) of galaxies in TNG and \textsc{L-Galaxies} as a function of halocentric distance. Each panel refers to central haloes with masses in the range indicated. Each solid line shows the local median value of the ratio for galaxies with stellar mass (in \textsc{L-Galaxies}) in a specific mass range, as indicated by the legend. Dashed horizontal lines denote the corresponding global medians for the simulations as a whole. In the top row, the $\rm 300\,Mpc$ box is used for the most massive stellar mass bin, while the $\rm \rm 100\,Mpc$ box  is  used for the other two bins. In the bottom row, all the results are taken from the $\rm 300\,Mpc$ box.}
\label{Fig: StellarMass_SSFR_ratio_dis}
\end{figure*}

The bottom row of Fig. \ref{Fig: StellarMass_SSFR_ratio_dis} shows the median $\rm sSFR^{TNG}/sSFR^{LGal}$ ratio of galaxies as a function of halocentric distance. In the vicinity of clusters with $\log_{10}(M_{200}/\rm M_{\odot})>14$ (right panel), the local ratio of sSFR is lower than the global ratio out to a distance of at least $3R_{200}$. There is a local minimum at $R/R_{200}\sim 1.5-2$, which is close to the splashback radius indicating the most distant objects that have passed through the halo. We note that this radius likely also corresponds to the outermost extent of the hot gas halo, hence of the region within which environmental processes like ram pressure stripping of hot gas (RPS) can be significant. \textsc{L-Galaxies} assumes hot gas haloes to extend only to $R_{200}$, so environmental stripping effects are limited to satellites within this radius, whereas in TNG such effects can act to larger distances.

The ratio of sSFRs in the vicinity of groups with mass \mbox{$13 < \log_{10}(M_{200}/\rm M_{\odot}) < 14$} (lower middle panel of Fig. \ref{Fig: StellarMass_SSFR_ratio_dis}) exhibits a different behaviour. Close to the centres of haloes, this ratio is more than 1 dex lower than its global value for low-mass (cyan line) and massive (brown line) galaxies, and by about 0.5 dex for intermediate mass galaxies (magenta line). The large differences between the two models in this halo mass bin arise from the fact that \textsc{L-Galaxies} only considers RPS for satellites within $R_{200}$ of clusters, and no RPS is implemented for satellites within or around groups and lower mass haloes (see \S \ref{subsec: important_processes_Env}). In TNG, on the other hand, RPS is included self-consistently according to the resolved gas dynamics in all objects. We note that $\rm sSFR^{TNG}/sSFR^{LGal}$ is suppressed at all stellar masses to distances of $\sim 3R_{200}$, implying that RPS in TNG influences star formation both in and outside of the virialised region of galaxy groups.

The most massive galaxies with $\log_{10}(M_{\star}/\rm M_{\odot})>10.5$ in the vicinity of clusters and groups are significantly less star-forming in TNG than in \textsc{L-Galaxies} (brown solid lines, lower right and middle panels). This is very likely the influence of AGN feedback prior to infall, rather than of environmental stripping effects. Intermediate mass galaxies within the groups (solid magenta line, lower middle panel), are much less affected by environment than massive galaxies, whereas the opposite would be expected if RPS was the dominant processes in these systems. Although the gas expelled by SMBH feedback in TNG could return to the halo at later epochs, this re-accretion would be inhibited for satellite subhaloes, as the gas ejected from these objects is less bound and thus can be stripped by host halo. As a result, feedback in massive satellites enhances the influence of RPS, and this environmental suppression of star formation is stronger in TNG than in \textsc{L-Galaxies}.

The sSFR of galaxies in lower mass haloes with $12 < \log_{10}(M_{200} / \rm M_{\odot}) < 13$ (lower left panel) has much smaller differences between the two models. The $\rm sSFR^{TNG}/sSFR^{LGal}$ ratio is up to 0.5 dex above its global value for low-mass galaxies within $R_{200}$. We note that \textsc{L-Galaxies} does not include RPS processes within these low-mass haloes. In addition, tidal stripping processes may be different in the two models. Tidal stripping in \textsc{L-Galaxies} depends on the gas and dark matter profiles of the satellite, both of which are assumed to be isothermal profiles with $\rho \propto r^{-2}$. This is an approximation that may not reflect the profiles of TNG satellites. 

\subsection{Environmental quenching}
\label{subsec: quenchedFrac_dis}

\begin{figure*}
    \includegraphics[width=0.33\textwidth]{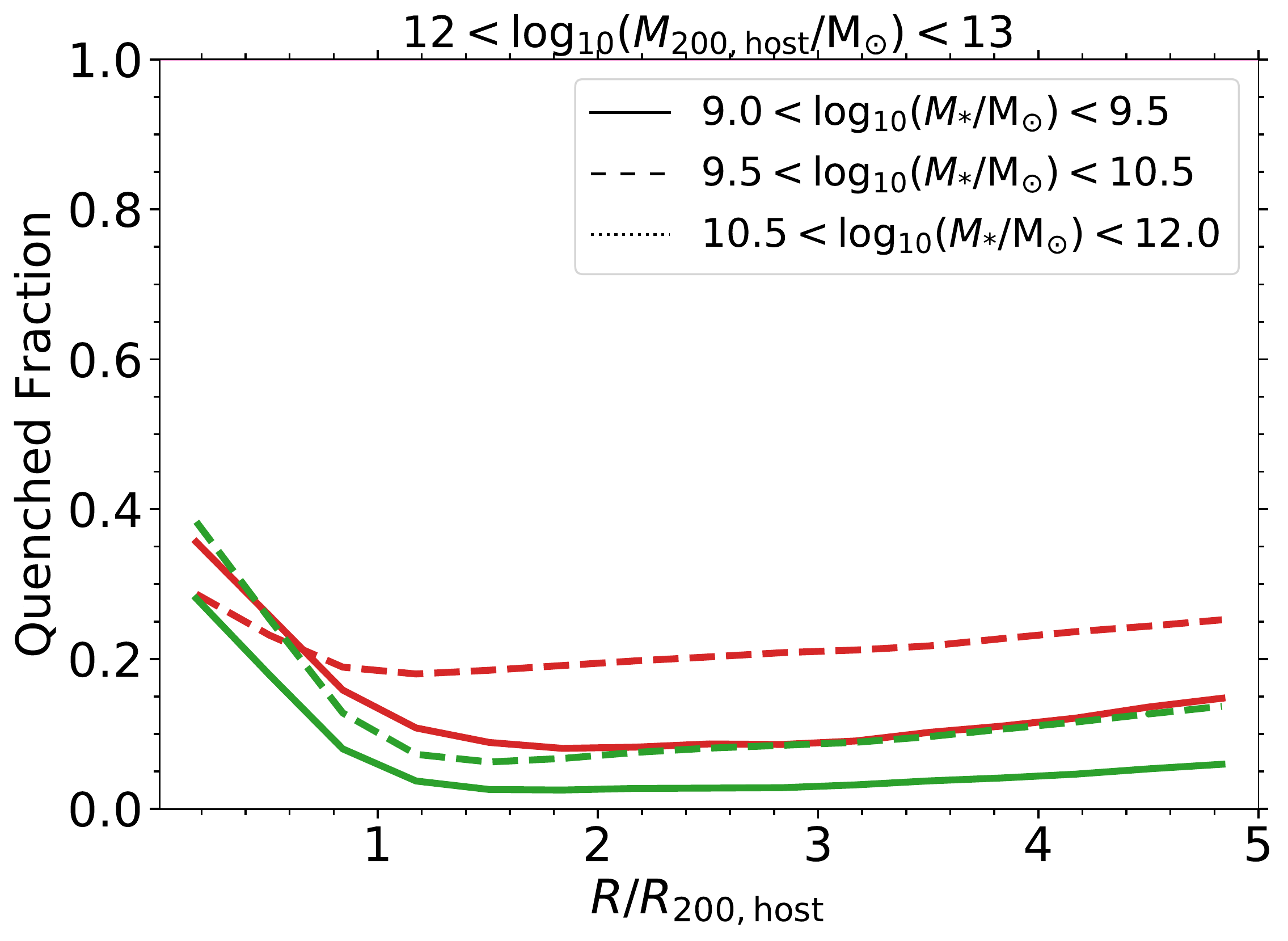}
    \includegraphics[width=0.33\textwidth]{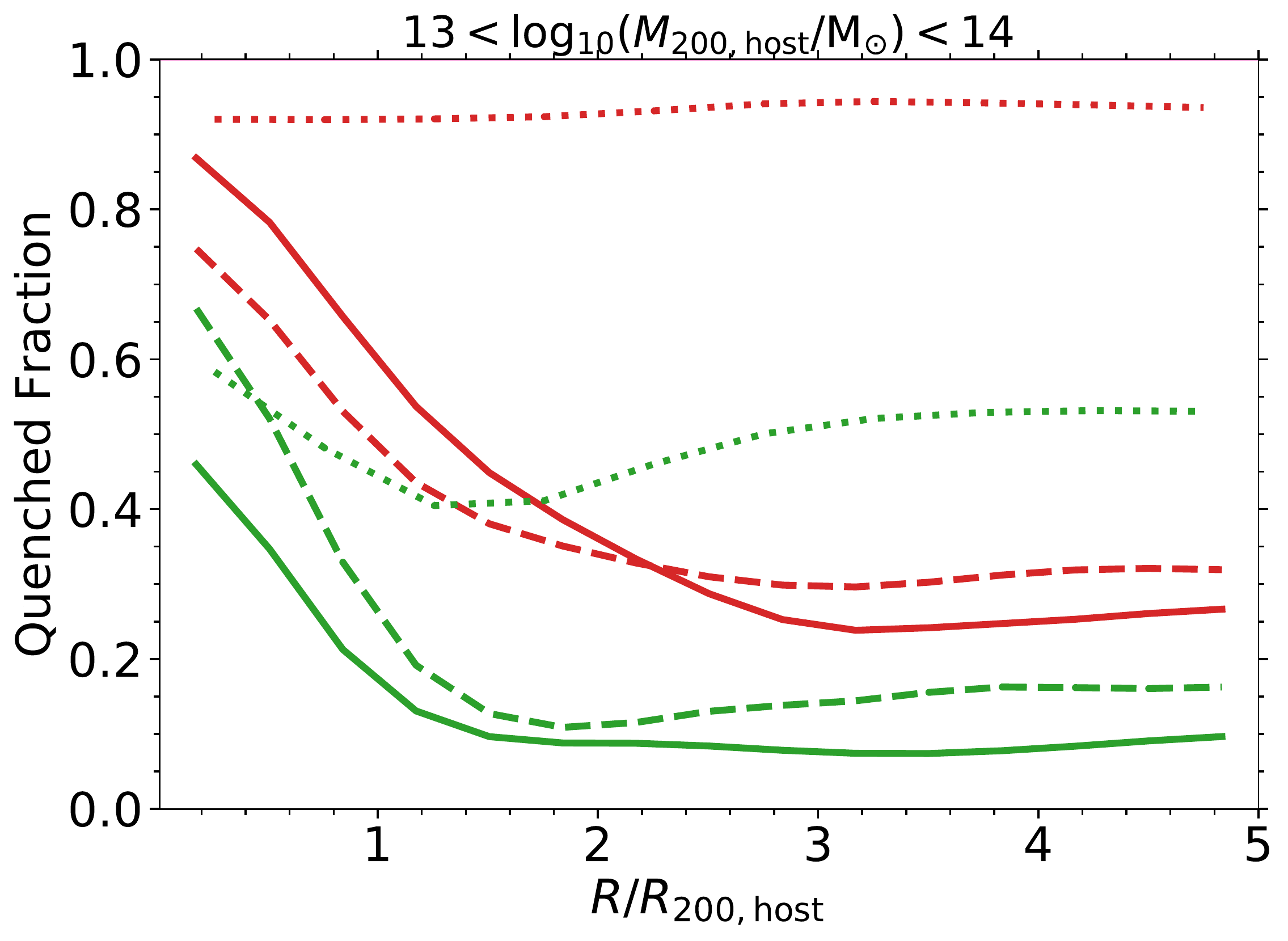}
    \includegraphics[width=0.33\textwidth]{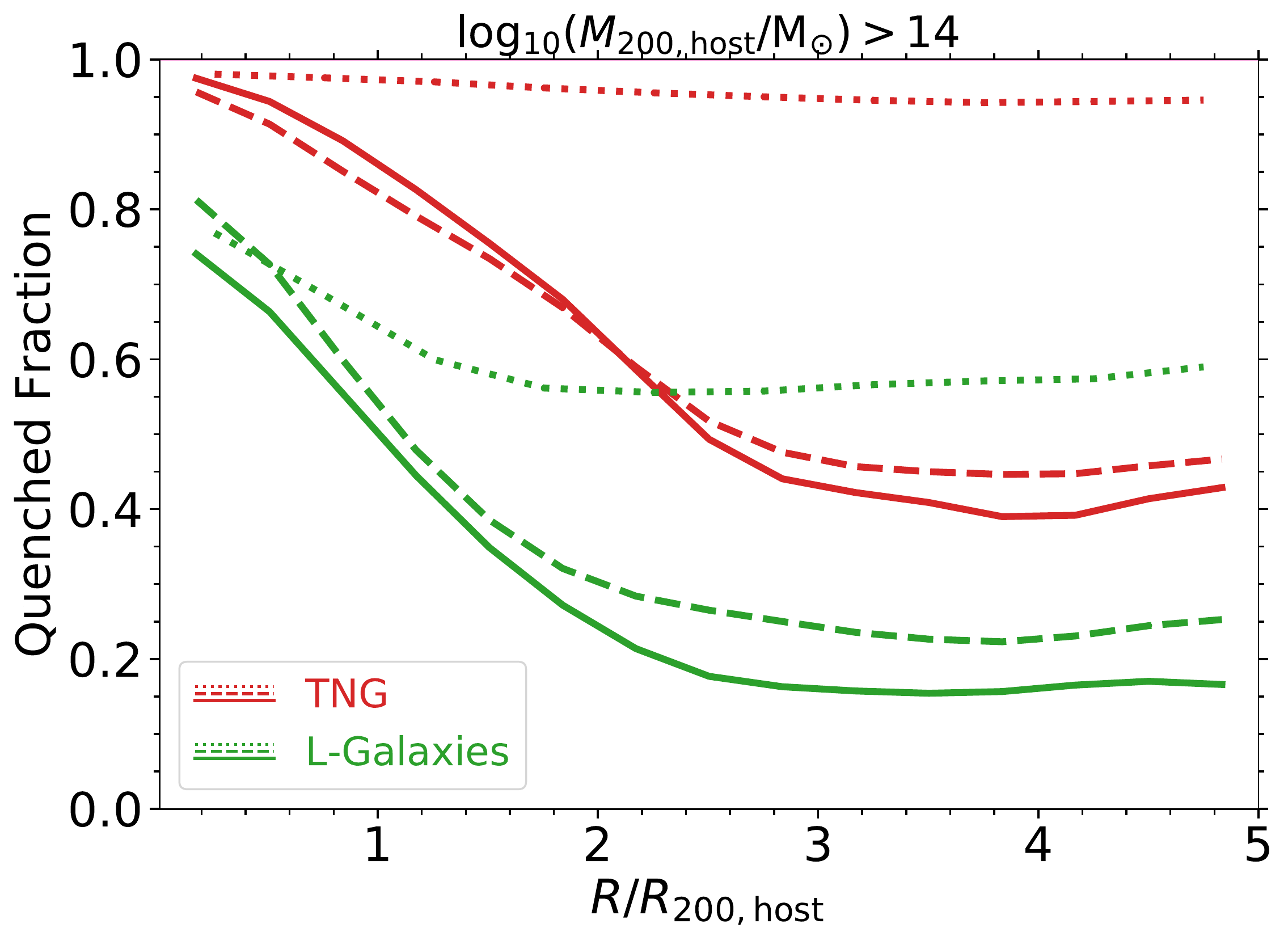}
    \includegraphics[width=0.95\columnwidth]{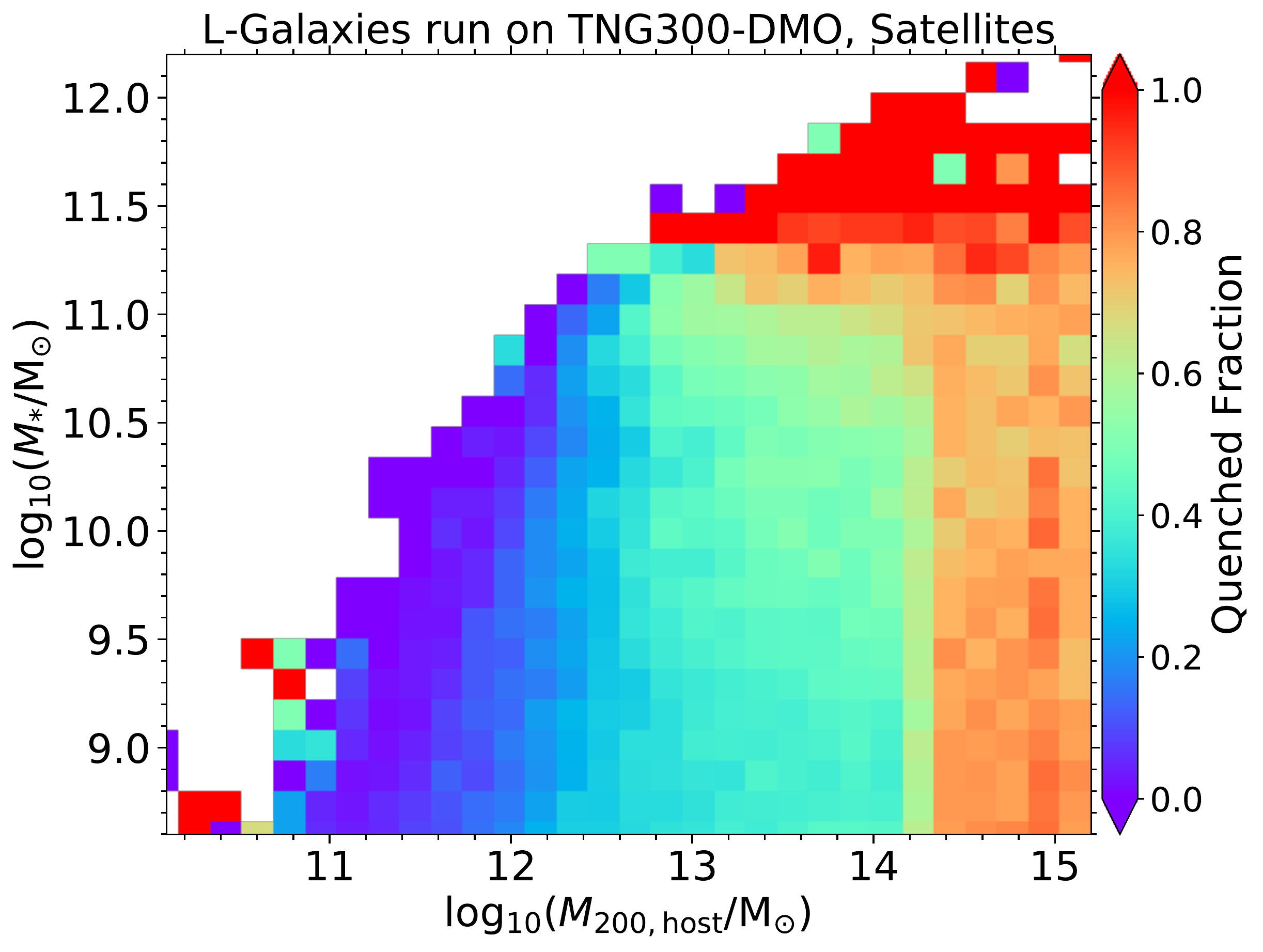}
    \includegraphics[width=0.95\columnwidth]{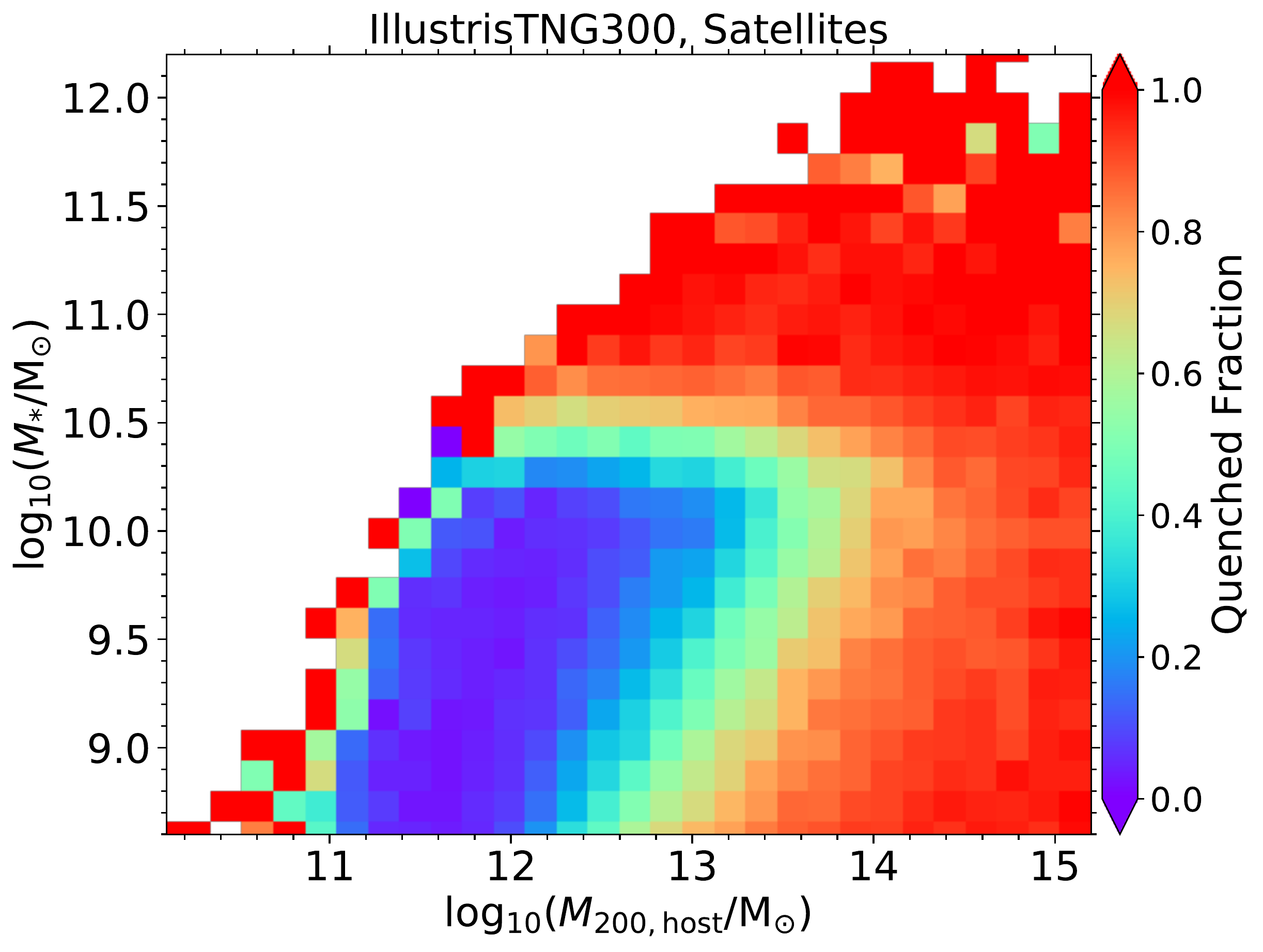}
    \caption{The top row shows the fraction of quenched galaxies in the vicinity of haloes in \textsc{L-Galaxies} and TNG as a function of halocentric distance. Each panel gives the quenched fraction around haloes in a particular mass range, from lower mass haloes (top left panel) to clusters (top right panel). The results in each panel are divided into three lines based on galaxy stellar masses (solid, dashed and dotted). All neighbouring galaxies, both centrals and satellites, are included. The bottom panels illustrate the 2D plane of satellite galaxy stellar mass versus host halo mass, where the colour shows the fraction of quenched satellite galaxies. The left and right panels show the results for \textsc{L-Galaxies} and TNG respectively. Galaxies with $\rm \log_{10}(SSFR[yr^{-1}]) < -11$ are defined as quenched.}
\label{Fig: quenchedFrac_dis}
\end{figure*}

As discussed in \S \ref{subsec: quenchedFrac}, TNG galaxies are generally more quenched than in \textsc{L-Galaxies}. Fig. \ref{Fig: quenchedFrac_dis} demonstrates that this also holds for galaxies in different environments, where we show the quenched fraction of galaxies versus halocentric distance (top row), and the quenched fraction of satellites in the two dimensional plane of stellar mass versus host halo mass (bottom row). In all clusters and groups, independent of distance from the centre, the majority of massive TNG galaxies are quenched (dotted red lines, top right and top middle panels). As discussed previously, AGN feedback rather than environment is the dominant effect. In contrast, the quenched fraction of similarly  massive galaxies in \textsc{L-Galaxies} near clusters show a clear trend with distance: galaxies closer to the centres of haloes are more quenched (dotted green line, top right panel). In general, the quenched fraction in \textsc{L-Galaxies} clusters decreases with halocentric distance up to $R/R_{200}\sim 2$ until reaches a constant value (also see \citealt{henriques17} for a comparison of \textsc{L-Galaxies} with observations). \footnote {We note that massive galaxies near lower mass haloes are excluded due to low number statistics (top left panel).}

The fraction of quenched intermediate and low-mass galaxies (dashed and solid lines) decreases with clustercentric distance up to $R\sim 1-3R_{200}$ in both models. Quenched fractions in TNG are always higher than in \textsc{L-Galaxies}. However, the difference between the models becomes smaller near the centres of clusters, implying that the environmental effects dominate over other general differences between the models. We see that the fraction of quenched galaxies around groups (top middle) is lower in both models compared to clusters (top right) and there is a clear trend with host halo mass.

The bottom panels of Fig. \ref{Fig: quenchedFrac_dis} show a two-dimensional view of quenched satellite fraction of haloes as a function of satellite stellar mass and host halo mass. The left panel illustrates this plane for \textsc{L-Galaxies}, and the right panel for TNG. We observe a very strong horizontal feature in $f_{\rm quench}$ at $\log_{10}(M_{\star}/\rm M_{\odot}) \sim 10.5$ in TNG ($\sim 11$ in \textsc{L-Galaxies}), which shows onset of the dominant, mass-dependent quenching mechanism of black hole feedback. The vast majority of galaxies above these stellar masses are quenched, regardless of environment (host halo mass). A second vertical feature in $f_{\rm quenched}$ occurs at host masses of $\sim 10^{14}\rm M_{\odot}$ for \textsc{L-Galaxies}. This feature is shifted to lower halo mass and also becomes diagonal in the case of TNG. This reflects the onset of environmental quenching processes. These two regimes are similar in spirit to those identified observationally by \cite{peng10}. Interestingly, for given stellar mass the transition between mainly star-forming and mainly quenched occurs more sharply in TNG than in \textsc{L-Galaxies} and at a halo mass which increases almost in proportion to the stellar mass.

In \textsc{L-Galaxies}, satellite galaxies in massive haloes of $\log_{10}(M_{200} / \rm M_{\odot}) \gtrsim 14$, are significantly more quenched than those in lower mass groups, because RPS is assumed not to act below this mass threshold. At lower stellar masses, the fraction of quenched satellite galaxies still increases with the host mass, however, as a result of tidal stripping processes. In TNG, low-mass satellite galaxies in group environments experience environmental stripping and are generally more quenched than in \textsc{L-Galaxies}. 

\subsection{Hot and cold gas content}
\label{subsec: gas_dis}
\begin{figure*}
    \includegraphics[width=0.329\textwidth]{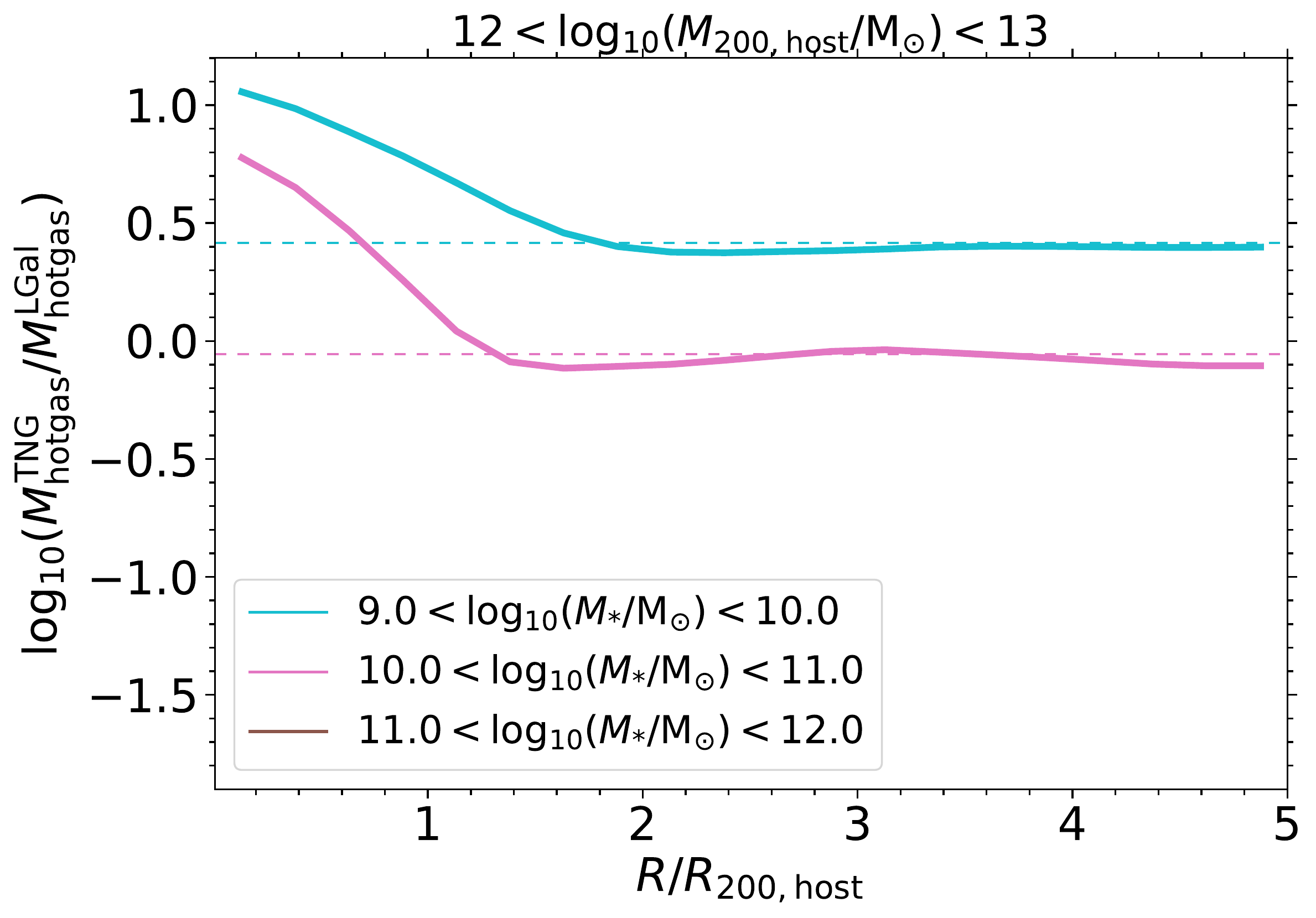}
    \includegraphics[width=0.329\textwidth]{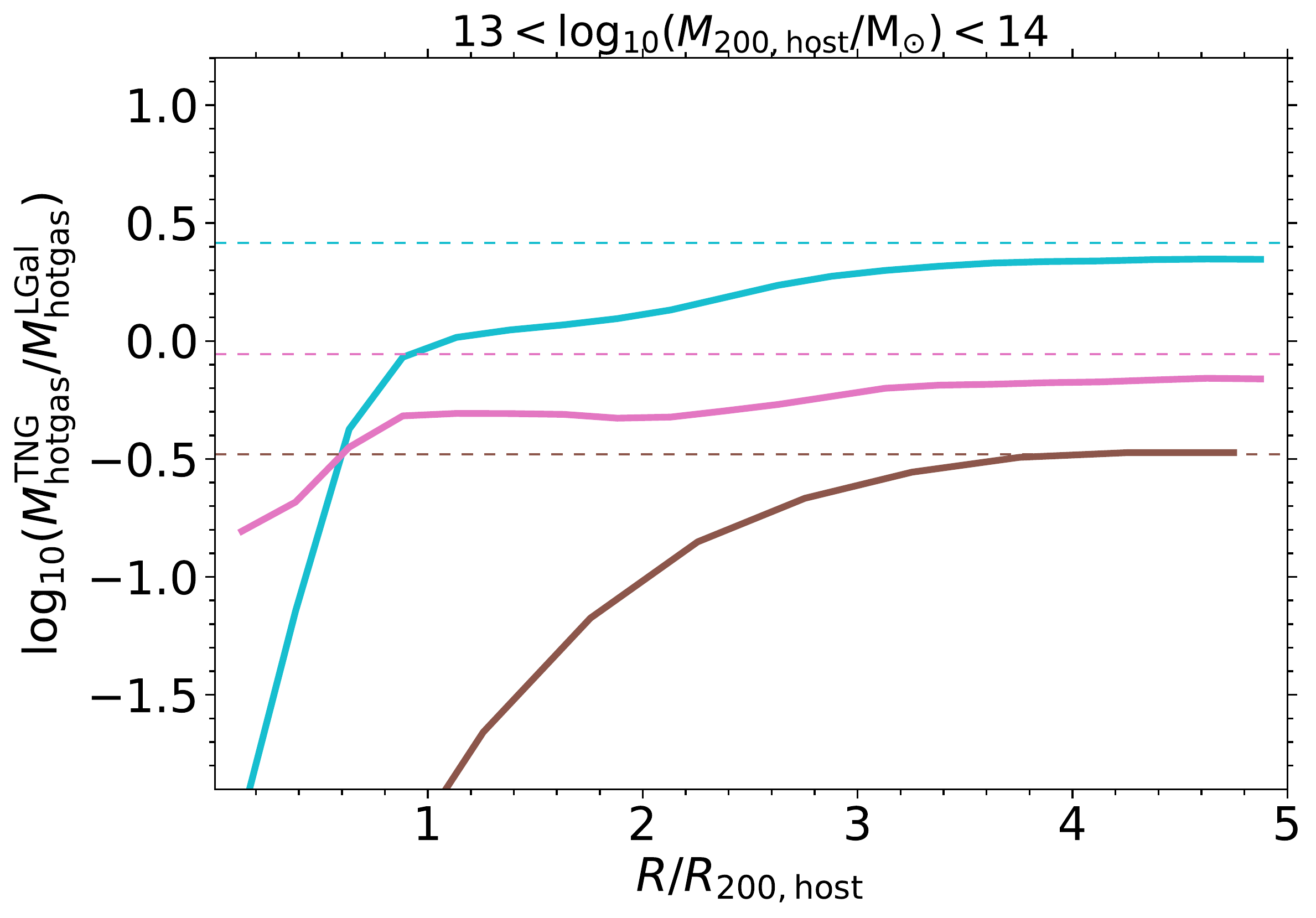}
    \includegraphics[width=0.329\textwidth]{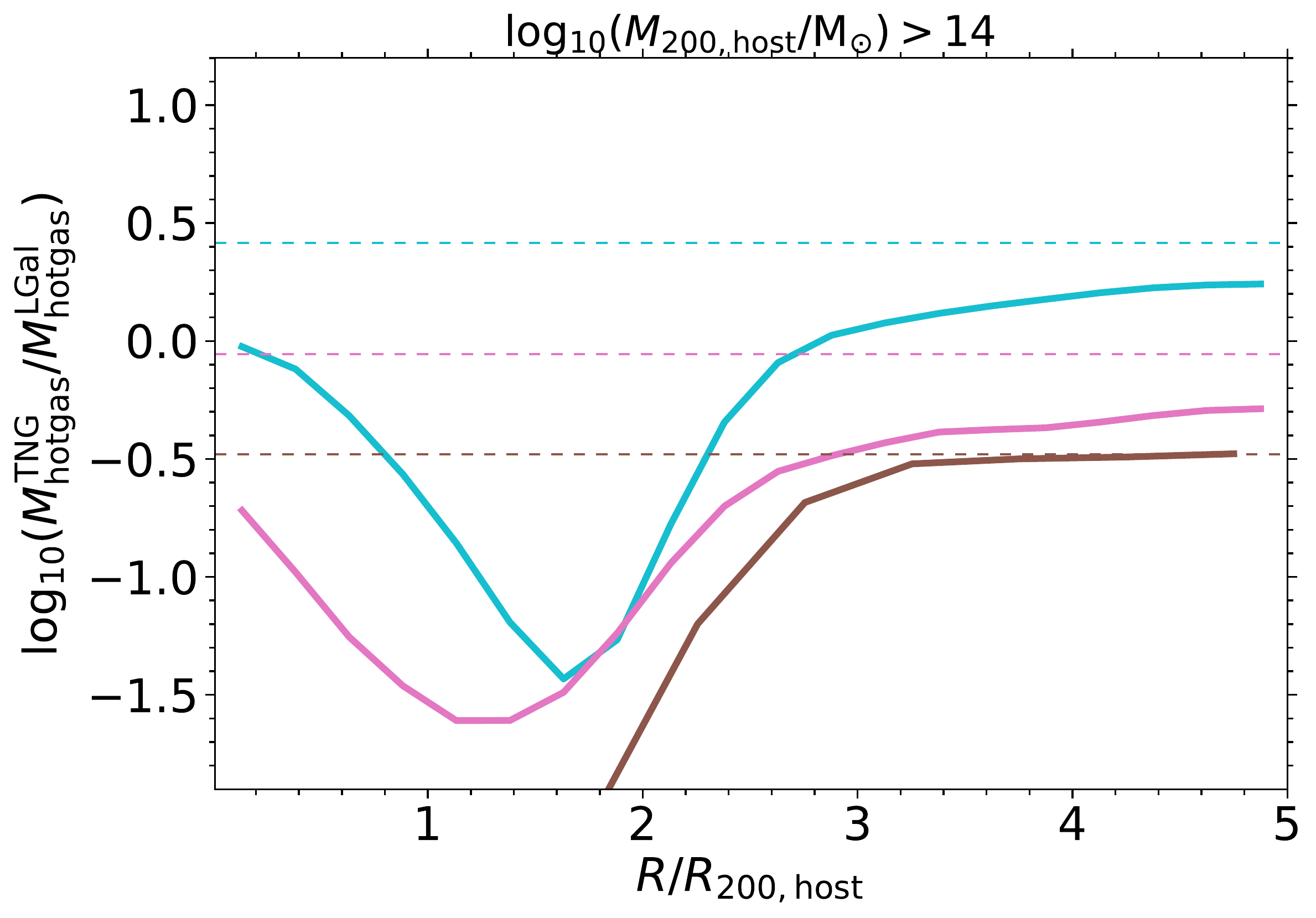}
    \caption{The ratio between hot gas masses in \textsc{L-Galaxies} and in \textsc{TNG} for galaxies in the vicinity of haloes as a function of halocentric distance. Each panel is made for central haloes in the mass range indicated. Solid coloured lines show the median ratios for various stellar mass ranges. The dashed lines denote  median ratios for the simulation as a whole.}
\label{Fig: hotgas_ratio_dis}
\end{figure*}

\begin{figure*}
    \includegraphics[width=0.33\textwidth]{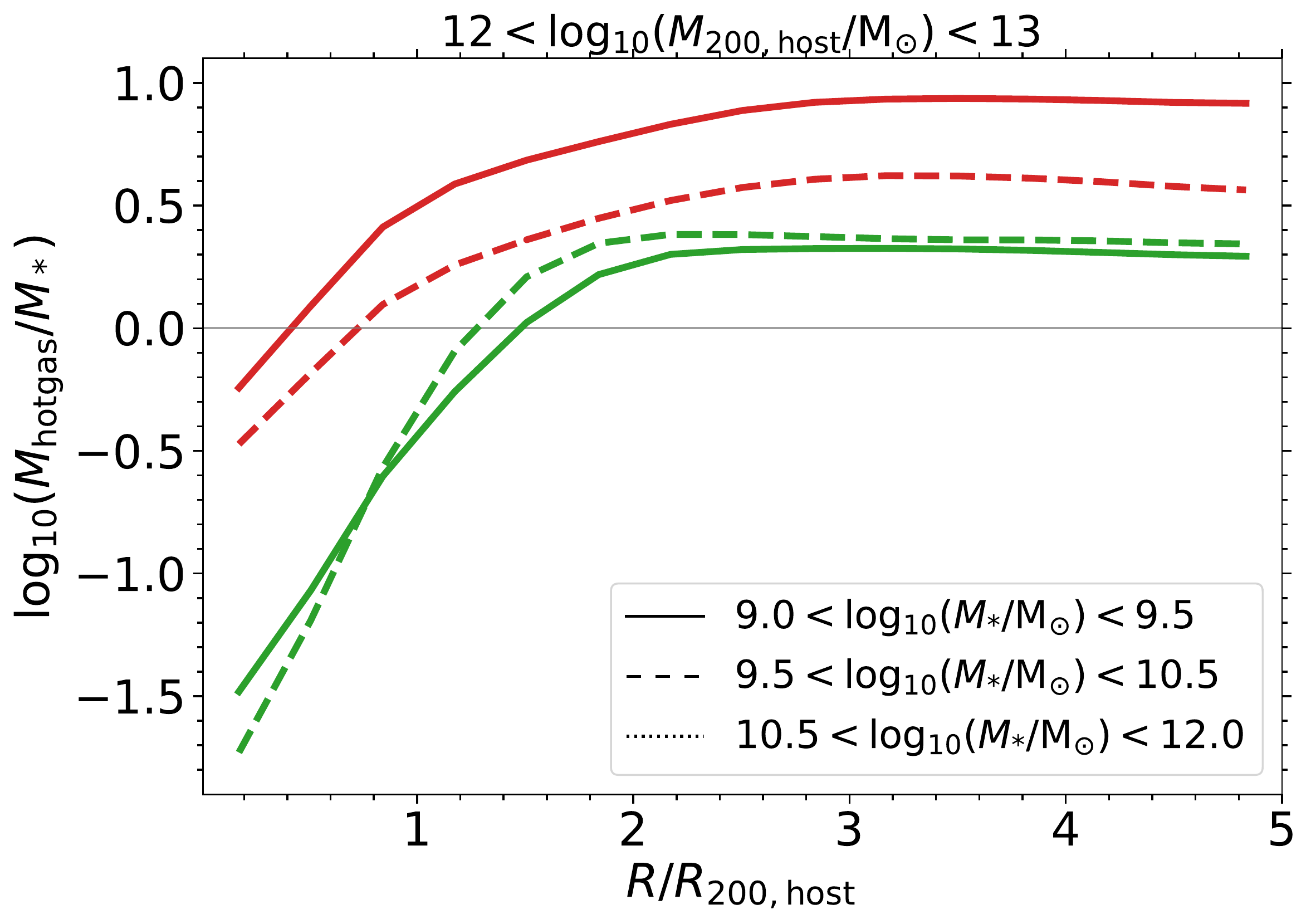}
    \includegraphics[width=0.33\textwidth]{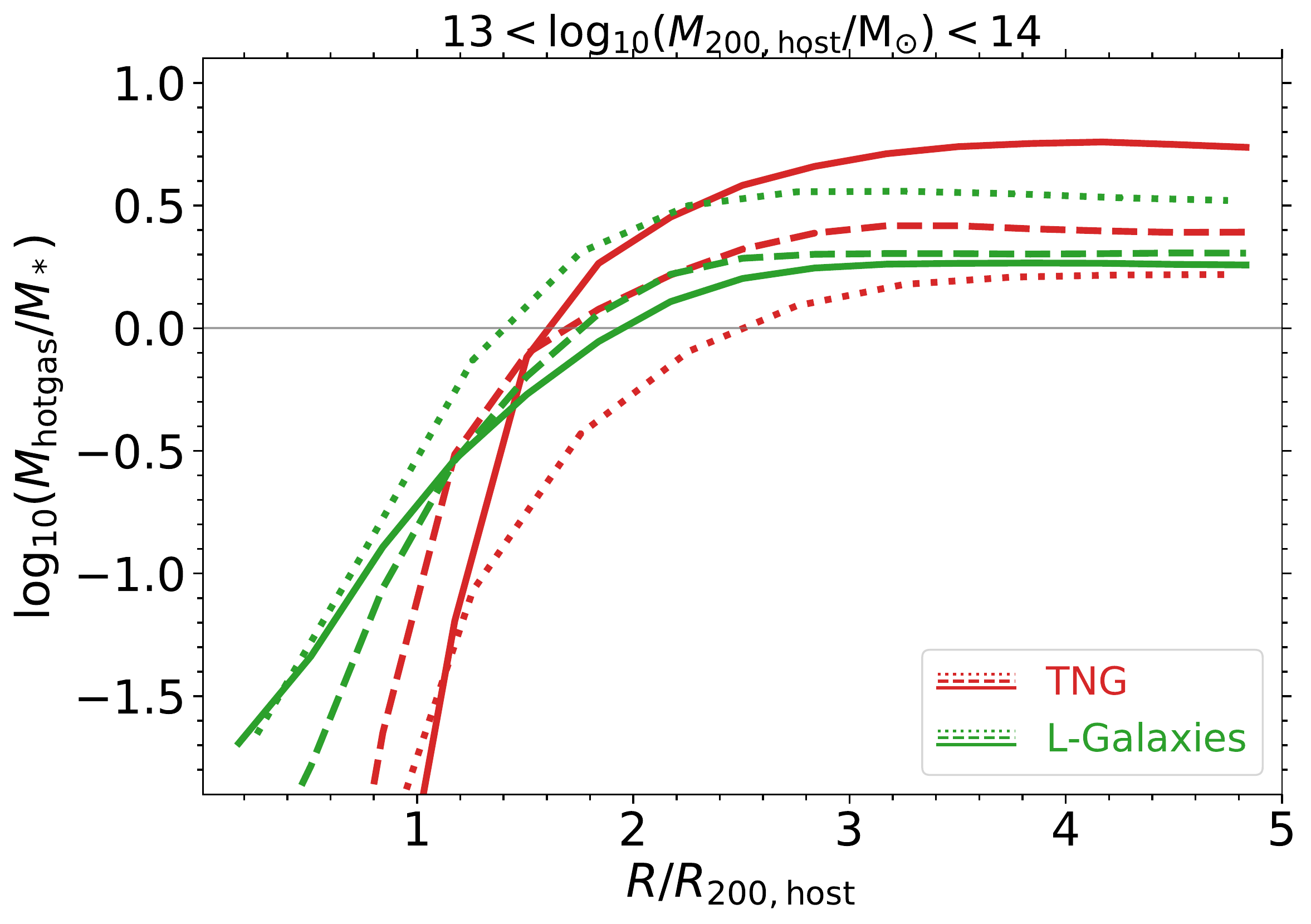}
    \includegraphics[width=0.33\textwidth]{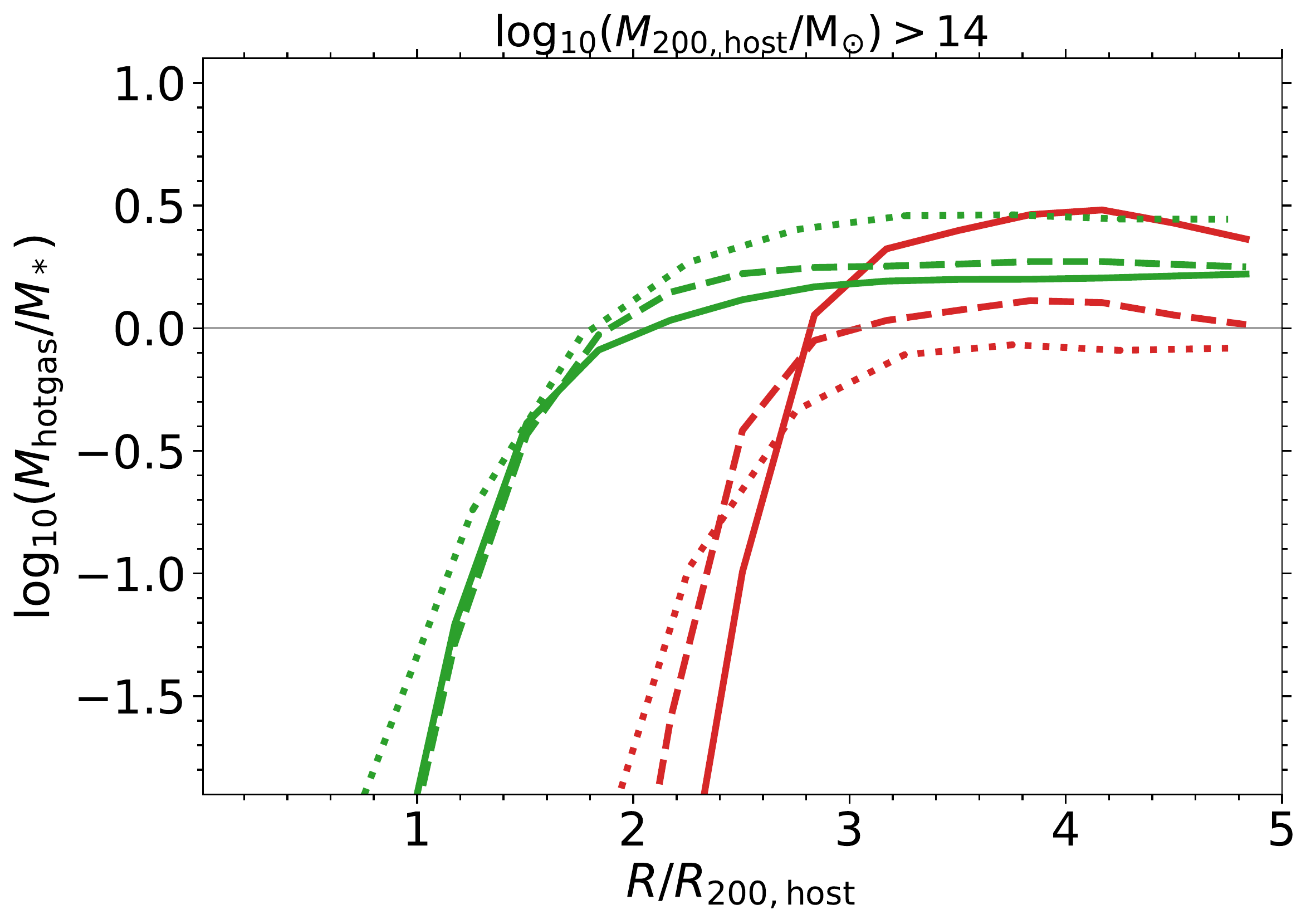}
    \includegraphics[width=0.85\columnwidth]{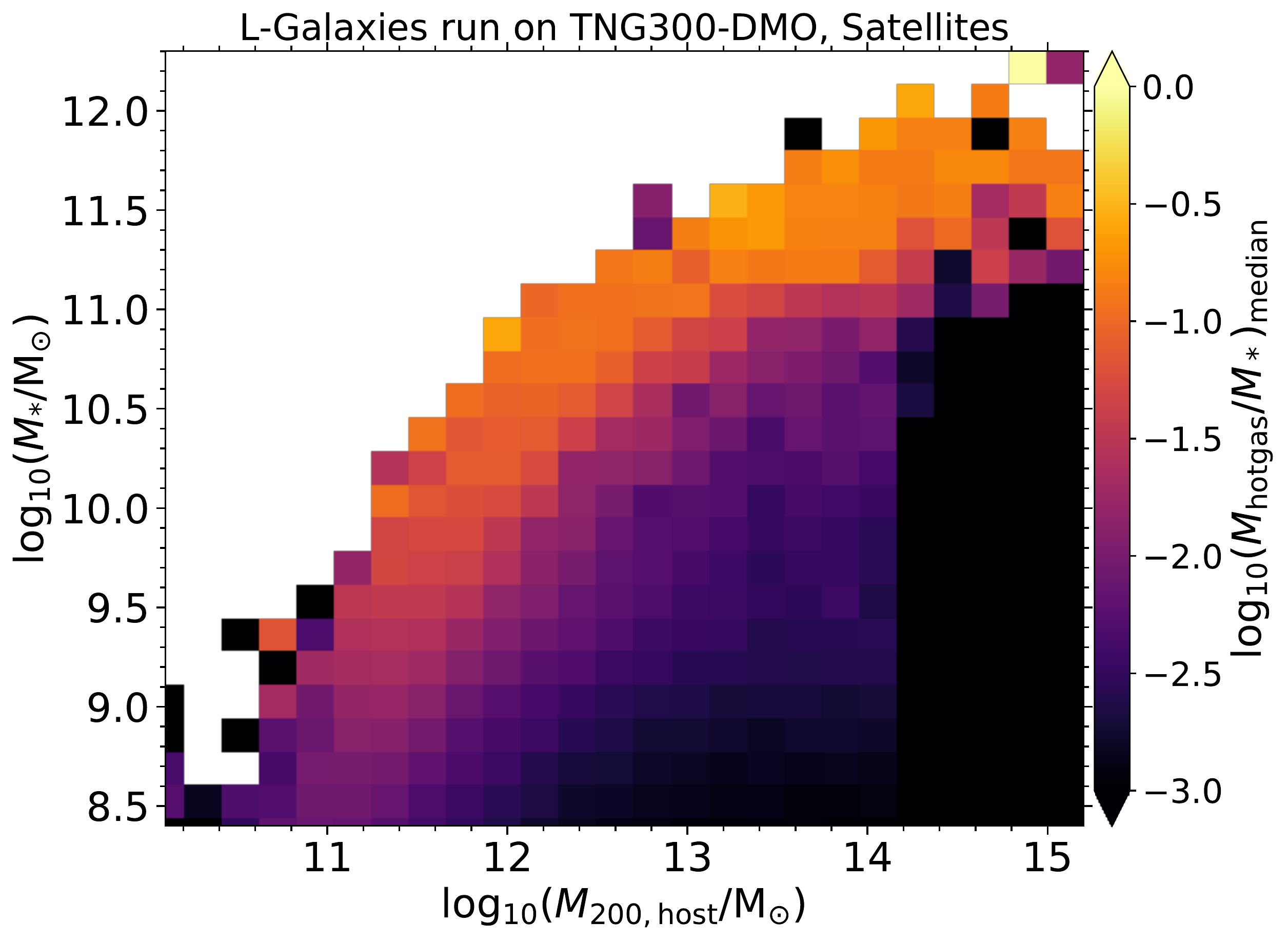}
    \includegraphics[width=0.85\columnwidth]{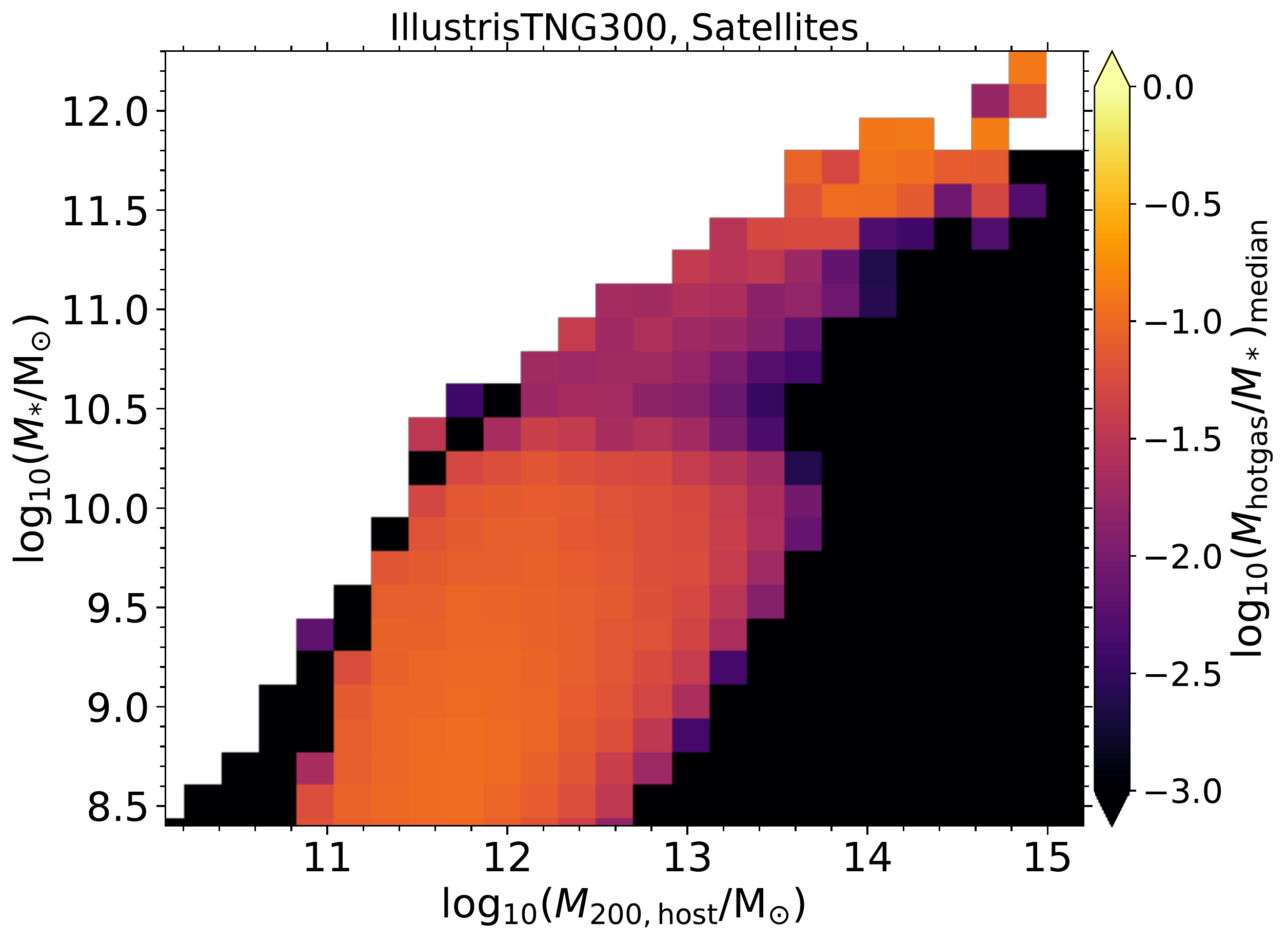}
    \caption{The top row shows the hot gas to stellar mass ratio of galaxies in the vicinity of haloes in \textsc{L-Galaxies} (green) and TNG (red) as a function of halocentric distance. Each panel is made for central haloes within the given mass range, from lower mass systems (top left panel) to clusters (top right panel). The results are split based on galaxy stellar mass as indicated by the linestyle. All neighbouring galaxies, both centrals and satellites, are included. The bottom panel shows the 2D plane of satellite galaxy stellar mass versus host halo mass, where the colour encodes the median hot gas to stellar mass ratio for satellite galaxies (left: \textsc{L-Galaxies}, and right: TNG).}
\label{Fig: hotgas_dis}
\end{figure*}

In this section we carry out a similar analysis of environmental influences on the hot gas content of galaxies in L-Galaxies and TNG. The results shown in  Fig. \ref{Fig: hotgas_ratio_dis} are qualitatively very similar to those obtained above for sSFR. In and around clusters (right panel), the ratios (solid lines) are lower than their global values (dashed lines). There is a local minimum at $R\sim 1.5-2 R_{200}$, for all stellar mass ranges, similar to that seen for the ratio of sSFRs (Fig. \ref{Fig: StellarMass_SSFR_ratio_dis}) and this is likely caused by the same phenomenon. 

In and around groups (middle panel), there is a clear trend with distance: the ratio increases from near zero to its global value with increasing distance away from the centres of haloes, reaching the average value by $R/R_{200} \sim 1-3$. The principal cause of this difference is again the lack of RPS in \textsc{L-Galaxies} groups, and the lack of any kind of stripping beyond $R_{200}$ in \textsc{L-Galaxies}. In addition, as we showed in \S \ref{subsec: halo_baryonFrac}, the gas fraction of TNG groups is below the cosmic value, as the gas is ejected beyond the halo by AGN feedback. As a result, galaxies in the infall regions must pass through relatively dense gas. RPS then causes them to lose a fraction of their hot gas even while they are still outside the nominal halo boundary, $R_{200}$. Finally, within lower mass haloes (upper left panel) $M_{\rm hotgas}^{\rm TNG}/M_{\rm hotgas}^{\rm LGal}$ increases by up to 0.5 dex above its global value similar to the trend seen for sSFR. TNG satellites are able to retain hot gas more effectively than assumed by  \textsc{L-Galaxies}.

In Fig. \ref{Fig: hotgas_dis} we show the hot gas to stellar mass ratio of galaxies as a function of distance to halo centre (top row). Comparing the different host halo masses, the hot gas-to-stellar ratio decreases more strongly near more massive hosts. In both models, satellites within clusters (top right panel) have little remaining hot gas. At $R/R_{200}\sim 1-1.5$ a strong increase is visible in \textsc{L-Galaxies}, while a similar rise occurs in TNG at $R/R_{200}\sim 2-2.5$. Indeed, the local minimum that can be seen at $R/R_{200}\sim 1-2$ in the upper right panel of Fig. \ref{Fig: hotgas_ratio_dis} is caused by the significant difference between the hot gas radii of TNG and \textsc{L-Galaxies} clusters. These correspond to the distance within which RPS environmental effects become significant. In \textsc{L-Galaxies}, this `hot halo radius' is assumed to equal $R_{200}$, while in TNG it is determined by gas dynamics and is influenced by a variety of physical processes including feedback. Such a scale exists for groups (top middle panel) and lower mass haloes (top left panel) as well, but the transition between hot gas-poor and hot gas-rich galaxies is more gradual with distance. 

The hot gas mass of massive galaxies (dotted lines) is more influenced by environment in TNG than in \textsc{L-Galaxies}. As the binding energy distribution of subhalo gas has been modified by baryonic feedback processes there is a larger fraction of gas weakly bound to the subhalo, where it can be more easily stripped. In contrast, gas is generally more strongly bound to subhaloes in \textsc{L-Galaxies} and thus harder to remove. As a result, RPS removes a considerable fraction of the hot gas from massive TNG galaxies in the vicinity of clusters and groups, leaving them significantly less gas-rich than the same objects in \textsc{L-Galaxies}. 

\begin{figure*}
    \includegraphics[width=0.329\textwidth]{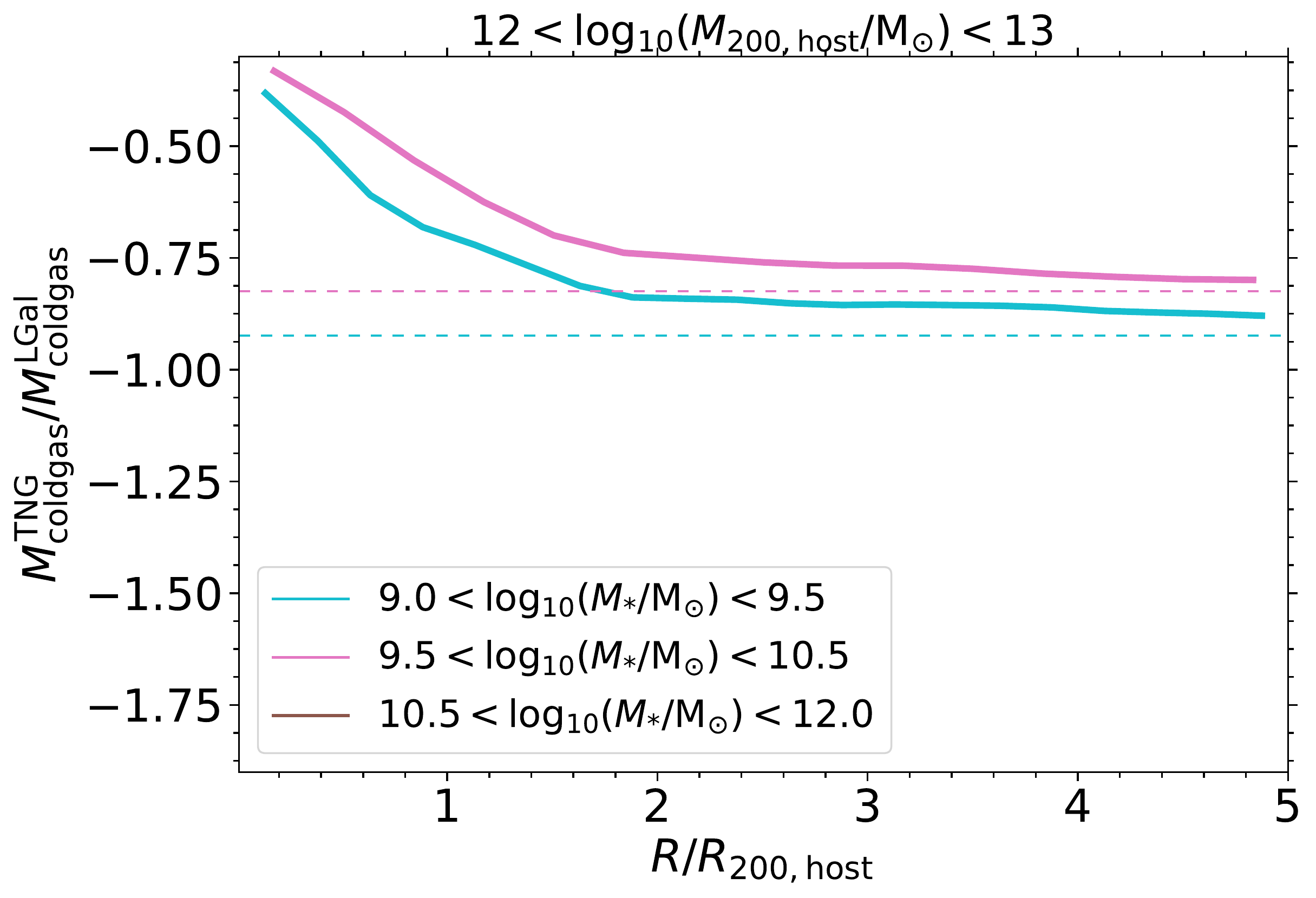}
    \includegraphics[width=0.329\textwidth]{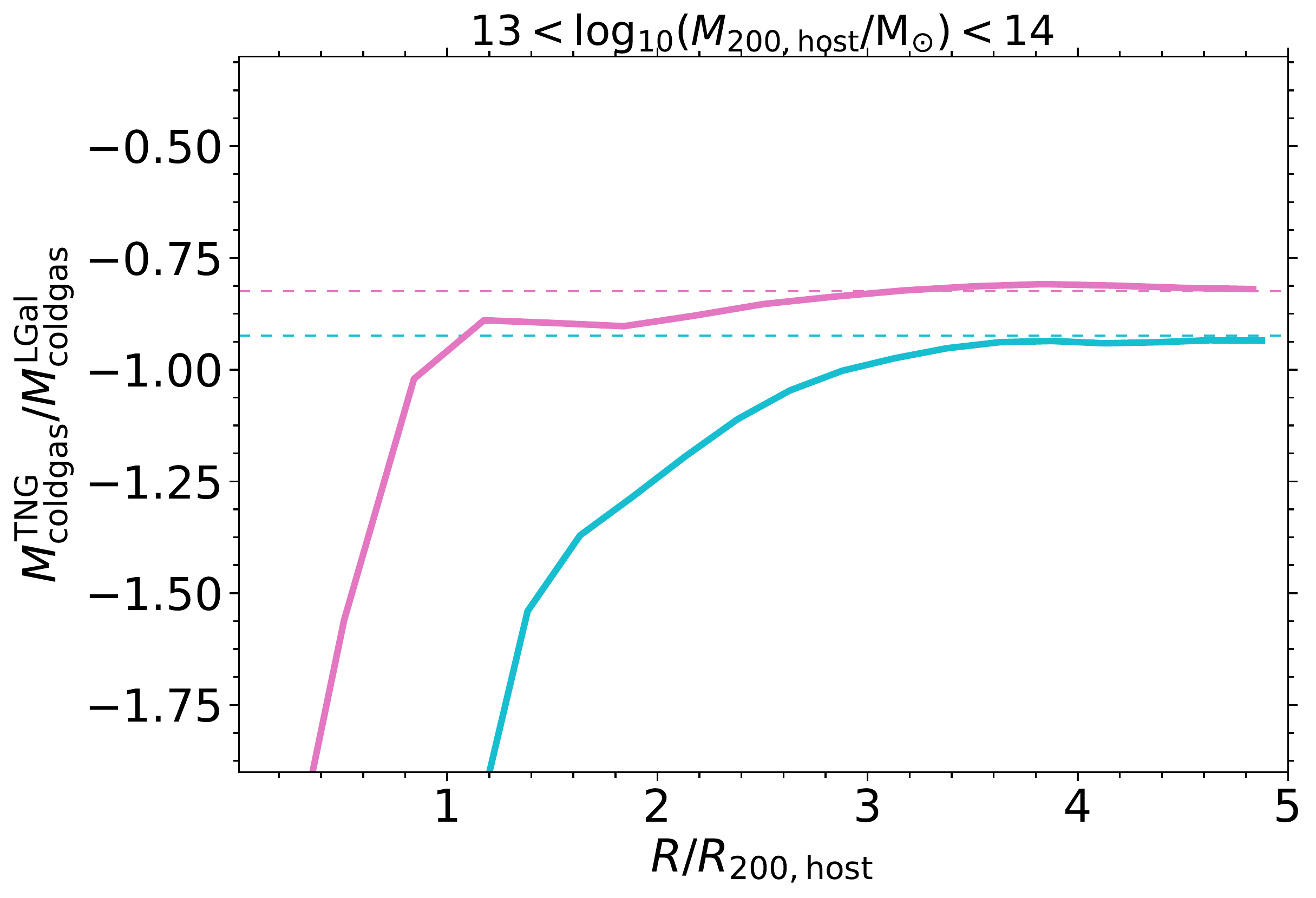}
    \includegraphics[width=0.329\textwidth]{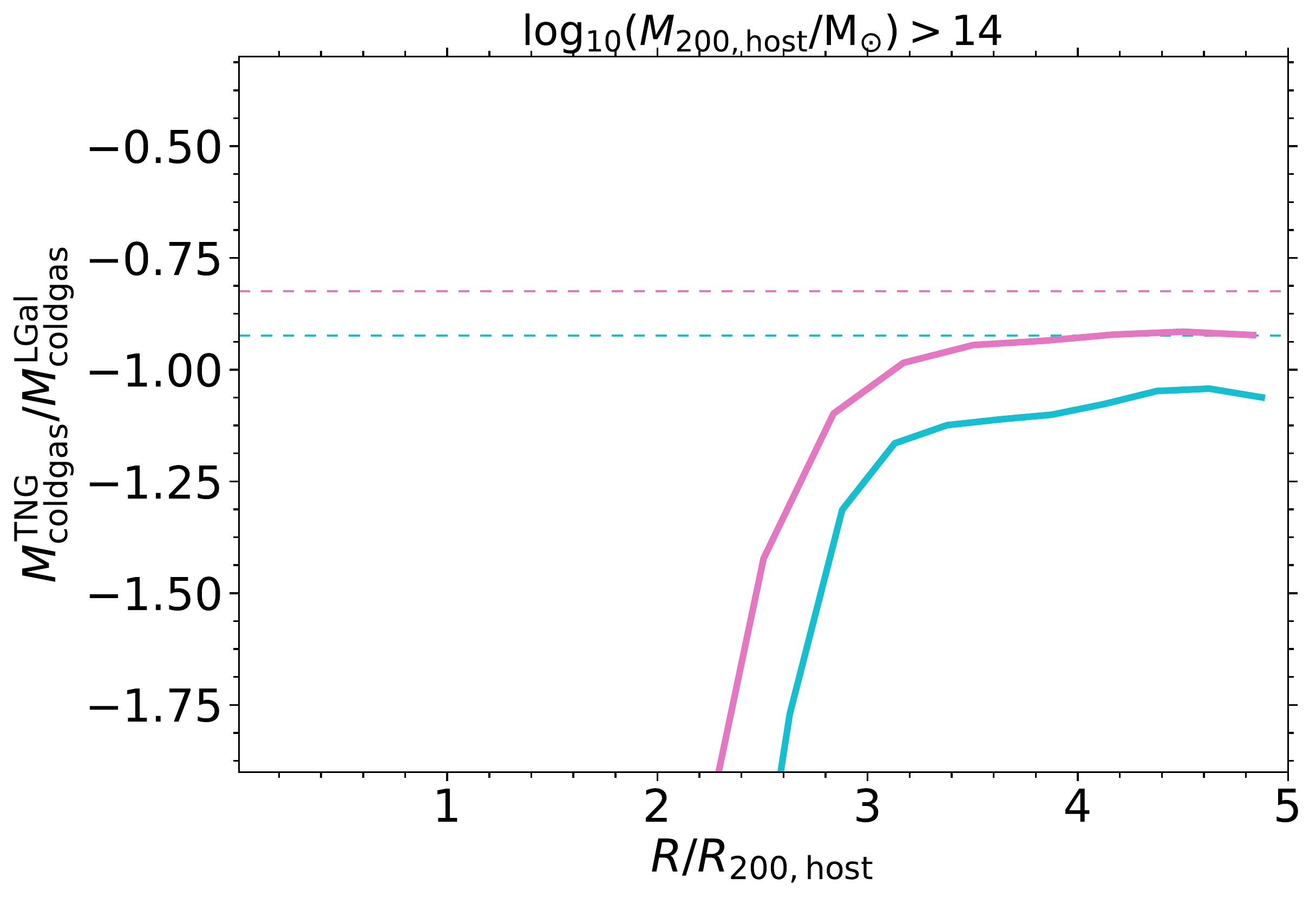}
    \caption{The ratio between cold gas masses in TNG and in  \textsc{L-Galaxies} for galaxies (both satellites and centrals) in the vicinity of haloes as a function of halocentric distance. Each panel is made for haloes in a different mass range as indicated. Solid coloured lines show the median ratios for different stellar mass ranges. Dashed lines denote the median ratio for each stellar mass range for the simulation as a whole.}
\label{Fig: coldgas_dis}
\end{figure*}

\begin{figure}
    \includegraphics[width=0.45\textwidth]{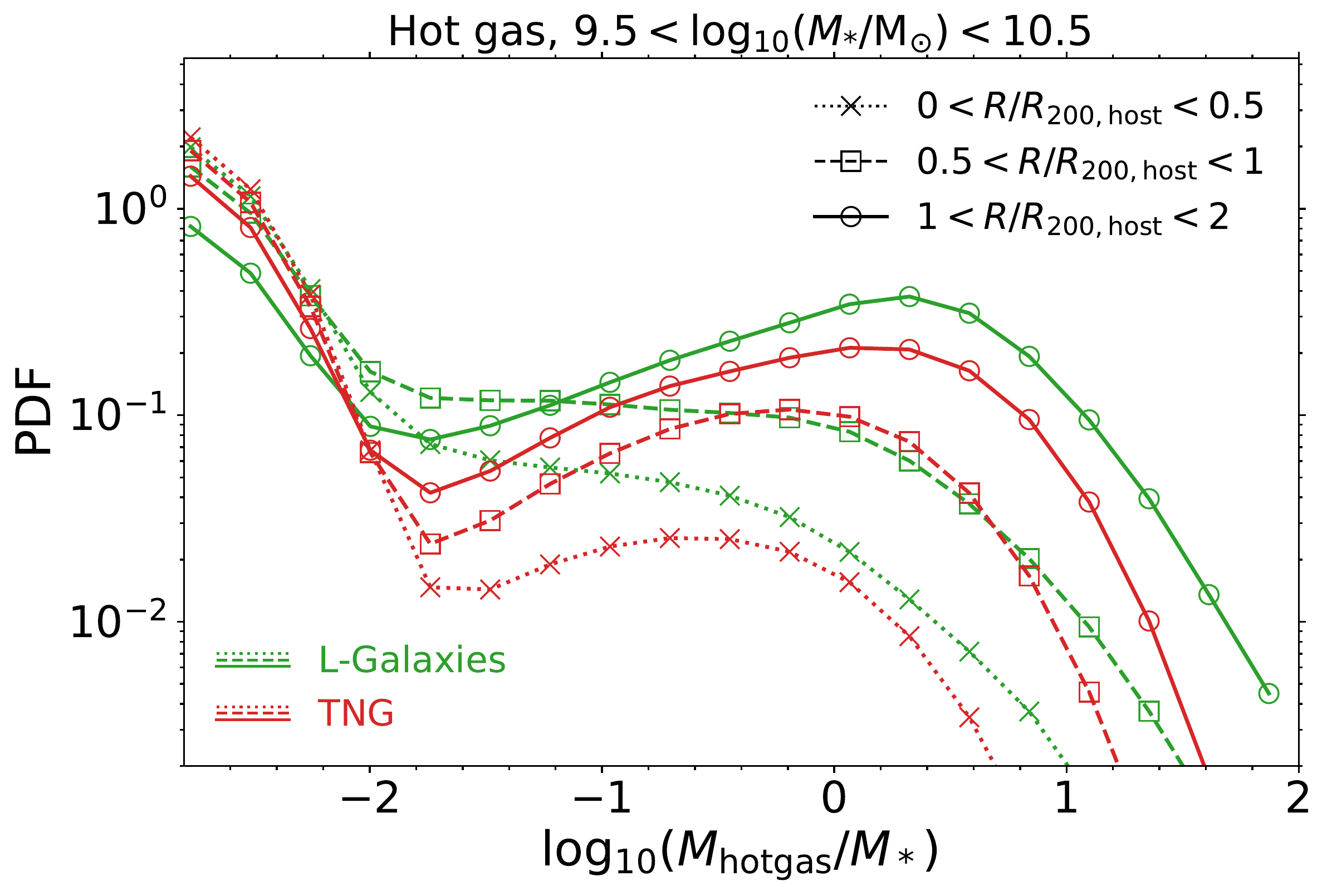}
    \includegraphics[width=0.45\textwidth]{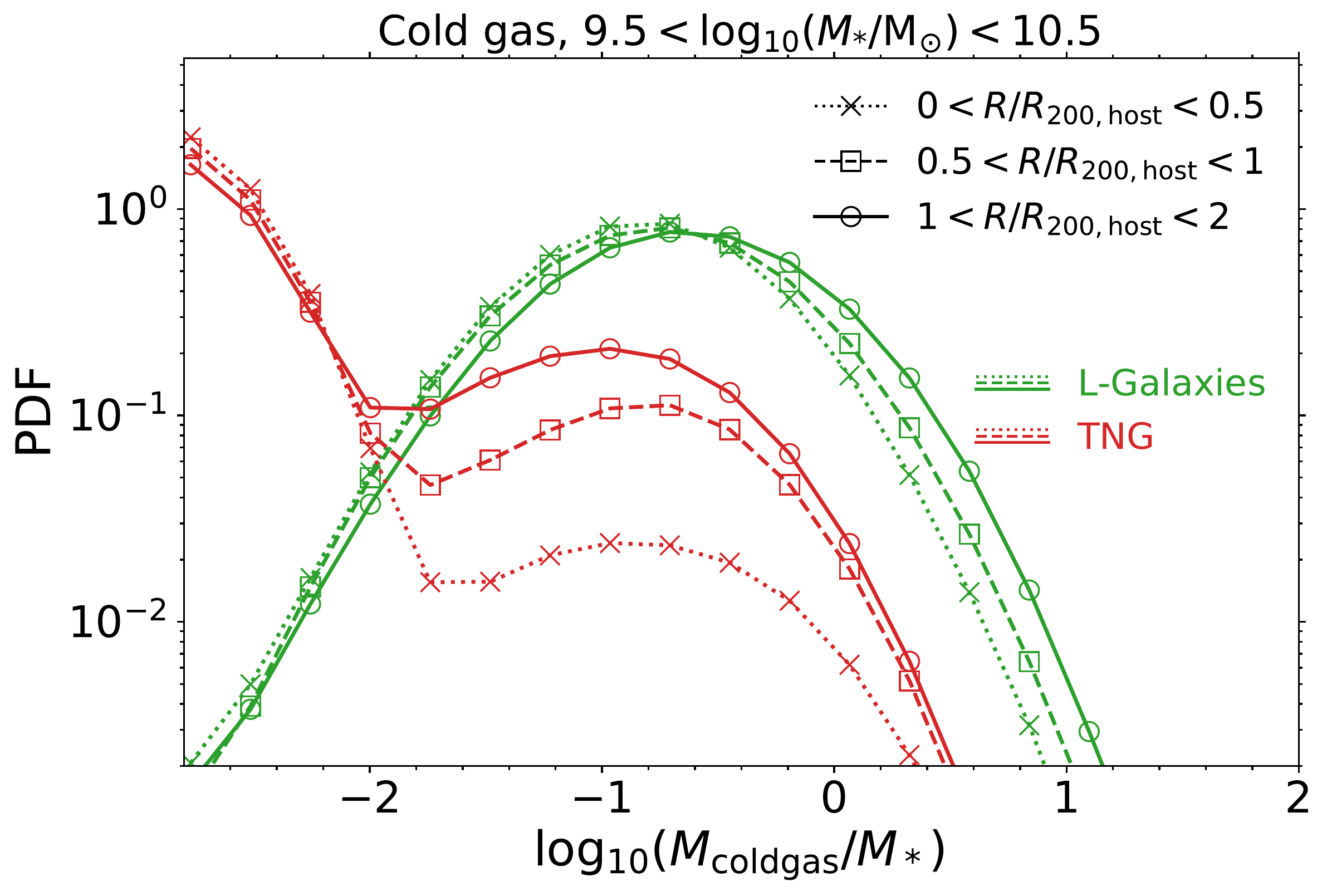}
    \caption{The ratio of hot (top) and cold (bottom) gas mass to stellar mass, for intermediate mass galaxies in the vicinity of massive clusters with $M_{200} / \rm M_{\odot} > 10^{14}$. Results are shown for \textsc{L-Galaxies} in green and for TNG in red. Different line styles refer to different distance ranges from cluster centre. Solid lines are for galaxies beyond $R_{200}$. These may not have passed through the cluster and may be central galaxies or satellites of other nearby haloes. All galaxies with zero gas or with logged ratios smaller than $-2.8$ are assigned to the leftmost bin.}
\label{Fig: lgaltng_sat_hotcoldgas_hist}
\end{figure}

The bottom panel of Fig. \ref{Fig: hotgas_dis} shows the same 2D view of the stellar mass versus host halo mass plane as previously, but now coloured by the hot gas to stellar mass ratio of galaxies. For \textsc{L-Galaxies} there is a sharp transition in the ratio at a mass of  $\log_{10}(M_{200} / \rm M_{\odot}) \gtrsim 14.08$.  This is due to  RPS acting only in haloes above this mass. In TNG (right) a large fraction of galaxies in groups are gas-poor, as they are affected by stripping both within the haloes and in the infall regions beyond the halo boundary. There is, however, a sharp boundary between stripped and non-stripped satellites which occurs at the same stellar mass-dependent value of halo mass as the quenching boundary in Fig. \ref{Fig: quenchedFrac_dis}.

Finally, Fig. \ref{Fig: coldgas_dis} shows, in the same format as Fig. \ref{Fig: hotgas_ratio_dis}, the variation with halocentric distance of the ratio of the cold gas mass in TNG to that in \textsc{L-Galaxies} for galaxies in the vicinity of haloes. In and around TNG clusters, most galaxies have low to no star-forming gas, while galaxies in groups and lower mass haloes are able to keep a fraction of their cold gas. On the other hand, galaxies in the vicinity of \textsc{L-Galaxies} haloes remain much more cold-gas-rich \footnote{We note that the ratios of the most massive galaxies are below the lower limit of Fig. \ref{Fig: coldgas_dis} and thus are not visible}. In the vicinity of clusters (right panel), the ratio is below its global value even for galaxies which reside further than $5R_{200}$ from halo centre. The ratio is also suppressed for galaxies near groups (middle panel), but extending out only to $\sim 1-3 R_{200}$. For satellites around low-mass hosts (left panel), the ratio is above its global value due to strong tidal stripping in \textsc{L-Galaxies}. We note that ram-pressure effects on the cold gas are not taken into account in \textsc{L-Galaxies} at any mass scale. In addition, \textsc{L-Galaxies} has a cold gas threshold, $M_{\rm crit}$, below which no star formation occurs, leading to a lower limit on the predicted cold gas fractions. However, both \textsc{L-Galaxies} and TNG have been shown to provide an adequate match to the observed HI mass function at $z=0$ \citep[see][]{henriques2015galaxy,diemer19}. 

So far, our analysis has focused on median gas content. Fig. \ref{Fig: lgaltng_sat_hotcoldgas_hist} shows for both models the full distributions of the ratios of hot (top) and cold (bottom) gas mass to stellar mass for intermediate-mass galaxies in the vicinity of massive clusters. We see that \textsc{L-Galaxies} (solid green lines) predicts more galaxies with a large hot gas mass fraction beyond the halo boundary than does TNG (solid red lines). Within $R_{200}$ (dashed and dotted lines) there are fewer gas-rich galaxies in both models, and more galaxies have little to no hot gas. The trends with distance imply that galaxies are more likely to be stripped during infall (i.e. at  $1 < R/R_{200} < 2$) in TNG than in \textsc{L-Galaxies}. 

The cold gas content of galaxies (the bottom panel of Fig. \ref{Fig: lgaltng_sat_hotcoldgas_hist}) differs more substantially between the two models, and also behaves differently as a function of clustercentric distance. A large fraction of TNG satellites have very little or no cold gas, producing a secondary population which piles up towards the left of this figure. The lack of cold gas stripping in \textsc{L-Galaxies}, together with the assumed cold gas mass threshold for star formation, implies that cold gas is not strongly depleted even in quenched satellites near cluster centre. The result is a uni-modal distribution with a well defined peak at $M_{\rm coldgas} / M_\star \sim 0.2$ which shifts only slightly to lower values near cluster centre. Thus in \textsc{L-Galaxies} this distribution changes little with distance (shown by the different line styles), while for TNG the amplitude of the cold-gas-rich peak drops substantially towards cluster centre.


\section{summary and discussion}
\label{sec: summary}

In this work we have compared, object-by-object and statistically, the \textsc{L-Galaxies} semi-analytical model with the IllustrisTNG magnetohydrodynamical simulation. To do so, we have run \textsc{L-Galaxies} on subhalo merger trees from the TNG-DMO simulation. Comparing the properties of galaxies and haloes between the two models, we find that:

\begin{itemize}
    \item The stellar mass functions (SMF) of the two models agree with each other and with observations (at a level better than $0.2$ dex) since $z=3$. The stellar masses of matched galaxies are also in good agreement with a typical scatter of $\sim 0.2$ dex, independent of redshift. The stellar mass to halo mass (SMHM) relation shows the stellar mass fraction of haloes to be generally larger in TNG than in \textsc{L-Galaxies} (up to 60\%, Fig. \ref{Fig: SMF}).
    
    \item By $z=0$ TNG galaxies become predominantly quenched for $\log_{10}(M_{\star} / \rm M_{\odot}) \geq 10.5$ whereas in \textsc{L-Galaxies} this characteristic mass scale is larger, $\log_{10}(M_{\star} / \rm M_{\odot}) \geq 11$. As a result, the specific star formation rate (sSFR) of galaxies is lower in TNG than in \textsc{L-Galaxies} over the transition regime, $10 \lesssim \log_{10}(M_{\star} / \rm M_{\odot}) \lesssim 11$ (Fig. \ref{Fig: ratio_SSFRs}). In addition, TNG predicts a higher fraction of quenched galaxies at higher redshifts, in particular at $z\geq2$ (Fig. \ref{Fig: quenchedfrac}). 
    
    \item The models agree on the amount of hot (non-star-forming) gas around massive galaxies ($\log_{10}(M_{\star} / \rm M_{\odot}) \gtrsim 11.5$). Intermediate-mass galaxies ($10.5 \lesssim \log_{10}(M_{\star}/\rm M_{\odot}) \lesssim 11.5$) are more hot-gas-rich in \textsc{L-Galaxies}. Low-mass galaxies ($\log_{10}(M_{\star} / \rm M_{\odot}) \lesssim 10.5$) are more hot-gas-rich in TNG (Fig. \ref{Fig: Hot_Cold_Gas_ratio}). 
    
    \item A large fraction of quiescent galaxies in TNG have little or no cold (star-forming) gas, whereas those in \textsc{L-Galaxies} are generally more cold-gas-rich at $z=0$. For galaxies with $\log_{10}(M_{\star} / \rm M_{\odot}) \lesssim 10.5$,  $M_{\rm coldgas}^{\rm TNG}/M_{\rm coldgas}^{\rm LGal}$ decreases with cosmic time (Fig. \ref{Fig: Hot_Cold_Gas_ratio}).
    
    \item In clusters with $\log_{10}(M_{200} / \rm M_{\odot}) \gtrsim 14$, both models predict halo baryon fractions near the cosmic value. For groups with $12\lesssim \log_{10}(M_{200} / \rm M_{\odot}) \lesssim 14$, TNG haloes have a lower baryon fraction than \textsc{L-Galaxies} haloes. At $\log_{10}(M_{200} / \rm M_{\odot}) \sim 12$, TNG haloes reach half the cosmic baryon fraction due to ejective feedback, whereas \textsc{L-Galaxies} haloes remain at the cosmic value (Fig. \ref{Fig: baryonFrac_LGal_TNG}).
\end{itemize}

The differences above are mainly a result of the differing treatments of stellar and black hole feedback between the two models. In TNG the mass ejected by both kinds of feedback is spatially resolved. In \textsc{L-Galaxies} only supernovae eject gas, and the exact location of the ejecta is not specified. This material is kept unavailable for cooling until it returns to the subhalo and is reincorporated in the hot gas component. Unlike in TNG, it does not contribute to processes like ram-pressure stripping at large radii from halo centre.

The characteristic stellar mass scale where the population transitions from star-forming to quenched reflects black hole feedback prescriptions and is 0.5 dex higher in \textsc{L-Galaxies} than in TNG. Such feedback can push gas well beyond the halo boundary in TNG but not in \textsc{L-Galaxies}. This redistribution of baryonic mass within and outside massive dark matter haloes is a key prediction of the TNG model.
    
Galaxies differ more strongly between the two models in and around dark matter haloes, due to differing treatments of environmental effects. For galaxies in the vicinity of haloes, we find that:

\begin{itemize}
    \item The stellar mass ratio, $M_{\star}^{\rm TNG}/M_{\star}^{\rm LGal}$, increases near the centres of massive haloes, whereas the ratio $\rm sSFR^{\rm TNG}/sSFR^{\rm LGal}$ decreases (Fig. \ref{Fig: StellarMass_SSFR_ratio_dis}). Galaxies in the vicinity of clusters and groups are more frequently quenched than field galaxies in both models, but more so in TNG. The influence of clusters on quenching extends to larger radii in TNG than in \textsc{L-Galaxies} (Fig. \ref{Fig: quenchedFrac_dis}). 
    
    \item The ratio between the hot gas masses predicted for galaxies by the two models, $M_{\rm hotgas}^{\rm TNG}/M_{\rm hotgas}^{\rm LGal}$, falls below its average value in the vicinity of clusters and groups (Fig. \ref{Fig: hotgas_ratio_dis}), as do the mass ratios for cold (star-forming) gas (Fig. \ref{Fig: coldgas_dis}). Overall, environment imprints a stronger signature on galaxies in the vicinity of haloes in TNG.  Gas stripping beyond $R_{200}$ is substantial in TNG but weak in \textsc{L-Galaxies}.
    
    \item Massive galaxies ($\log_{10}(M_{\star} / \rm M_{\odot}) \gtrsim 10.5$) in the vicinity of clusters and groups are less gas-rich and less star-forming in TNG than in \textsc{L-Galaxies} (Figs. \ref{Fig: StellarMass_SSFR_ratio_dis}, \ref{Fig: hotgas_ratio_dis}, \ref{Fig: hotgas_dis}). This reflects pre-existing differences in the field, together with an enhancement of ram-pressure stripping due to AGN feedback, which makes extended subhalo gas less bound and thus more easily stripped.
\end{itemize}

These environmental differences arise primarily because not all relevant stripping processes are implemented in \textsc{L-Galaxies} (e.g. ram-pressure stripping of cold gas is not included) or are intentionally restricted (e.g. ram-pressure stripping of hot gas is assumed to be negligible in galaxy groups). Our comparisons with TNG suggest that these assumptions need to be re-evaluated, and may need to be modified if the stripping levels seen in TNG are realistic. In recent modelling we have updated \textsc{L-Galaxies} to include a new method to extend ram-pressure stripping of hot gas to all galaxies, even those outside clusters, based on local measurements of the particle density and velocity fields \citep{ayromlou2019new}. In the future, we will study the degree to which such prescriptions can align the environmental effects found in semi-analytic versus  hydrodynamical simulations, as explored in this work.

Careful comparison with observations of the sSFR and gas fractions of galaxies in and around clusters will highlight the strengths and limitations of each simulation technique and allow both to achieve a new degree of physical realism. In addition, we expect that our comparisons between SAMs and hydrodynamical simulations  techniques will become more important in the future development of galaxy evolution models to be run in Gigaparsec simulation volumes.


\section*{Acknowledgements}

MA thanks Volker Springel, Annalisa Pillepich, Ruediger Pakmor and Bruno Henriques for useful discussions and assistance. MA also thanks Seunghwan Lim for providing SZ halo gas fraction observational data and useful discussions. MR acknowledges support by Argonne National Laboratory supported under the U.S. Department of Energy contract DE-AC02-06CH11357.
The analysis herein was carried out on the compute cluster of the Max Planck Institute for Astrophysics (MPA) at the Max Planck Computing and Data Facility (MPCDF). The TNG simulations were run with compute time granted by the Gauss Centre for Supercomputing (GCS) under Large-Scale Projects GCS-ILLU and GCS-DWAR on the GCS share of the supercomputer Hazel Hen at the High Performance Computing Center Stuttgart (HLRS).

\vspace{-1em}
\bibliographystyle{mnras}
\bibliography{refbibtex}

\label{lastpage}
\end{document}